\documentclass[12pt]{report}

\usepackage[utf8]{inputenc}
\usepackage[french]{babel}
\usepackage[T1]{fontenc}
\usepackage{graphicx}
\usepackage{hyperref}
\usepackage{float}

\usepackage{amsmath,mathrsfs,amssymb}
\usepackage[french,linesnumbered,plain,fillcomment]{algorithm2e}
\usepackage{lipsum,xcolor}
\everymath{\displaystyle}

\SetKw{WEntree}{\textcolor{red}{Entrée}}
\SetKw{WEntrees}{\textcolor{red}{Entrées}}
\SetKw{WSaisir}{Saisir}
\SetKw{WInitialisation}{\textcolor{red}{Initialisation}}
\SetKw{WTraitement}{\textcolor{red}{Traitement}}
\SetKw{WAssigne}{\textcolor{blue}{prend la valeur}}
\SetKw{WSortie}{\textcolor{red}{Sortie}}
\SetKw{WSorties}{\textcolor{red}{Sorties}}
\SetKwIF{Si}{SinonSi}{Sinon}{Si}{alors}{sinon si}{sinon}{FinSi}
\SetKwFor{Tq}{Tant que}{faire}{FinTantque}
\SetKwFor{Pour}{Pour}{faire}{FinPour}
\SetKw{WAfficher}{Afficher}

\makeatletter
\newcounter {subsubsubsection}[subsubsection]
\renewcommand\thesubsubsubsection{\thesubsubsection .\@alph\c@subsubsubsection}
\newcommand\subsubsubsection{\@startsection{subsubsubsection}{4}{\z@}%
                                     {-3.25ex\@plus -1ex \@minus -.2ex}%
                                     {1.5ex \@plus .2ex}%
                                     {\normalfont\normalsize\bfseries}}
\newcommand*\l@subsubsubsection{\@dottedtocline{3}{10.0em}{4.1em}}
\newcommand*{\subsubsubsectionmark}[1]{}
\makeatother

\usepackage{multirow}
\usepackage{algpseudocode}
\usepackage{slashbox}
\usepackage{url}

\usepackage{geometry}
\begin{document}

\thispagestyle{empty}
\begin{center}
\textsc{République Tunisienne} \\
\textsc{Ministère De l'Enseignement Supérieur Et De La Recherche Scientifique} \\
\textsc{\textbf{Université De Tunis El Manar}} \\
\textsc{\textbf{Faculté Des Sciences De Tunis}} 

\begin{figure}[!ht]
\begin{center}

\includegraphics[scale=0.4]{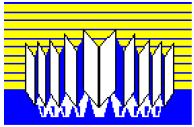}

\end{center}
\end{figure}

\textsc{Département Des Sciences de l'Informatique} \\

\vspace{2\baselineskip}

\LARGE \textbf{\textbf{MÉMOIRE DE MASTÈRE}} \normalsize \\

\vspace{2\baselineskip}

\textit{Présenté en vue de l'obtention du }\\
\textbf{Diplôme de Mastère de recherche en Informatique} \\ 
\vspace{2\baselineskip}

\textbf{Emna JEBRI} \\
\emph{Licence Fondamentale en Informatique de Gestion} \\
\textsc{Faculté des Sciences Économiques et de Gestion de Tunis}

\vspace{\baselineskip}

\rule{0.95\textwidth}{1.75pt} \\ \vspace{0.9\baselineskip}
\LARGE \textbf{PROPOSITION D'UNE NOUVELLE MÉTHODE DE GÉNÉRATION DE REQUÊTES \textbf{SPARQL} A PARTIR DE GRAPHES \textbf{RDF}} \normalsize  
\rule{0.95\textwidth}{1.75pt}

\end{center}
\begin{center}
Soutenu le 21/02/2018 devant le jury composé de : 
\end{center}
\begin{center}
\begin{tabular}{ll}

\emph{\textbf{Président}} & M.  \textsc{Faouzi MOUSSA}, Professeur à la FST  \\
\emph{\textbf{Rapporteur}} & M.  \textsc{Sami ZGHAL}, Maître-Assistant à la FSEGJ \\ 
\emph{\textbf{Directeur}}  & M.  \textsc{Sadok BEN YAHIA}, Professeur à la FST \\ \\

\end{tabular}
\end{center}

\begin{center}
\textsc{Laboratoire en Informatique en Programmation Algorithmique et Heuristique} \textbf{- LIPAH -} \\
 2017 - 2018 \\

\end{center}

\newpage
\thispagestyle{empty}

\newpage
\strut 
\newpage

\thispagestyle{empty}
\vspace*{\stretch{1}}
{\fontfamily{pzc}\selectfont
\hfill À moi-même

\hfill À mon père

\hfill À ma mère

\hfill À toute ma famille

\hfill Et mes amis

}
\vspace*{\stretch{1}}

\begin{titlepage}
\begin{center}
\line(1,0){300} \\
\huge{\textbf{REMERCIEMENTS }} \\ 

\line(1,0){200} \\
\end{center}

Je remercie Monsieur Sadok \textsc{BEN YAHIA}, Professeur à la Faculté des Sciences de Tunis, de m'avoir guidé tout au long de ce mémoire, de m'avoir apporté ses précieuses lumières, et surtout de m'avoir appris à mener par moi-même un travail de recherche. Sans ses conseils, ce mémoire n'aurait tout simplement pas existé. \\ \\

Je remercie également Monsieur Faouzi \textsc{MOUSSA}, Professeur à la Faculté des Sciences de Tunis, de m'avoir fait l'honneur de présider le jury de ma soutenance. \\ \\

Je tiens à remercier Monsieur Sami \textsc{ZGHAL}, Maître-Assistant à la Faculté des Sciences juridique Économique et de Gestion de Jendouba, d'avoir accepté d'être membre du jury en tant que rapporteur.

\end{titlepage}

\pagenumbering{roman}
\tableofcontents
\thispagestyle{empty}
\cleardoublepage
\addcontentsline{toc}{chapter}{Table des figures}
\listoffigures

\listoftables
\addcontentsline{toc}{chapter}{Liste des tableaux}
\cleardoublepage



\pagenumbering{arabic}
\addcontentsline{toc}{chapter}{Introduction Générale}
\chapter*{Introduction Générale}

\parindent =0.5cm
Le Web sémantique permet aux machines de comprendre la sémantique, le sens de l'information sur le Web. Il permet aux agents logiciels d'accéder plus intelligemment aux diverses sources de données existantes sur le Web. Ainsi, la sémantique du contenu des ressources dans le Web tente à rendre les données sous une représentation formelle et standardisée. La description de la sémantique dans le contexte du Web sémantique est effectuée par l'intermédiaire des ontologies (ou bases de connaissances). L'ontologie est la base et l'appui de ce que nous appelons la description de connaissances. \\

Les ontologies ont été reconnues comme une composante essentielle pour le partage des connaissances et la concrétisation de la vision du Web sémantique. En définissant les concepts associés à des domaines particuliers, elles permettent à la fois de décrire le contenu des sources à intégrer et d'expliciter le vocabulaire à travers des requêtes lancées par des utilisateurs. \\ 

\parindent=0.5cm
Les objets ainsi que les relations dans l'univers du discours sont conceptualisés dans le but de les décrire sous une forme utilisable par des machines. Une des caractéristiques importantes du \textit{WS}, est son hétérogénéité qui décrit à la fois sa richesse et son ambiguïté. Cette hétérogénéité s'explique par le fait que les informations et les connaissances peuvent être produites à partir de plusieurs sources diverses. De même, elles peuvent être exprimées dans des différents formats. \\

\parindent=0.5cm
Le nombre croissant des ensembles de données publiés sur le Web, offre à la fois des possibilités de disponibilité élevée des données, ainsi que des défis inhérents lors de l'interrogation de ces données dans un environnement sémantiquement hétérogène et distribué. \\
De même, RDF s'est imposé comme un modèle de données standard. De considérables quantités de données RDF sont désormais disponibles. \\

\parindent=0.5cm
Pour interroger les données liées sur le Web aujourd'hui, les utilisateurs doivent d'abord savoir quels ensembles de données exposés contiennent potentiellement les données qu'ils souhaitent avoir, et quel modèle de données décrit ces ensembles de données, pour qu'ils puissent utiliser cette information afin de créer des requêtes structurées. \\
En plus de la connaissance du modèle de données, les utilisateurs qui interrogent des informations liées, doivent maîtriser la syntaxe des langages d'interrogation hautement spécifiques tels que \textit{SPARQL}.  \\

\parindent=0.5cm
Ces difficultés d'interrogation des données peuvent nous amener au problème suivant : alors que les usagers souhaitent avoir des réponses à leurs besoins, et même des réponses de qualité, il n'en obtiennent aucunes. \\ 

\parindent=0.5cm
Vis-à-vis de ce problème, l'utilisateur ne peut pas réellement savoir si sa requête est mal formulée, trop sélective ou bien si le résultat souhaité n'existe tout simplement pas dans l'ontologie. Il peut ainsi se lancer dans un processus d'essai et d'erreur pendant lequel il tente de modifier et de remettre optionnelles les conditions de sa requête. Ainsi, il analyse les résultats obtenus, afin d'aboutir à une reformulation lui permettant d'avoir des résultats satisfaisants. \\
Ce processus semble long et fastidieux, ainsi que les résultats retrouvés peuvent ne pas être ceux souhaités. Cela peut également causer une frustration des utilisateurs étant donné que ces derniers ne comprennent même pas pourquoi, avec un aussi grand volume de données, leurs requêtes ne retournent aucun résultat. Alors, nous estimons que les utilisateurs ont besoin d'un appui afin de formuler des requêtes significatives et de retrouver des réponses satisfaisantes. Pour l'utilisateur, les requêtes en langage naturel apparaissent comme une alternative simple et intuitive. \\

\parindent=0.5cm
D'un autre côté, avec l'avènement des bases de données \textit{NoSQL}, d'autres pistes de prises en charge ont été conçues. En particulier, les BD graphiques sont maintenant utilisées dans des secteurs aussi divers que la santé, le commerce de détail, le pétrole et le gaz, les médias, les jeux vidéo et bien d'autres. \\

Parmi les avantages des BD, nous pouvons citer : évolutivité, efficacité de stockage, structuration idéale permettant un temps optimal d'exécution de requêtes utilisateurs, gestion des droits d'accès. Les BD offrent aussi des mécanismes de contrôle et de sécurité. \\ \\

\parindent =1cm
\textbf{Structure de ce mémoire} \\

Les résultats de nos travaux de recherche sont synthétisés dans ce mémoire qui est composé de quatre chapitres : \\

\parindent =0.5cm
\textbf{Le premier chapitre} introduit les notions de base pour limiter le champ d'étude. En fait, nous allons présenter une étude bibliographique sur les différentes étapes de l'évolution du Web, et plus précisément le Web sémantique, l'ontologie, ses composants ainsi que les langages avec lesquels elle est représentée. La dernière section élucide la notion d'hétérogénéité et liste brièvement ses divers types. \\

\textbf{Le deuxième chapitre} décrit les standards, protocoles et outils utilisés pour le stockage et l'interrogation des données RDF. Ensuite, une description détaillée sur les\textit{ BD-RDF }et le langage \textit{SPARQL}, a été présentée. Enfin, nous avons présenté une revue sur les différents travaux de comparaison de \textit{BD-RDF }ainsi que sur quelques méthodes existantes pour l'interrogation \textit{SPARQL} de graphes RDF. \\

\textbf{Le troisième chapitre} est consacré à la description de notre nouvelle vision d'interrogation de graphe RDF généré à partir d'une ontologie d'entrée. Il présente les différentes phases constituants notre méthode, avec quelques exemples d'exécution. \\

\textbf{Le quatrième chapitre} décrit les expérimentations effectuées ainsi que l'environnement de réalisation de la méthode développée. Enfin, une description détaillée sur les mesures retrouvées, a été présentée. \\ 

\parindent=0.5cm
Le mémoire se termine par une conclusion générale qui récapitule l'ensemble de nos travaux et présente également quelques perspectives futures de recherche.

\chapter{Fondement du Web sémantique} 

\section{Introduction}

\parindent =0.5cm Avec l'évolution du Web et l'explosion du volume d'information, les ontologies sont devenues un concept primordial pour le Web sémantique. Elles cherchent à exporter les données sous une forme structurée (limite semi-structurée) avec une sémantique bien définie. Cela permet aux agents de consommer ces données et de les utiliser d'une manière plus intelligente. Les ontologies ont pour objectif, aussi, de permettre aux programmes d'assurer des inférences basées sur des règles afin de déduire des faits, faire des raisonnements ou examiner la cohérence de l'ensemble de données. \\

Le reste de ce chapitre est organisé comme suit: d'abord, nous allons définir le Web sémantique, voir ses défis et ses objectifs. Ensuite, nous rappellerons la définition du terme ontologie. Enfin, nous détaillerons les langages de représentation de cette dernière.

\section{L'évolution du Web sémantique}

\par Dés sa création, au début des années 90, l'objectif du Web se basait essentiellement sur l'accès et le partage d'un grand volume d'information, sans aucune maîtrise/connaissance du contenu. \\
Sa nature, sa structure et son usage ont évolué au fil du temps. Dans ce contexte, nous citons les quatre étapes de son évolution \cite{weeb} : \\
\parindent =0cm
\\ - \textbf{Web 1.0,} ou encore Web traditionnel. C'est un Web statique (ou de documents) qui se caractérisait par des sites web orientés produits et par un contenu plutôt limité (en général de type texte et multimédia), créé surtout par des utilisateurs professionnels. Il nécessite aucune intervention utilisateur. L'utilisateur n'était qu'un consommateur passif.

En effet, le Web « read only » était très consultatif et à sens unique. Le lecteur n'intervenait que d'une façon très minime et ne pouvait pas réagir en temps réel ou contribuer au contenu qu'il abordait. C'était une sorte de magazine énorme en ligne à consulter. \\

- \textbf{Web 2.0,} appelé également Web social \textit{(Vickery andWunsch-Vincent (2007))}, et change totalement de perspective. Il favorise plutôt la dimension du partage et d'échange d'informations et de contenus. Le Web se dynamise. Il se caractérise par un contenu plutôt illimité créé par des utilisateurs professionnels et amateurs. \\

Ce Web est devenu à double sens. En effet, l'usager devient co-développeur, \textit{i.e.} à la fois consommateur et acteur. Mais pour être précis, seule une minorité prend le rôle du producteur (read et write), car la majorité se contente de consulter et partager (read et share). Ainsi l'utilisateur intervient énormément : il peut consulter, réagir en temps réel, partager, et contribuer au contenu qu'il aborde\textit{, etc.} Il devient la source du contenu (même professionnel), et éventuellement la source de l'évolution de ce contenu (il peut modifier le contenu). Citons comme exemple, \textit{Youtube} qui, avec lequel, l'utilisateur peut envoyer des vidéos (du contenu), et également, les modifier (adapter le contenu). \\

Il y a aussi une émergence des applications participatives, surtout les \textit{UGC}s (les contenus créés par les utilisateurs) pour communiquer, partager, tagger, et stocker du contenu (comme les blogs, les sites e-commerce, les forums, et les réseaux sociaux, \textit{etc}). \\
L'avis de l'utilisateur est constamment sollicité, ainsi, nous avons vu apparaître les commentaires et les notes sur les articles, preuve qu'il est fondamental de tenir en considération l'impression du lecteur. \\
le Web s'est transformé d'un espace pour la simple consultation des données, à un lieu d'expression riche. \\

- \textbf{Web 3.0,} initié à l'époque par \textit{Berners-Lee et al.(2001)} comme le Web sémantique (ou Smart Web). C'est un Web focalisé sur le savoir. Il consiste à rendre les ressources Web, non seulement compréhensibles par les humains, mais également par les machines. Il tente de fournir un sens aux données afin de répondre aux besoins des utilisateurs mobiles toujours connectés à travers une multitude de supports d'application. \\

- \textbf{Web 4.0,} et également connu comme étant le Web symbiotique ou Web intelligent. C'est un Web qui connectera l'intelligence. Il est orienté interaction individus/objets, et ses données vont évoluer vers des standards ouverts et un langage universel.
Il vise à tout innover grâce aux connexions intelligentes, et à immerger l'individu dans un environnement Web de plus en plus prégnant.
Il est appelé Web symbiotique car il est prévu que l'esprit humain et les machines vont interagir en symbiose, \textit{i.e.} l'utilisateur deviendra créateur en constante symbiose avec son environnement. \\
Il représente une continuité de la voie du Web 3.0, tout en posant de considérables questions telles que la protection de la vie privée, le contrôle des données\textit{, etc}. \\ \\

\begin{figure}[H]
\begin{center}

\includegraphics[scale=0.3]{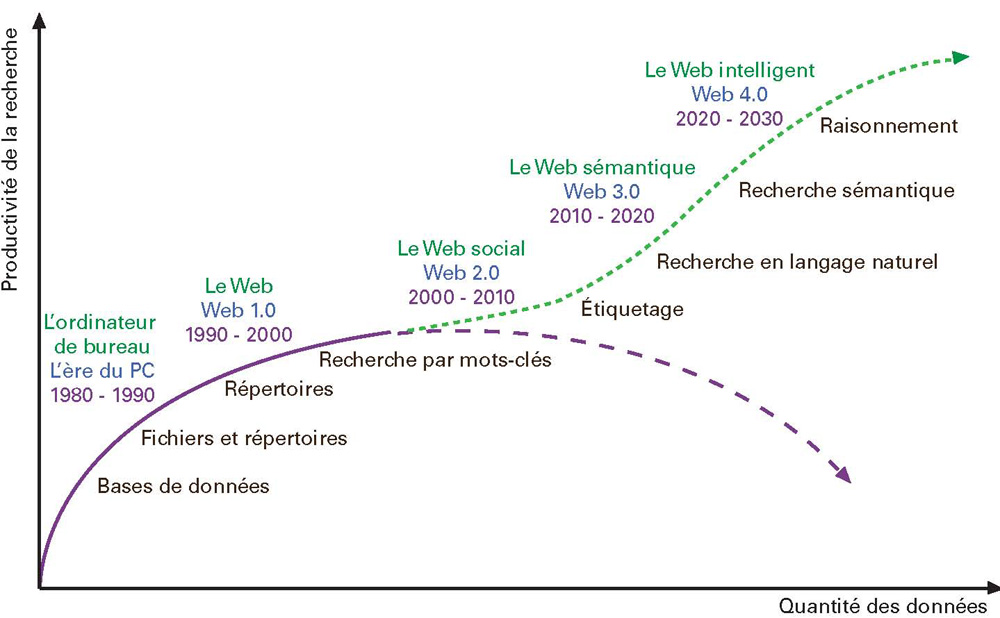}

\end{center}
\caption{L'évolution du Web \label{figure 1.1}}
\end{figure}

\newpage
\parindent =0.5cm La figure \ref{figure 1.1} \footnote{http ://www.camillejourdain.fr/schemas-evolution-du-web/ Consulté le 11/10/2017} montre les différentes étapes précitées de l'évolution du Web. 

\parindent =0cm 
Au stade du \textit{Web 1.0}, nous remarquons que le volume d'information reste minime et proportionnel à la productivité, ainsi qu'au nombre d'utilisateurs. 

A l'arrivée du \textit{Web 2.0}, le volume d'information produite accroît suivant l'accroissement du nombre d'internautes. 

Avec l'apparition du \textit{Web 3.0}, nous constatons l'augmentation du nombre d'utilisateurs et du volume de données produites (\textit{e.g.} un accroissement énorme du nombre de sites (555 millions de sites + 300 millions nouveaux en 2011), une augmentation du nombre d'usagers \textit{Facebook} (200 millions nouveaux au même date), \textit{etc}). Par conséquent, les moteurs de recherche traditionnels ne suffisaient plus. 

Finalement, avec l'avènement du \textit{Web 4.0}, nous remarquons l'existence des agents intelligents.
Cette vision reste futuriste. \\

\subsection{Définition}

\parindent=0.5cm
Le Web sémantique est un web intelligent qui permet, non seulement de stocker les informations, mais également de les rendre compréhensibles par les ordinateurs et réutilisables par différentes applications. C'est une nouvelle génération du Web où l'information est structurée par les connaissances. Ceci permet les agents logiciels d'effectuer des actions complexes pour leurs utilisateurs, grâce à une inter-opération des services Web. \\

De la même manière que le Web actuel, le Web sémantique est construit principalement autour des identifiants (\textit{URI}s : \textbf{U}niform \textbf{R}esource \textbf{I}dentifier) et du protocole \textit{HTTP} (\textbf{H}yper\textbf{T}ext \textbf{T}ransfer \textbf{P}rotocol), mais il est par contre basé sur le langage \textit{RDF} (\textbf{R}esource \textbf{D}escription \textbf{F}ramework) et non plus sur \textit{HTML} (\textbf{H}yper\textbf{T}ext \textbf{M}arkup \textbf{L}anguage). Cela est pour but de séparer l'information qui décrit le sens et le contexte des données, de l'information qui décrit la présentation des données. Il est basé principalement sur les bases de connaissances et non plus sur les bases de données. La recherche aussi va s'affecter et devenir une recherche par concept, non plus par mot-clé. \\

Dans le Web sémantique, toutes les données du Web, textuelles ou multimédia, doivent être annotées sémantiquement par des méta-données pertinentes, car les machines ne peuvent traiter les ressources Web qu'à travers une explication plus spécifique de leur contenu et cela en utilisant un mark-up sémantique nommé « méta-données » qui aidera les agents logiciels à comprendre les données et à prendre des décisions. L'annotation de ces ressources d'information repose sur l'accès à des représentations de connaissances (des ontologies) échangées sur le Web et qui sont en train de s'intégrer de plus en plus dans les applications Web, sans même que les utilisateurs ne le sachent.
\\

Pour conclure, l'objectif principal du Web sémantique est de transformer la masse non-maîtrisable des pages Web, en un énorme index hiérarchisé afin de garantir un réel partage de connaissances. \\

\subsection{Historique}
\par Le Web sémantique n'est pas pour autant une nouvelle idée. La notion a été décrite premièrement dans les années 1990. En 1993, \textit{Tim Berners-Lee} fournit une solution au problème du partage de connaissances entre les applications Web, à l'aide d'un mécanisme à base d'ontologies qui structure les données d'une manière compréhensible par la machine. \\

Le vrai départ du Web sémantique a eu lieu en 2001 par \textit{Berners-Lee et al}. \\
Le mot sémantique tente de faire référence à la définition traitée par le domaine de la logique de descriptions, dont le but est de faire apparaître du sens à travers la structuration des informations et l'expression de la logique qui les relie. \\
 \\ Alors, quelles sont les défis relevés par le Web sémantique ?

\subsection{Les défis du Web sémantique}

\parindent=0.5cm
Afin d'être à la hauteur de ses promesses, le Web sémantique doit faire face à toutes ces questions \cite{defis} : \\

 \parindent =0cm \textbf{Immensité :} le \textit{World Wide Web} contient plusieurs milliards de pages. A l'arrivée du Web sémantique, tous les systèmes de raisonnement automatisé devraient gérer une quantité de données vraiment immense. \\
 
\textbf{Imprécision :} il existe des notions imprécises comme "grand" ou "jeune". Ceci est produit de l'imprécision des requêtes des utilisateurs. La logique floue est le remède pour faire face à ce défi. \\

\textbf{Incertitude :} il peut exister des concepts précis avec des valeurs incertaines. A titre d'exemple, un patient expose un ensemble de symptômes qui conviennent avec un certain nombre de diagnostics différents, chacun avec une probabilité différente. \\ 

\textbf{Incohérence :} ce sont des contradictions logiques qui apparaissent fatalement au cours du développement des grandes ontologies venant de sources diverses. 


\subsection{Les objectifs du Web sémantique}

De nos jours, le souci du Web n'est plus vraiment l'augmentation continuelle de sa taille d'informations, mais plutôt l'amélioration de la recherche dans cette énorme masse d'informations, et la réalisation de systèmes permettant de filtrer et délivrer les informations de façon « intelligente ». \\

Le Web actuel est un ensemble de documents (données et pages) dédié aux humains, stocké et manipulé d'une façon purement syntaxique. Les deux principaux problèmes du Web actuel sont : \\
- D'une part, il y a énormément de sources de données, du fait que n'importe qui peut facilement publier un contenu (sachant qu'il n'a pas la moindre idée sur la probabilité que ce contenu soit trouvé par autrui) ; il n'a qu'à l'annoter. Ainsi, les moteurs de recherche à base de mot-clé auront la tâche de l'indexer pour pouvoir l'afficher aux utilisateurs lorsqu'ils font une recherche. Par conséquent, l'information sur Internet est tellement énorme que l'utilisateur a beaucoup de difficultés à la retrouver. \\

\parindent=0cm
- D'une autre part, les résultats de recherche sont imprécis, très sensibles au vocabulaire, et assez longs à trouver. En effet, les moteurs de recherche ne sont capables de répondre qu'à deux questions principales : 

\parindent=0.5cm
 Quelles sont les pages contenant ce terme ? et ;

\parindent=0.5cm
 Quelles sont les pages les plus populaires à ce sujet ? \\

\parindent=0cm
Le Web est essentiellement syntaxique, et l'Homme est le seul à pouvoir interpréter le contenu (des documents et des ressources) inaccessible et non interprétable par la machine. Lui seul doté de la capacité de comprendre ce qu'il a trouvé et décider en quoi cela se rapporte à ce qu'il veut vraiment chercher. Finalement, nous ne pourrons pas se passer de l'intervention humaine pour naviguer, chercher, faire le tri de documents manuellement, interpréter et combiner les résultats. \\

 \parindent =0.5cm L'affaire principale du Web sémantique est de diriger l'évolution du Web afin de permettre aux utilisateurs de trouver, partager et arranger l'information plus aisément et sans avoir recours à un intermédiaire. Il autorise également l'accès aux données entre différentes applications.  \\
Les machines ne peuvent pas effectuer toutes les tâches telles que, la réservation d'un billet du train, l'allocation d'une voiture \textit{etc}, sans assistance humaine, parce que les pages Web sont faites pour être lues uniquement par les humains. \\ Par conséquent, avec le Web sémantique, il s'avère possible d'effectuer des travaux rigoureux et automatiques par les machines. Il représente un système permettant aux agents de "comprendre" et de répondre aux requêtes complexes des utilisateurs. Cette "compréhension" exige que la sémantique des sources d'informations existantes, soit structurée à l'avance. 
\\

\parindent=0.5cm
Le but ultime du Web de la $3\up{ème}$ génération est de permettre aux utilisateurs d'exploiter tout le potentiel du Web en s'aidant par les machines qui pourront accomplir les tâches encore réalisées par l'Homme comme la recherche ou l'association d'informations, et ainsi atteindre un Web intelligent qui regroupera l'information de manière utile et qui apportera à l'utilisateur ce qui cherche vraiment.

\section{Ontologies}
\par Les ontologies existent afin de pouvoir présenter et partager les connaissances d'un domaine, ainsi que pour créer un consensus.

\parindent=0cm Elles sont les concepts clés du Web sémantique. Leur apparition a eu lieu au début des années 1990 dans la communauté de l'ingénierie de connaissances \textit{(IC)}. \\

Les ontologies font partie intégrante des normes du W3C pour le Web sémantique, car elles sont indispensables pour représenter la sémantique des documents, \textit{i.e.} les connaissances qui coexistent dans le Web, en structurant et en définissant la signification des termes actuellement collectées et normalisées. \\

\parindent=0.5cm
De nombreuses applications telles que la recherche d'informations, la réponse aux requêtes, la recherche documentaire, la synthèse de texte, \textit{etc,} sont réalisées en utilisant les ontologies (de domaine). Elles peuvent être utilisées aussi dans des services pour publier des bases de connaissances réutilisables, ou pour faciliter l'interopérabilité entre plusieurs systèmes hétérogènes et bases de données. Ainsi, nous pouvons considérer les ontologies comme une représentation pivot qui a pour but d'intégrer les sources de données hétérogènes pour décrire le contenu du Web. \\

Les ontologies sont nommées avec un IRI \textit{(\textbf{I}nternationalized \textbf{R}esource \textbf{I}dentifier)}, et puisqu'elles sont des documents Web, elles sont ainsi référencées par un URI \textit{(\textbf{U}niform \textbf{R}esource \textbf{I}dentifier} : IRI physique) qui doit pointer sur la localisation de l'URL \textit{(\textbf{U}niform \textbf{R}esource \textbf{L}ocator)} choisi pour les publier. \\

\subsection{Revue sur les définitions d'une ontologie}

Une des premières définitions de l'ontologie a été attribuée par \textit{Neches et al. (1991)} : \textit{ "Une ontologie définit les termes et les relations de base comportant le vocabulaire d'un domaine, aussi bien les règles pour combiner ces termes et ces relations, afin de définir des extensions du vocabulaire."}

\parindent =0cm
D'après cette définition, nous comprenons qu'une ontologie désigne les termes et les relations de base d'un domaine donné. La combinaison de ces derniers s'effectue en se basant sur des règles bien déterminées. Selon cette définition descriptive, nous saisissons également qu'une ontologie inclut de plus un volume informationnel, inféré des termes qu'elle peut caractériser. \\

\parindent = 0.5cm
Toutefois, la définition la plus adoptée et la plus utilisée dans le contexte du Web sémantique, est celle de \textit{Gruber (1993)} \cite{guarino2009ontology} : \textit{" Une ontologie est une spécification explicite et formelle d'une conceptualisation d'un domaine de connaissances"}. La conceptualisation est le dénouement d'une analyse ontologique du domaine exploré. Cette conceptualisation est exprimée dans une forme concrète, dite spécification. En fait, une spécification n'est qu'une représentation formelle des concepts et des relations, ainsi que les contraintes imposées dessus. \\

\parindent =0.5cm
\textit{Guarino et Giaretta (1995)} dans \cite{sept}, ont étudié et collecté sept définitions d'une ontologie : \\

\parindent=0cm
- \textit{Une ontologie est vue comme un exercice philosophique;} \\
- \textit{Une ontologie est un système conceptuel informel;} \\
- \textit{Une ontologie est un compte, ou récit sémantique formel;} \\
- \textit{Une ontologie est une spécification d'une conceptualisation;} \\
- \textit{Une ontologie est une représentation d'un système conceptuel à partir d'une théorie logique;} \\
- \textit{Une ontologie est un vocabulaire utilisé par une théorie logique;} \\
- \textit{Une ontologie est considérée comme un méta-niveau d'une spécification, d'une théorie logique.} \\

\parindent =0.5cm \textit{Borst (1997)} dans \cite{guarino2009ontology}, a légèrement modifié la définition de \textit{Gruber (1993)} et il a proposé la suivante : \textit{"Les ontologies sont définies comme une spécification formelle d'une conceptualisation partagée"}. Les définitions de \textit{Gruber (1993)} et \textit{Borst (1997)} ont été alliées par \textit{Studer et al.(1998)} comme suit : \textit{"Une ontologie est une spécification formelle et explicite d'une conceptualisation partagée"}. Selon \textit{Borst (1997)}, une conceptualisation se rapporte à un modèle abstrait retraçant
un phénomène quelconque du monde réel. L'abstraction se fait grâce à la détermination des concepts de ce phénomène. Le terme "explicite", provient du fait que les types de concepts et les contraintes de leur usage, doivent être explicitement définis. Ainsi, le terme "formelle" signifie qu'une ontologie doit être compréhensible et lisible par la machine. Finalement, le mot "partagée", montre qu'une ontologie doit couvrir un centre d'intérêt de façon à la rendre abordable, par tous les usagers de toute communauté. \\

\par  Nous avons récapitulé les définitions les plus pertinentes du terme ontologie, et qui reflètent des points de vue à la fois complémentaires et différents. En effet, nous pouvons conclure que les ontologies sont crées et inventées à la base pour modifier les connaissances consensuelles d'une façon générique. \\
En somme, les ontologies matérialisent les principes du partage et de la réutilisation de l'information. \\

Dans la littérature, nous listons plusieurs types d'ontologies selon différentes classifications. Malgré cette variété, toutes ces ontologies partagent quasiment les mêmes composants afin de décrire un certain domaine, et qui sont : connaissances et domaine de connaissances, concepts, relations et axiomes.

\subsection{Constituants d'une ontologie}

\par Les ontologies ne peuvent être construites, qu'à travers leurs éléments de base. Cette sous-section décrit les principaux constituants d'une ontologie.

\subsubsection{Connaissances et domaine de connaissances}

\parindent=0.5cm La signification de plusieurs termes diffère d'un domaine à un autre. C'est pourquoi, une ontologie ne peut être bâtie que dans le cadre d'un domaine particulier de connaissances \textit{(Bachimont
(1999))}. En plus, il faut associer une sémantique claire à l'ontologie construite. Un domaine de connaissances est composé par des termes du domaine et par un contexte d'utilisation de ces termes
\textit{(Bachimont (1999))}.  
En outre, certaines connaissances, qui peuvent constituer en elles-mêmes un domaine, sont employées dans tous les autres domaines. De plus, les connaissances humaines s'étendent et se développent suivant plusieurs dimensions. Elles peuvent être développées, non seulement sur la réalité, mais aussi sur un domaine de connaissances (\textit{i.e.} méta-connaissances). \\
\par Tout objet sur le Web possède une méta-donnée (étiquette) qui le représente fidèlement, et qui peut être lue par les agents logiciels aussi bien que par les agents humains. Les méta-données ou bien les méta-connaissances sont des informations sur les données décrivant les connaissances contenues dans les ressources.
Ainsi, cerner le domaine de connaissances à représenter, suscite une délimitation précise et explicite de l'objectif opérationnel de l'ontologie. Dans le domaine de l'ingénierie de connaissance, la représentation de connaissances se traduit par une description à base des faits et des règles.
 Nous pouvons considérer, qu'il n'y a de connaissances que lorsque l'information présentée dans la machine prenne un sens, et que ce sens soit le même pour tous les utilisateurs \textit{(Charlet (2001))}.
 Ainsi, une ontologie doit être représentée à partir d'un champ de connaissances bien déterminé pour un objectif opérationnel non ambigüe. Elle doit se baser sur des connaissances objectives, dont la sémantique puisse être scrupuleusement et formellement exprimée. Ces connaissances sont représentées dans l'ontologie sous forme de concepts, de relations et finalement d'axiomes. Dans ce qui suit, nous décrivons la notion de concepts.

\subsubsection{Concepts}

\parindent=0.5cm
Les concepts constituent le centre d'intérêt de plusieurs ontologies, et représentent en réalité des objets sur qui porte une connaissance.  Ils peuvent représenter un objet matériel, une notion ou une idée \textit{(Ushold and Jasper (1999))}. Un concept peut être décomposé en trois parties qui sont : un ou plusieurs termes, une notion et un ensemble d'objets. La notion est appelée également \textit{intention du concept}. Elle contient la signification du concept exprimée en termes de règles et de contraintes, de propriétés et d'attributs. L'ensemble d'objets est appelé aussi \textit{extension du concept} ou encore instance du concept. Il regroupe les objets manipulés à travers un concept. \\
Les deux aspects d'un concept (intention et extension) sont assez différents dans plusieurs mesures. En effet, deux extensions peuvent ne pas être disjointes, alors que deux intentions s'excluent réciproquement par au moins une propriété. Un concept peut avoir une extension vide, c'est le cas d'un concept générique, cela est généralement pour se référer à une notion abstraite, \textit{e.g.} le concept "fonctionnel" pris dans le sens de "ce qui est fonctionnel" et non pas du "degré de fonctionnalité". Deux concepts peuvent avoir la même extension sans pour autant posséder la même intention. Par exemple, le cas des concepts "directeurs" et "employés" qui peuvent partager quelques extensions. \\

Dans le langage naturel, il existe de nombreux termes désignant plusieurs concepts sémantiquement différents, et qui ne sont pas gérables par la machine. Par exemple, la "souris" est un animal et la même expression représente aussi un matériel d'ordinateur. La limitation à un domaine de connaissances bien déterminé permet de remédier à ce problème et alors d'éviter les homonymes des concepts. Néanmoins, il est préférable de gérer la synonymie et l'hyponymie pour garantir plus de souplesse dans l'usage de l'ontologie \textit{(Gómez Pérez et al. (1996))}. \\ 

\parindent =0.5cm
Les concepts sont reliés par des propriétés conceptuelles. Ces propriétés se divisent en deux critères \cite{concept}. La première catégorie regroupe les propriétés associées à un seul concept. La seconde rassemble les propriétés associées à deux concepts. La majorité des propriétés portant sur un seul concept sont proposées par \textit{Guarino and Giaretta (1995)} et elles sont constituées par : \\

\parindent =0cm
- \textbf{Généricité :} un concept est dit générique s'il n'admet pas d'extensions, \textit{i.e.} d'instances (\textit{e.g.} la vérité est un concept générique) ;

- \textbf{Identité :} un concept a une propriété d'identité si cette dernière permet de différencier une instance d'une autre (\textit{e.g.} le concept "étudiant" porte une propriété d'identité liée au numéro de l'étudiant, deux étudiants sont identiques s'ils ont le même numéro) ;

- \textbf{Rigidité :} un concept est considéré rigide si toute instance de ce concept en demeure instance dans tous les domaines possibles (\textit{e.g.} humain est un concept rigide) ;

- \textbf{Anti-rigidité :} un concept est dit anti-rigide si toute instance de ce concept est principalement définie par son appartenance à l'extension d'un autre concept (\textit{e.g.} "enseignant" est un concept anti-rigide car l'enseignant est avant tout un humain) ;

- \textbf{Unité :} un concept s'avère unité si pour chacune de ses instances, les différentes parties de l'instance sont attachées par une relation qui ne lie pas d'autres instances de concepts (\textit{e.g. }les deux parties d'un couteau, manche et lame sont liées par une relation "emmanché" qui ne relie que cette lame et cette manche). \\

Les propriétés portant sur deux concepts sont :\\

- \textbf{Équivalence :} deux concepts sont équivalents s'ils aient les mêmes extensions ; \\
- \textbf{Disjonction :} deux concepts sont disjoints si leurs extensions sont disjointes (\textit{e.g.} homme et femme); 

- \textbf{Dépendance :} un concept \textit{C1} est dépendant d'un concept \textit{C2}, si pour toute instance de \textit{C1}, il existe une instance de \textit{C2} qui ne soit ni partie ni constituant de l'instance de \textit{C2} (\textit{e.g.} le concept "parent" est un concept dépendant du concept "enfant" et inversement). \\

Les concepts, dans une ontologie, sont bien évidemment liés. Cette liaison se fait grâce à des relations. 

\subsubsection{Relations}

\parindent = 0.5cm Les relations permettent d'associer des instances de concepts, ou des concepts génériques. Elles sont particularisées par un terme (ou plusieurs) et une signature \textit{(Gómez-Pérez et al. (1996))}. La signature décrit le nombre d'instances de concepts que la relation lie, leurs types et l'ordre des concepts (la direction dont la relation doit être lue). Par exemple, la relation "développe" lie une instance du concept "développeur" à une instance du concept "programme", dans cet ordre. Elles sont formellement définies comme une fonction N-aire tandis que les ontologies contiennent généralement des relations binaires afin d'exprimer la sémantique qui relie chaque deux concepts, tel que le premier argument est le domaine et le deuxième argument correspond au Co-domaine (il peut être un type de donnée défini, tel que, \textit{« String », « Integer »},\textit{ etc.}).\\

\parindent =0.5cm
Les relations peuvent également être caractérisées par des propriétés \cite{concept}, et qui sont de trois types : les propriétés intrinsèques à une relation, les propriétés qui lient deux relations et les propriétés qui lient une relation à un concept. \\

 Nous citons les propriétés intrinsèques à une relation: \\

\parindent = 0cm
- \textbf{Propriétés algébriques :} comme la transitivité, symétrie, \textit{etc.} ; 

- \textbf{Cardinalité :} elle précise le nombre possible de relations de ce type entre les mêmes concepts (ou instances de concepts).\\

\parindent = 0.5cm
 Les propriétés liant deux relations sont comme suit :\\

\parindent = 0cm
- \textbf{Incompatibilité :} deux relations sont considérées incompatibles si elles ne peuvent pas lier les mêmes instances de concepts (\textit{e.g.} les relations "être à l'extérieur" et "être à l'intérieur" sont incompatibles); 

- \textbf{Inverse :} deux relations binaires sont inverses l'une de l'autre si, lorsque l'une lie deux instances \textit{I1} et \textit{I2}, l'autre lie \textit{I2} et \textit{I1} (\textit{e.g.} les relations "a pour père" et "a pour enfant" sont inverses); 

- \textbf{Exclusivité :} deux relations sont dites exclusives si, quand l’une lie des instances de concepts, l'autre ne lie pas ces instances, et vice-versa. L'exclusivité soulève l'incompatibilité. \\

\parindent=0.5cm
 Les propriétés liant une relation à un concept ont été proposées par \textit{Kassel and Perpette (1999)} : \\

\parindent=0cm
- \textbf{Lien relationnel :} il existe un lien relationnel entre une relation \textit{L} et deux concepts \textit{C1} et \textit{C2} si, pour tout couple d'instances des concepts \textit{C1} et \textit{C2}, il existe une relation de type \textit{L} qui lie les deux instances de \textit{C1} et \textit{C2} ;

- \textbf{Restriction de relation :} il peut avoir une restriction de relation entre une relation \textit{R} et un concept \textit{C}, si pour tout concept de type \textit{C}, et toute relation de type \textit{R} liant \textit{C}, les autres concepts liés par la relation sont d'un type imposé. La relation \textit{"mange"} portant sur une \textit{"personne"} et un \textit{"aliment"}, lie alors une instance de \textit{"végétarien"}, qui est un concept subsumé par \textit{"personne"}, et l'instance de \textit{"aliment"}, qui est forcément une instance de \textit{"végétaux"}. \\

\parindent = 0.5cm Les concepts et les relations permettent de représenter une ontologie légère basée sur une simple taxonomie de classes. Cependant, les ontologies lourdes contiennent de plus des concepts et des relations, des axiomes \textit{(Fürst and Trichet (2005))}. Dans ce qui suit, nous expliquons la notion d'axiomes, comme étant l'un des composants d'une ontologie.

\subsubsection{Axiomes}

\parindent = 0.5cm
Décrire les connaissances dans une ontologie ne suffit pas pour atteindre l'objectif opérationnel d'un \textit{\textbf{S}ystème à \textbf{B}ase de \textbf{C}onnaissances} \textit{(SBC)}. En effet, le plus important n'est pas d'avoir des connaissances sur un domaine, mais plutôt de leur mise en œuvre dans une action, afin d'atteindre un objectif particulier \textit{(Teuliet and Girard (2001))} \cite{defis}. Il s'agit, également, d'avantager les systèmes informatiques de la possibilité d'inférer de nouvelles connaissances à partir d'autres déjà existantes \textit{(Bachimont (2000))}. \\

Les axiomes logiques constituent des assertions liées aux entités. Au lieu de compter sur les labels et les termes des entités (qui sont destinés aux humains) pour transmettre la sémantique, le concepteur d'ontologies doit contraindre l'interprétation possible des entités à travers une utilisation judicieuse d'axiomes logiques pour rende leurs sens beaucoup plus précis. \\
Ils sont aussi utilisés pour vérifier la consistance de l'ontologie, car ils permettent à un raisonneur d'inférer des connaissances additionnelles qui ne sont pas déclarées directement. Plus les axiomes exprimées dans les ontologies sont complexes, plus elles transportent des connaissances implicites qui peuvent être inférées par le raisonneur. \\

 Les connaissances opérationnelles peuvent être des faits, des règles, ou des contraintes. \\
Un fait peut être une affirmation vraie et cognitive ou, un axiome qui aide à la description du monde cognitif dans lequel s'inscrit le système à base de connaissances. L'énoncé \textit{"la classe contient 32 étudiants"} est un exemple de fait. \\
Une règle permet d'inférer de nouvelles connaissances, elle incarne une implication.\\
Les contraintes spécifient les impossibilités ou les obligations. \\

Le choix entre l'utilisation d'une règle ou d'une contrainte pour la représentation d'une connaissance, n'est pas toujours facile. Il dépend de la manière dont les connaissances vont être utilisées au niveau opérationnel. Certains considèrent que ces types de connaissances doivent être incorporés dans les propriétés des concepts, des relations ou des instances de concepts.\\
Une règle apparaît plutôt comme une propriété d'un concept ou d'une règle, elle ne porte pas sur les instances.\\
Les contraintes peuvent se porter aussi bien sur les concepts et les relations que sur les instances. \\

Alors, les axiomes servent d'une part, à modéliser les énoncés qui sont toujours vrais, et d'autre part, à vérifier la consistance de l'ontologie elle-même. En effet, ils sont très avantageux pour inférer de nouvelles connaissances. \\

Pour résumer, selon \textit{Kalfoglou, et al.} (2003), une approche algébrique plus formelle, identifie une ontologie comme étant une paire \textit{<S, A>}, où \textit{S} est la signature des entités de l'ontologie (modélisée par une structure mathématique comme un treillis ou un ensemble non structuré) et \textit{A} est l'ensemble des axiomes ontologiques qui spécifient l'interprétation voulue de la signature dans un domaine donné. \\

Selon \textit{Udrea, et al.} (2007), les ontologies modélisent la structure des données (\textit{i.e.} les ensembles de classes et des propriétés), la sémantique des données (sous la forme de contraintes (axiomes) telles que les relations d'héritage ou les contraintes sur les propriétés), et les instances des données (les individus). Ainsi, les ontologies se composent d'une partie « structure », et d'une partie « donnée ». \\

D'après \textit{Zhang, et al.} (2017), une ontologie est un modèle à base d'arbre, à cause du principe de l'hyponymie (la subsomption \textit{is-a}) qui fait que chaque entité (classe ou propriété) soit héritée d'une seule super-entité directe, formant ainsi une structure de graphe acyclique enracinée.

Afin que les données d'une ontologie puissent être utiles et traitables par des machines, elles doivent être décrites à travers des notations concrètes en utilisant des langages formels. Dans la section suivante, nous allons présenter ces langages ontologiques. \\

\section{Langages de représentation des ontologies}
\parindent=0.5cm
Nous avons fait une revue sur les différentes définitions d'une ontologie ainsi que ses constituants de base. Les ontologies constituent le principe général qui régit le Web sémantique. Leur représentation repose sur des standards et des schémas de données bien définis, on note \textit{RDF}, \textit{RDFS} et \textit{OWL}\textit{, etc}. \\

Les langages de représentation des ontologies, sont les appuis pour la mise en place des ontologies dans le Web. Ils constituent les standards qui régissent l'architecture du Web sémantique et qui, également, assurent une interopérabilité sémantique entre les différents intervenants.

\subsection{XML et XML Schema}

Le langage \textit{XML} \cite{langage} procure une syntaxe pour les documents structurés d'une manière hiérarchique.
Par contre, ce langage n'exige aucune contrainte sémantique et alors il n'apporte aucune signification aux documents. Il n'existe pas de normes pour la recherche intelligente, l'échange de données et la présentation adaptable. \\
En outre, XML est un langage à balises pouvant contenir également des informations supplémentaires (méta-données) qui sont incorporées dans le texte à présenter. \\

\parindent=0cm
\textbf{Un schéma XML}, noté \textit{XML(S)}, est une description du type d'un document XML. Il comporte un ensemble de contraintes sur la structure et le contenu du document. Ainsi, un document XML doit respecter cet ensemble de règles afin d'être considéré valide. \\

\textbf{L'espace de noms }(\textit{namespace}) est un contexte ou un conteneur abstrait contenant des noms, des termes ou des mots qui représentent des objets, des concepts dans le monde réel. Un nom défini dans un espace de noms correspond à un et seulement un objet, alors que deux objets ou concepts différents sont référencés par deux noms différents dans un même espace de noms.

\subsection{RDF et RDF Schema}

\parindent =0.5cm 
RDF \cite{langage}, est un modèle de méta-données ayant pour objectif de référencer des objets (ressources) et décrire les liens entre eux afin de rendre plus "structuré" les données nécessaires aux moteurs de recherche. Dans ce modèle, les ressources sont référencées par les URIs. \\

L’\textbf{URI} ressemble à une étiquette/label numérique (chaînes de caractères) permettant de représenter à la fois l'adresse d'une ressource sur un réseau, quand il est considéré sous sa forme d’\textit{URL (Uniform Ressource Locator)}, et/ou le « nom propre » de cette ressource, quand il est estimé sous sa forme d’\textit{URN (Uniform Ressource Name)}. Une \textit{URL} est alors un \textit{URI} qui sert à identifier l'emplacement d'une ressource, alors qu'un \textit{URN} est un \textit{URI} qui aide à désigner ou à nommer n'importe quelle ressource sur le Web. \\

\textbf{Linked Data} permet d'utiliser le Web pour faciliter la construction de liens entre des données dépendantes qui n'étaient pas encore liées, ou qui l'étaient auparavant via d'autres méthodes. On y retrouve un ensemble de pratiques recommandées pour exposer, échanger et relier des fragments de données, informations et connaissances sur le Web sémantique grâce à l'usage des \textit{URI} et de \textit{RDF}.\\
\textit{Linked Data} définit alors un moyen de publication de données semi-structurées sur le Web de données, dans le but de lier les données et de les rendre plus accessibles. \\
\parindent=0cm
Elles sont un espace de données global, composé de sources de données structurées et distribuées publiées dans le Web en langage RDF, sous forme de triplets RDF (sujet -> prédicat -> objet), et inter-liées explicitement. \\

\parindent=0.5cm
\textbf{Linked Open Data} (LOD) : ont connu une croissance rapide au cours des dernières années. Elles se constituent de 1 163 data sets lisibles par machine (en 2017) ayant plus de 31 milliards de triplets RDF (en 2011) inter-liés par environ 504 millions liaisons (en 2011). \\
Les data sets du LOD sont principalement classés en neuf domaines : le domaine transversal (\textit{e.g.} DBpedia : la version « Linked Data » de Wikipedia), la géographie (\textit{e.g.} Geonames), le gouvernement, les sciences de la vie, la linguistique, les médias (\textit{e.g.} NYTimes, et LinkedMDB), les publications, les réseaux sociaux (\textit{e.g.} FOAF), et le contenu généré par les utilisateurs. \\

\begin{figure}[H]
\begin{center}

\includegraphics[scale=0.6]{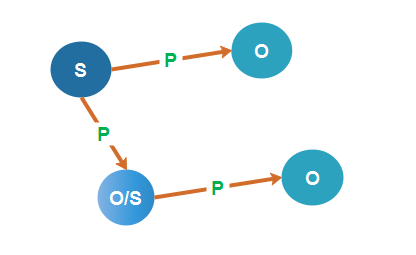}

\end{center}
\caption{La structure d'un triplet RDF \label{figure 1.2}}
\end{figure}

\parindent=0.5cm
RDF est basé sur la théorie de graphes, qui permet de représenter les données sous forme de \textbf{triplets RDF}. C'est le modèle standard pour le partage et l'échange des informations sur le Web.\\
Un triplet RDF est un ensemble de trois informations liées \textbf{<sujet, prédicat, objet>} comme l'indique la figure \ref{figure 1.2}. Ce type de représentation est facile à interpréter parce qu'il est un modèle mental assez simple. \\
- \textbf{ Le sujet} est la ressource ou l'objet à modéliser, il est représenté par une \textit{URI}. \\
- \textbf{ Le prédicat} exprime un lien ou une relation entre le sujet et l'objet, représenté par une \textit{URI} également. \\
- \textbf{ L'objet} est la valeur du prédicat du sujet, il peut être une \textit{URI}, un identifiant local, ou un littéral. \\  

 \parindent =0.5cm
\textbf{Entailment :} une caractéristique importante est la représentation de triplets implicites qui ne sont pas explicitement modélisés dans un graphe mais ils font partie de celui-ci. Le W3C appelle \textit{RDF entailment}, le mécanisme avec lequel les triplets implicites sont engendrés (ou dérivés) grâce aux triplets déjà existants (explicites) d'un graphe, ainsi que grâce à des règles d'entailment. \\

\textbf{Sérialisation RDF:} comme RDF a été conçu principalement pour représenter des ressources Web, sa syntaxe formelle utilise ordinairement XML (on appelle alors cette forme \textit{RDF/XML}). Cependant, cette dernière devient vite fastidieuse à lire pour les agents humains.
Pour rendre un document RDF plus accessible, le W3C offre des notations plus abstraites que sont \textit{N3} et \textit{N-Triples} (N-Triples étant un sous-ensemble de N3). \\

\parindent =0.5cm

En ce sens, bien que RDF soit très flexible (donne la possibilité d'assembler n'importe quoi, n'importe comment, à la description d'une ressource), les représentations dans ce format sont assez abstraites, platoniques et sans grande profondeur sémantique (faible degré de
hiérarchisation des descriptions). Pour couvrir cette lacune, un vocabulaire élémentaire a été développé : \textit{\textbf{RDFS} (RDF Vocabulary Language)}.  \\

\parindent=0.5cm
\textit{RDF Schema} est un langage pour décrire de simples vocabulaires dans le modèle RDF. Ainsi, il n'est qu'une extension sémantique de RDF. Grâce à ce vocabulaire, la sémantique des prédicats et les classes d'entités peuvent être définis. Cela laisse, entre autre, la possibilité de déterminer des règles d'inférence afin de générer une grande quantité de triplets dérivés de ceux existants, ce qui augmente, par conséquent, la capacité du raisonnement que peut avoir les agents logiciels et les programmes qui traitent les données sur le Web sémantique. 

RDFS permet notamment de définir des liens de \textit{"Subsomption"} entre des classes et des relations en utilisant les primitives \textit{rdfs :sub ClassOf} et \textit{rdfs :subPropertyOf}. Ces primitives sont les liens de "Spécialisation" ou "is-a", qui permettent aux classes et aux relations d'hériter des caractéristiques de leurs classes (ou de leurs relations) mères.

\subsection{OWL}

\parindent= 0.5cm 
Le langage \textit{OWL} \textit{(Schreiber et Dean)} a vu le jour en 2004, dans la même période que \textit{RDF} et \textit{RDFS}. Il a, par la suite, été étendu, dans l'année 2009, dans la norme \textit{OWL 2} \textit{(OWL Working Group)}.

\parindent =0.cm
Il est, tout comme \textit{RDF}, un langage profitant de l'universalité syntaxique de \textit{XML}. \\

\parindent=0.5cm

Ce standard a été conçu également pour réutiliser des concepts et
des propriétés déjà définis dans \textit{RDF} et \textit{RDFS}. En ce sens, une base de connaissances \textit{OWL} peut alors être écrite comme une collection de triplets \textit{RDF}. \\
\textit{OWL} permet aux applications, qui l'utilisent, de pouvoir effectuer des inférences sur les données cumulées. Pour ce faire, il offre des opérateurs issus de la logique descriptive. \\

A part sa capacité de définir et de décrire des classes, des propriétés, et des individus de classes, OWL permet aussi de définir des relations entre les classes (union, intersection, disjonction, équivalence, subsomption etc.), des contraintes de cardinalité pour les valeurs des propriétés (minimum, maximum, nombre exact), des relations spéciales pour les propriétés (transitive, symétrique, fonctionnelle, inverse, réflexive, etc.), et des restrictions sur le domaine et le co-domaine des propriétés, etc. Par conséquent, OWL possède une logique très développée qui permet le raisonnement sémantique sur ces règles à l'aide d'un raisonneur. \\

En outre, \textit{OWL} est capable de décrire formellement la signification de la terminologie employée dans les documents Web, constituant ainsi le premier niveau indispensable du Web sémantique après \textit{RDF}. Ainsi, apprêter les machines d'une capacité à mener des raisonnements appropriés sur les documents, revient à fournir un langage de qui la sémantique dépasse celle du schéma \textit{RDF}. \\

Le langage \textit{OWL} diffère du couple RDF/RDFS en ceci que, contrairement à RDF, il est précisément un langage d'ontologies. Les langages RDF et RDFS fournissent à l'utilisateur la capacité de représenter des classes (\textit{i.e.} avec des constructeurs) et des propriétés. \\
$\Rightarrow$ OWL offre aux machines une plus grande capacité d'interprétation du contenu Web que RDF et RDFS, grâce à une vraie sémantique formelle et à un vocabulaire plus riche. \\

\parindent=0cm
OWL est composé de trois parties : 

\parindent = 0.2cm
- \textbf{Classe} : est un groupe d'individus (instances) possédant des caractéristiques similaires. Les classes peuvent être organisées hiérarchiquement selon une taxonomie. \\
Les classes définies par l'utilisateur sont toutes des enfants de la super-classe \textit{”owl:Thing”} et des parents de la sous-classe \textit{”owl:Nothing”}. \\

\parindent = 0.2cm
- \textbf{Propriété} : permet de définir des faits ou des relations entre les individus. Il existe en \textit{OWL} deux types de propriétés : les propriétés d'objet et celles de types de données. \\

\parindent = 0.2cm
- \textbf{Instance} : est un objet particulier instancié par les classes à l'aide de la relation prédéfinie \textit{"Instance-Of"}. Elles peuplent les classes et véhiculent les connaissance à propos du domaine en question. \\

\subsubsection{OWL 1}

\parindent = 0.5cm
Le langage \textit{OWL 1} détermine trois sous-langages, du moins expressif au plus expressif, à savoir : \textit{OWL-Lite},\textit{ OWL-DL} et \textit{OWL-Full} \cite{langage}. \\

\parindent=0cm
- \textbf{\textit{OWL-Lite}} est le premier sous langage de \textit{OWL} et le plus simple. Il est dédié aux usagers qui n'ont besoin que d'une hiérarchie simple de concepts. \textit{OWL-Lite} est adapté, par exemple, aux migrations rapides depuis un ancien thésaurus vers une représentation sous forme d'ontologies. \\
\\ - \textbf{\textit{OWL-DL}} est plus complexe qu'\textit{OWL-Lite}, fournissant une expressivité bien plus importante. \textit{OWL-DL} \textit{(OWL Description Logic)} est fondé sur la logique descriptive, assurant alors une adaptation au raisonnement automatisé. Malgré sa complexité relative face à \textit{OWL-Lite}, \textit{OWL-DL} garantit la complétude des raisonnements (toutes les inférences sont calculables) et leur décidabilité (leur calcul se fait en une durée finie).\\ 

- \textbf{\textit{OWL-Full}} est la version la plus complexe et complète d'\textit{OWL}, mais également celle qui permet d'avoir le plus haut niveau d'expressivité. Il garantit la complétude et la décidabilité
des calculs liés à l'ontologie. \textit{OWL-Full} fournit toutefois des mécanismes intéressants, comme la possibilité d'étendre le vocabulaire par défaut de \textit{OWL}. Il existe, entre ces trois sous langages, une dépendance de nature hiérarchique : toute ontologie \textit{OWL-Lite} valide est également une ontologie \textit{OWL-DL} valide, et toute ontologie \textit{OWL-DL} valide est alors une ontologie \textit{OWL-Full} valide. \\

Ainsi, les développeurs d'ontologies choisissant \textit{OWL}, devraient estimer quel sous-langage convient le mieux à leurs besoins.

\subsubsection{OWL 2}

\parindent =0.5cm
\textit{OWL 2} reprend les caractéristiques de sa version précédente en ajoutant quelques composants supplémentaires. La principale contribution d'\textit{OWL 2} est de déterminer des profils, des sous-langages de \textit{OWL 2}, qui imposent et limitent les expressions exploitables selon l'utilisation voulue de l'ontologie. \\
A l'opposé des sous-ensembles \textit{OWL-Lite, DL et FULL}, les profils de \textit{OWL 2} sont indépendants. 

\parindent =0cm
La norme \textit{OWL 2} détermine trois profils : \textit{OWL-EL, OWL-RL et OWL-QL} \cite{owl2}. L'expressivité de ces langages proposés par \textit{OWL 2} est réglée en fonction de l'exploitation voulue, de la complexité et également du niveau de raisonnement désirés.

\section{Types d'hétérogénéité}

La diversité du monde réel est une source de richesse et d'hétérogénéité. En effet, dans les systèmes ouverts et distribués, tel que le Web sémantique, l'hétérogénéité ne peut pas être évitée parce que les acteurs ont des intérêts et des habitudes différentes, et utilisent des connaissances et des outils différents, avec des niveaux de détails différents. Toutes ces raisons mènent à diverses formes d'hétérogénéité. \\
L'hétérogénéité ne repose pas seulement sur les différences entre les buts des applications pour lesquelles elles ont été désignées, ou sur les formalismes d'expression par lesquelles les ontologies ont été codées.  \\

Prenons l'exemple du domaine biomédical. Il y a neuf ontologies qui décrivent une maladie neurologique, allant des ontologies très spécifiques couvrant une seule maladie (\textit{e.g.} l'épilepsie, l'Alzheimer) à des ontologies couvrant toutes sortes de maladies telles que la « Disease Ontology ». Il en résulte plusieurs ontologies qui décrivent les mêmes concepts sous des modèles légèrement différents. \\

On distingue quatre types d'hétérogénéité principales : \\ 

\parindent = 0.5cm
$\bullet$ L'\textbf{hétérogénéité syntaxique} se produit quand deux ontologies ne sont pas exprimées avec le même langage. Cela se produit également quand deux ontologies sont construites en utilisant différents formalismes de représentation de connaissances, \textit{e.g.} \textit{OWL} et \textit{F-logic}. Ce type de différence (\textit{mismatch}) est généralement résolu au niveau théorique quand on établit des équivalences entre les constructeurs / les primitifs des différents langages. Ainsi, il est parfois possible de traduire les ontologies en d'autres langages tout en préservant leurs sens. \\ 

$\bullet$ L'\textbf{hétérogénéité terminologique} se produit à cause des variations des noms utilisés pour faire référence aux mêmes entités dans différentes ontologies. Ceci peut être causé par l'utilisation de différents langages naturels, \textit{e.g.} \textit{Paper} vs. \textit{Articulo}, ou différents sous-langages techniques spécifiques à un domaine de connaissances bien déterminé, \textit{e.g.} \textit{Paper} vs. \textit{Memo}, ou l'utilisation de synonymes, \textit{e.g.} \textit{Paper} vs. \textit{Article}. \\

$\bullet$ L’\textbf{hétérogénéité conceptuelle}, nommée aussi hétérogénéité sémantique ou différence (mismatch) logique, concerne les différences dans la modélisation d'un même domaine. Cela peut se produire à cause de l'utilisation de différents axiomes dans la définition des concepts, ou à cause de l'utilisation de concepts totalement différents. Voici les trois différences conceptuelles majeures : \\ 

\parindent = 1cm

- La différence de convergence : elle survient lorsque deux ontologies décrivent différentes connaissances avec le même niveau de détail pour une unique perspective. 

\parindent = 1cm
- La différence de granularité : elle se produit quand deux ontologies décrivent le même domaine avec une même perspective mais avec différents degrés d'expression des détails. 

\parindent = 1cm
- La différence de perspectives : elle se manifeste quand deux ontologies décrivent un même domaine, avec un même degré d'expression des détails mais avec des points de vue et des perspectives différents. \\

\parindent =0.5cm

$\bullet$ L’\textbf{hétérogénéité sémiotique}, nommée aussi l’hétérogénéité pragmatique, concerne la manière dont les entités sont interprétées par les utilisateurs. En effet, les entités, qui ont exactement la même interprétation sémantique, sont souvent interprétées par les humains relativement au contexte et au domaine d'application, \textit{e.g.} relativement à la manière dont elles vont être utilisées au final. Ce type d'hétérogénéité est difficile à détecter par la machine, et encore plus difficile à résoudre, parce qu'il est hors de sa portée. L'utilisation future des entités a toujours un grand impact sur leur interprétation.  

\section{Conclusion}

\parindent=0.5cm
Enfin, le volume des ontologies ne cesse pas d'augmenter avec les connaissances dans différents domaines. Cette croissance, fait apparaître les limitations des anciennes méthodes de gestion et de
manipulation des ontologies et de graphes RDF, ce qui nous oblige de concevoir des nouvelles méthodes pour cet effet. 

\chapter{Revue sur l'interrogation des graphes RDF} 

\section{Introduction}
 
\parindent = 0.5 cm 
La sémantique du contenu des ressources dans le Web sémantique, vise à rendre l'information explicite pour les machines dans une représentation formelle et standardisée. Cette standardisation peut aider différents programmes et divers intervenants à inter-opérer ainsi qu'à échanger des données. La description de la sémantique, dans le cadre du Web sémantique, est effectuée par l'intermédiaire des ontologies. Les objets ainsi que les relations, dans l'univers du discours, sont conceptualisés afin d'être exploitables par les machines. Une des caractéristiques du Web sémantique, est son hétérogénéité, qui décrit à la fois sa richesse et son ambiguïté. Cette hétérogénéité reflète la diversité du monde réel, où les connaissances et les informations peuvent provenir de plusieurs sources différentes. Ainsi, elles peuvent être représentées dans divers formats. De même, dans le Web sémantique, l'hétérogénéité ne sera pas une caractéristique exceptionnelle. Alors, le recours à des ontologies s'avère une bonne solution, mais il ne permet pas d'atteindre une interopérabilité complète à cause de certains handicaps, tels que le volume, la langue naturelle, \textit{etc}. La récupération des données devient de plus en plus difficile, dans le cadre du Web sémantique, en raison de l'hétérogénéité ainsi que de l'énorme quantité de données qui circulent sur le Web. D'autre part, les utilisateurs ordinaires ne peuvent pas gérer les langages d'interrogation hautement spécifiques (\textit{e.g. SPARQL}) ou les techniques basées sur la gestion des connaissances.

\section{Définitions : Standards, Protocoles et Outils}

\parindent = 0.5 cm 
Il y a, de plus en plus, un réel besoin d'outils efficaces pour le stockage et l'interrogation des connaissances utilisant des ontologies et des ressources connexes. Dans ce contexte, le stockage des données non structurées est devenu une nécessité afin d'augmenter l'efficacité du traitement des requêtes.

\subsection{Stockage des données RDF}

\parindent = 0.5cm

Avec l'énorme volume des données à sauvegarder, les fichiers (ceux qui stockent les représentations textuelles des données) deviennent vite illisibles pour les utilisateurs et difficilement exploitables. C'est dans ce cadre que \textbf{les bases de données RDF} \textit{(BD-RDF)}, ou \textit{Triple-Stores}, sont apparus. 
 
\subsubsection{Bases de données RDF (BD-RDF)}

\parindent =0.5cm

Nous allons commencer cette partie en expliquant \textbf{la théorie de graphes} \cite{graphtheory}. \\
La naissance de la théorie de graphes est attribuée au mathématicien suisse \textit{Leonhard Euler}, qui a d'abord résolu le problème des \textit{Sept Ponts de} \textit{Konigsberg} en 1736. \\ Formellement, un graphe est une collection de sommets et d'arcs et réellement, il représente des entités, en tant que des nœuds, ainsi que la manière dont ces entités reflètent (relient) le monde réel, ceci est représentée comme des relations. Tous cela forme ce qu'on appelle des \textbf{triplets}. \\ \textit{"Les graphes ont l'avantage de pouvoir conserver toutes les informations sur une entité, dans un seul nœud et de représenter des informations connexes par des arcs connectés"} \textit{(Paredaens et al. (1995))}.

\begin{figure}[H]
\begin{center}

\includegraphics[scale=0.6]{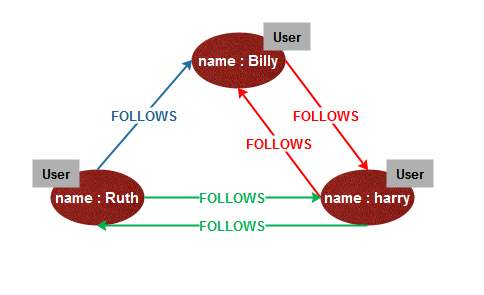}

\end{center}
\caption{Un échantillon d'un graphe social \label{figure 2.1}}
\end{figure} 

\parindent =0.5cm
Cette structure expressive polyvalente nous permet de modéliser tout sorte de scénarios, de la construction d'une fusée spatiale à un système de routes, de la chaîne d'approvisionnement ou de la provenance des aliments, à l'histoire médicale des populations, \textit{etc} (\textit{e.g.} les données de \textit{Twitter} sont facilement représentées en tant qu'un graphe). La figure \ref{figure 2.1} montre un petit réseau des utilisateurs \textit{Twitter} dont chaque nœud est étiqueté \textit{User}, indiquant son rôle. Ces nœuds sont alors connectés grâce à des relations, ce qui aide à établir, davantage, le contexte sémantique: à savoir que \textit{Billy FOLLOWS Harry et que Harry, à son tour, FOLLOWS Billy, etc.} \\  \\  \\

\begin{figure}[H]
\begin{center}

\includegraphics[scale=0.9]{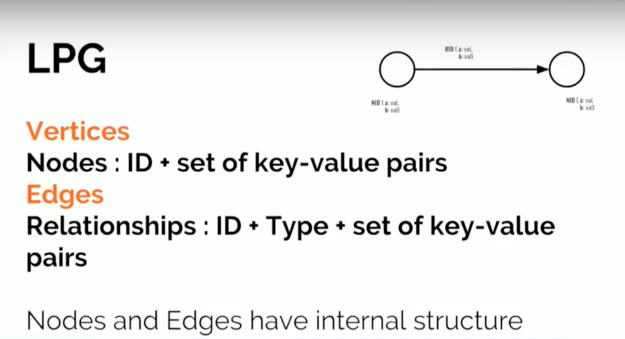}

\end{center}
\caption{Modèle graphique étiqueté par des propriétés \label{figure 2.2}}
\end{figure}

\parindent = 0cm 
\textbf{$\triangleright$}\textbf{ Modèle graphique étiqueté par des propriétés \textit{(i.e. Labeled Property Graph Model)}} (figure \ref{figure 2.2}) \footnote{https://neo4j.com/blog/rdf-triple-store-vs-labeled-property-graph-difference/ Consulté le 12/10/2017}, est le plus populaire des modèles. Il suit les caractéristiques suivantes \cite{modele} : \\
- Il contient des nœuds et des relations (dont chaque nœud possède un ID unique); \\
- Les nœuds contiennent des propriétés (des paires clé-valeur); \\
- Les nœuds peuvent être étiquetés par un ou plusieurs labels (ces labels caractérisent les différents rôles d'un domaine); \\
- Les relations sont nommées et orientées, et toujours, possèdent un nœud de début et un autre de fin; \\
- Les relations peuvent aussi contenir des propriétés. \\

\parindent = 0.5cm 

Ce modèle a été développé par un groupe suédois, dont leur principale motivation ne se basait pas essentiellement sur l'échange et la publication de données, mais plutôt sur un stockage efficace de ces données connectées afin d'assurer une interrogation et un parcours rapides du graphe. \\
Pour comparer un modèle \textit{RDF} simple et un modèle graphique étiqueté par des propriétés, nous pouvons dire que le premier est conçu spécialement pour l'échange des données, alors que le deuxième est  conçu principalement pour le stockage et l'interrogation \cite{modele}. 

\begin{figure}[H]
\begin{center}

\includegraphics[scale=0.8]{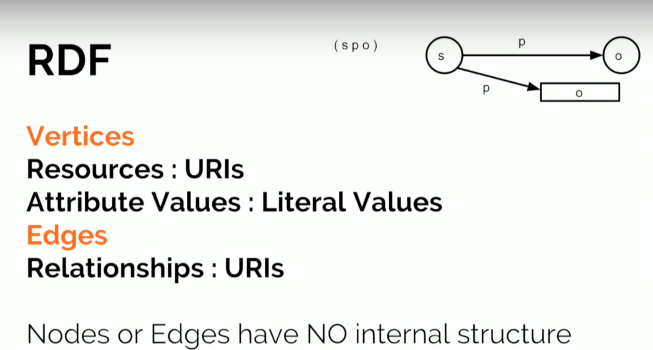}

\end{center}
\caption{Modèle RDF simple \label{figure 2.3}}
\end{figure}

Pour le modèle \textit{RDF} simple, un sommet est soit un sujet soit un objet, ainsi qu'un arc décrit le prédicat qui les relie. Ce modèle ne possède pas une structure interne (\textit{i.e.} un ensemble de propriétés), comme l'indique la figure \ref{figure 2.3} \footnote{https://neo4j.com/blog/rdf-triple-store-vs-labeled-property-graph-difference/ Consulté le 12/10/2017}. \\ 

\subsubsubsection{Historique des BD-RDF}

\parindent =0.5cm
L'activité autour des bases de données graphiques s'est épanouie dans la première moitié des années 90, puis le sujet a presque disparu. Les raisons de ce déclin sont multiples : la communauté de base de données s'est déplacée vers des données semi-structurées, l'émergence de XML a capturé toute l'attention, \textit{etc}. \\

\parindent =0.5cm
En 2008, \textit{Angles et Gutierrez} ont réalisé une enquête sur l'historique des modèles de bases de données graphiques \cite{historiqueBD}.

\parindent =0cm
Dans une première approche, \textit{Roussopoulos et Mylopoulos (1975)}, face à l'échec des systèmes de cette époque à prendre en compte la sémantique des bases de données, ont proposé un réseau sémantique pour le stockage des données. Une structure implicite des graphes pour les données elles-mêmes, a été présentée dans le modèle de données fonctionnelles par \textit{Shipman (1981)}, dont le but était de fournir une interface de base de données « conceptuellement naturelle ». \\

Une approche différente décrit le modèle logique de données \textit{(MLD) (Kuper et Vardi (1984))}, où un modèle explicite de base de données graphique avait pour but de généraliser les modèles relationnels, hiérarchiques et de réseau. Plus tard, \textit{Kunii (1987)} a proposé un modèle de base de données graphiques appelé \textit{G-Base} permettant de représenter des structures de connaissances complexes. \\

\parindent =0cm
\textit{Lécluse et al. (1988)} ont présenté \textit{O2} qui est  un modèle de base de données orientées objet, basé sur une structure de graphe. Dans le même contexte, \textit{GOOD (Gyssens et al. (1990))} est un modèle d'objet orienté graphe, destiné à devenir une base théorique pour les systèmes dans lesquels la manipulation ainsi que la représentation sont basées sur des graphes transparents. \\

\parindent = 0cm
\textit{Guting (1994)} a proposé \textit{GraphDB}, destiné à modéliser et interroger des graphes dans des bases de données orientées objet, et motivé par la gestion de l'information dans les réseaux de transport. \\ 

\parindent =0cm 
\textit{Silberschatz et al. (1996)} a défini un modèle de base données (\textit{db-model}) comme \textit{"une collection d'outils conceptuels utilisés pour modéliser des représentations des entités du monde réel, ainsi que les relations entre elles"}. Pour les BD graphiques, les outils conceptuels qui font un \textit{db.model}, doivent au moins s'adresser à la structuration des données, la description, la maintenance et la manière avec laquelle s'effectue la récupération ou l'interrogation des données.  \\

\parindent =0cm
Le projet \textit{GRAS} \textit{(Kiesel et al (1996))} utilise les graphes attribués afin de modéliser des informations complexes à partir de projets d'ingénierie logicielle. L'\textit{OEM} \textit{(Papakonstantinou et al. (1995))} vise à fournir un accès intégré à des sources d'information hétérogènes, en mettant l'accent sur l'échange des informations. \\

\textit{Newman (2003)} \cite{historiqueBD} a divisé les bases de données graphiques en quatre catégories : \\
- \emph{\textbf{Réseaux sociaux} }: où les nœuds sont des personnes ou des groupes, tandis que les liens représentent les relations ou les flux entre les nœuds (\textit{e.g.} l'amitié, les relations d'affaires, les réseaux de recherche, \textit{etc}). \\
- \emph{\textbf{Réseaux d'information}} : où les relations représentent le flux d'information, comme les citations entre les papiers académiques, les relations entre les classes de mot dans un thésaurus, \textit{etc}. \\
- \emph{\textbf{Réseaux technologiques}} : dont les aspects géographiques et spatiaux de la structure sont les plus importants, \textit{e.g. }Internet (comme un réseau d'ordinateurs), réseaux téléphoniques, réseau de livraison (bureau de poste), \textit{etc}. \\
- \emph{\textbf{Réseaux biologiques}} : représentent des informations biologiques dont le volume, la gestion et l'analyse sont devenus un problème, en raison de l'automatisation du processus de collecte de données (\textit{e.g.} réseaux alimentaires, réseaux de neurones, \textit{etc}). 

\subsubsubsection{RDF stores}

\parindent = 0.5 cm 
Un \textit{RDF Store} (magasin \textit{RDF}/ \textit{Triples Store}) est une base de données qui sert à stocker et à interroger des données \textit{RDF}, ainsi que des méta-données \textit{RDF}  en utilisant un langage d'interrogation. Parmi les Triples Stores nous citons \textit{Filament, G-Store, Redis-graph, VertexDB, CloudGraph, etc.} Ces derniers possèdent une capacité de stockage même de milliards de triplets. \\

\parindent = 0.5 cm 
Récemment, il y a eu des développements majeurs au niveau des techniques de stockage et de traitement des requêtes à partir des magasins \textit{RDF}.
Certains magasins \textit{RDF} sont construits à partir de zéro (section \ref{native}, page \pageref{native}), tandis que d'autres s'appuient sur des \textit{SGBDR} existants (section \ref{non-native}, page \pageref{non-native}). \\

\parindent = 0.5cm 
Un \textit{Triple Store} est considéré comme un système de gestion de base de données \textit{RDF}, dont une de ses caractéristiques clés est sa capacité de mener des inférences. Les Triples Stores (encore non pas tous) offrent, en plus du chargement et du stockage de données, la possibilité de traiter l'occurrence, la sécurité, la récupération, les mises à jour, \textit{etc}. \\

\parindent = 0.5cm 
Les systèmes de gestion de données \textit{RDF} peuvent être divisés en deux catégories en fonction de leur capacité à être conformes au modèle RDF : les natives (\textit{resp.} non natives) qui sont (\textit{resp.} ne sont pas) conformes à un modèle \textit{RDF}. Nous allons détailler ces deux types de \textit{BD-RDF} en indiquant leurs différences dans ce qui suit.

\subsubsubsection{BD-RDF non natives}
\label{non-native}

\parindent = 0.5cm
La principale fonctionnalité des \textit{BD-RDF} non natives est leur architecture. Ces bases de données \textit{RDF} sont implémentées au-dessus d'un système de gestion de base de données relationnel (\textit{SGBDR}), et c'est pour cette raison qu'elles sont appelées non natives. En effet, les \textit{BD-RDF} non natives permettent de bénéficier de l'efficacité des \textit{SGBDR}. Pour ce faire, il y a trois techniques de stockage pour les \textit{BD-RDF} non-natives : \textit{la représentation verticale, la représentation binaire et la représentation horizontale} \cite{native}. Les performances de ces représentations peuvent changées dépendamment des données \textit{RDF} et des requêtes appliquées sur ces données. \\  

\parindent= 0.5cm
\textit{\textbf{- Représentation verticale}} \\
Les données sont stockées dans une unique table de trois colonnes qui correspond chacune au \textit{sujet, prédicat,} ou \textit{objet}, respectivement. \\ 
Alors, afin d'interroger ces données, les requêtes doivent être traduites en \textit{SQL}. Pour améliorer la traduction d'un langage d'interrogation \textit{RDF} (tel que SPARQL; section \ref{sparql}, page \pageref{sparql}) vers le langage \textit{SQL}, il est possible de créer d'autres tables auxiliaires. \\
Ce type de \textit{BD-RDF} emploient également des outils d'optimisation de requête,\textit{ e.g.} les vues matérialisées et les indexs, afin de précipiter l'exécution des requêtes \textit{SQL}. Parmi les \textit{BD-RDF} utilisant cette représentation, nous citons \textit{3Store, RDFStore, etc.} \\ 
\textbf{$\ominus$} \emph{Inconvénient:} cette représentation est coûteuse pour l'interrogation des bases de données, vu le besoin d'effectuer un grand nombre d'auto-jointures. De même, les mises à jour nécessitent, dans la majorité du temps, la modification de l'ensemble de triplets. \\

\parindent= 0.5cm
\textit{\textbf{- Représentation binaire}} \\
Cette représentation fait correspondre chaque prédicat à une table de deux colonnes \textit{(sujet, objet)}, contenant les sujets possédant au moins une valeur pour ce prédicat, ainsi que pour chaque sujet, le ou les objets correspondants pour ce prédicat. Alors, une table est créée pour chaque propriété. Parmi les \textit{BD-RDF} qui utilisent cette représentation, nous pouvons citer \textit{4Store} et \textit{JenaSDB} (qui peut être couplé avec \textit{PostgreSQL} et \textit{MySQL}). \\

\parindent= 0.5cm
\textit{\textbf{- Représentation horizontale}} \\
Permet de créer une table à chaque classe, possédant pour colonnes les propriétés de la classe. Elle utilise le constructeur \textit{rdfs:Class} afin de construire les tables associées aux différentes classes (\textit{e.g. ontoDB}). Afin d'identifier les propriétés des classes, nous utilisons le constructeur \textit{ rdfs:Domain.}   \\

\parindent = 0.5cm
\textbf{$\Rightarrow$} Pour conclure, les relations dans les \textit{BDR} existent seulement pour la modélisation du temps comme un moyen de joindre des tables. Cependant, nous avons besoin souvent de désambiguïser la sémantique des relations qui relient les entités. Et comme les données aberrantes se multiplient et que la structure globale de l'ensemble de données devient plus complexe et moins uniforme, la connectivité également se traduit, dans le monde relationnel, par des jointures accrues. Tous cela influence les performances et rend difficile l'évolution d'une base de données existante (en réponse à l'évolution des besoins).
De même, la plupart des \textit{BD NoSQL} (\textit{i.e.} clé-valeur, de document ou de colonnes) stockent des ensembles de documents/valeurs/colonnes, déconnectés. Cela rend difficile leur utilisation pour des données et des graphes connectés.

\subsubsubsection{BD-RDF natives}
\label{native}

\parindent = 0.5cm
Ces BD-RDF offrent un stockage spécifique qui ne repose pas sur un SGBDR (\textit{e.g. AllegroGraph, HyperGraphDB, DEX, InfiniteGraph, Neo4j, Sones, etc}). D'après \cite{native}, elles sont groupées en deux types : \\
$\bullet$ Les \emph{\textbf{BD-RDF in-memory}} : dès le lancement de l'application, l'ontologie est stockée en mémoire principale. Cette méthode peut ne pas être efficace dans le cas du stockage d'un de données extrêmement énorme. \\
$\bullet$ Les \emph{\textbf{BD-RDF non-memory}} : l'ontologie ainsi que les données sont stockées sur disque. \\

\parindent =0cm
Nous pouvons citer quelques exemples de \textit{BD-RDF} natives : 

\parindent =0.5cm
\emph{AllegroGraph} \textit{(2005)} : est une \textit{BD} graphique moderne et persistante pour le stockage et l'interrogation des données RDF. Elle utilise le stockage sur disque et donc elle peut stocker des milliards de triplets. \textit{AllegroGraph} fournit une \textit{API} Java très puissante. \\

\parindent =0.5cm
\emph{DEX} \textit{(2007)} : son objectif est d'assurer de bonne performance dans la gestion de très grands graphes. Elle permet même l'intégration de diverses sources de données. Elle a un bon modèle d'intégrité pour la gestion de graphes temporaires et persistants. Les opérations comme l'analyse des liens, l'analyse du réseau social, la reconnaissance des patrons et la recherche par mots-clés, peuvent être réalisées grâce à son \textit{API} Java. \\

\parindent =0.5cm
\emph{Neo4j} \textit{(2007}) : est une \textit{BD} graphique open source, implémentée en Java. Elle est intégrée et basée sur disque. Elle est capable de représenter les données d'une manière facilement accessible. Dans Neo4j, un graphe peut être considéré comme une structure richement inter-connectée. \\

\parindent=0cm
\textbf{La puissance des bases de données graphiques (\textit{i.e.} \textit{BD-RDF} natives) \cite{native} :} 

\parindent = 1cm

$\bullet$ \textbf{Performance :} les performances des \textit{BD} graphiques restent relativement constantes, même lorsque l'ensemble de données augmente. C'est parce que les requêtes sont localisées sur une partie du graphe. Par conséquent, le temps d'exécution pour chaque requête est proportionnel uniquement à la taille de la partie du graphe traversée pour traiter cette requête, et non pas à la taille du graphe global. \\ 

\parindent = 1cm

$\bullet$ \textbf{Flexibilité :} les graphes sont de nature additifs, ce qui signifie que nous pouvons ajouter de nouveaux types de relations, de nouveaux nœuds, de nouvelles étiquettes et de nouveaux sous-graphes à une structure existante, sans perturber les fonctionnalités d'une application. Nous n'avons pas à modéliser notre domaine de manière exhaustive à l'avance. 
La flexibilité du modèle graphique signifie également que nous pouvons effectuer moins de migrations, ce qui réduit alors les risques et les frais  généraux de maintenance. \\

\parindent = 1cm
$\bullet$ \textbf{Agilité :} la nature sans schéma du modèle de données graphiques, associée à la nature testable de l'\textit{API} d'une base de données graphiques, ainsi qu'au langage d'interrogation, nous permet de faire évoluer notre application de manière contrôlée en fonction des environnements métier changeants. \\

 La description de données dans les \textit{BD-RDF} natives, différencie d'une \textit{BD-RDF} à une autre. En effet, la représentation de données est interne et propre à chaque \textit{DB-RDF }native. Nous pouvons ainsi distinguer les \textit{BD-RDF }à partir des structures de données utilisées. 

\subsubsubsection{Différences entre RDF Store et BD-RDF}

\parindent = 0.5cm
Même si ses derniers ont été conçus afin de stocker des données inter-connectées (Linked Data), sachant que \textit{RDF} est un type particulier de Linked Data, il y a quelques différences entre eux et qui sont comme suit : \\ 

\parindent=0cm
- Les bases de données graphiques sont plus polyvalentes avec les langages d'interrogation ; \\
- Les bases de données graphiques peuvent stocker divers types de graphes, notamment les graphes non orientés, les hypergraphes, \textit{etc}. Par contre, les \textit{Triples Stores} se concentrent uniquement sur le stockage de rangées de triplets \textit{RDF}. Ils ont pour but de stocker des données RDF qui représentent un type particulier de graphes : graphe étiqueté orienté ; \\
- Les bases de données graphiques sont basées sur les nœuds et les propriétés alors que les \textit{Triples Stores} sont centrés arcs ; \\ 
- Les \textit{Triples Stores} fournissent également des inférences sur les données tandis que les bases de données orientées graphe ne le font pas ; \\
- Les bases de données graphiques peuvent assurer une très haute performance au niveau de l'accessibilité et la navigation (\textit{e.g.} la récupération du chemin le plus court entre deux nœuds, ou bien la détermination de la manière dont n'importe quels deux nœuds sont connectés). \\

\parindent=0.5cm
D'après \cite{angles2012comparison} et en considérant le niveau de maturité des bases de données graphiques en termes d'équipements offerts par le système de gestion de base de données, il existe deux types de développements : \\
- Bases de données graphiques : elles doivent fournir les composants majeurs d'un système de gestion de base de données ; \\
- Magasins graphiques (graph stores) : regroupe des implémentations offrant des fonctionnalités de base pour stocker et interroger des graphes. \\

\parindent = 0.5cm
Puisque \textit{RDFS} permet de raisonner sur les données \textit{RDF}, quelques \textit{BD-RDF} utilisent un moteur de raisonnement. 
Nous allons voir dans la sous-section suivante, les techniques existantes destinées à réaliser des raisonnements sur les \textit{BD-RDF}.

\subsection{Moteurs de raisonnement et saturation de données}

\parindent = 0.5cm
Appelés également moteurs d'inférence ou raisonneurs, ils permettent d'effectuer des inférences sur un ensemble de données \textit{RDF}. Leur principal rôle est de générer de nouvelles données sous forme de triplets, ou également de vérifier la validité de l'ensemble de données, tout cela en appliquant des règles d'inférences. La plupart des \textit{BD-RDF} possèdent un raisonneur (\textit{e.g. Hermit, Pellet, Racer}, \textit{etc}). \\

\parindent=0cm
- \textbf{Validation de données} : consiste à vérifier s'il n'existe pas de données inconsistantes. Une donnée inconsistante peut, par exemple, être une restriction qui spécifie qu'une classe doit hériter de deux classes disjointes. Ainsi, cette information ne peut être jamais effectuer en suivant les règles établies par le schéma d'ontologies. Le moteur d'inférence affichera alors une incohérence. \\

\parindent =0cm
- \textbf{Inférence} : la seconde caractéristique du raisonneur consiste à créer un modèle de données dérivées (inférées), tout en se basant sur le modèle de base ainsi que sur des restrictions et des règles définies par l'ontologie. \\

\parindent = 0.5cm
Les moteurs d'inférence procèdent avec deux principales techniques : \textit{le chaînage avant} et \textit{le chaînage arrière}. Ces dernières sont très coûteuses en termes du temps par rapport aux règles d'inférence utilisées et à la taille des données à inférer. Par conséquent, ce raisonnement peut être effectué avant l'exécution des requêtes, on appelle cela \textit{saturation des données}, ou bien lors de l'interrogation de données, c'est la \textit{reformulation des requêtes} \cite{native}. 

\subsubsection{Saturation des données}

\parindent =0.5cm
C'est la production de quasiment tous les triplets possibles en appliquant toutes les règles d'inférence sur les données \textit{RDF}. Ces triplets sont appelés alors \textit{triplets inférés}, et c'est le rôle du moteur de raisonnement de les produire. Ces triplets dérivés seront rajoutés à l'ensemble de données de départ, et à ce moment là, nous pouvons dire que l'ensemble de données \textit{RDF} est saturé. \\

\parindent = 0.5cm
Pendant la saturation, chaque triplet inféré peut participer à l'inférence d'autres triplets jusqu'à ce qu'un point fixe soit atteint (\textit{i.e.} pas de nouveaux faits peuvent être inférés). Ensuite, la réponse à la requête est réduite à une simple évaluation de la requête sur cet ensemble de données saturé. \\
Par conséquent, la saturation fournit un support efficace pour le traitement de requêtes. Cependant, elle nécessite beaucoup du temps pour qu'elle soit calculée, et beaucoup d'espace de stockage pour le sauvegarde des données inférées, également un effort important pour le maintien ou le re-calcul de la fermeture après les mises à jour.

\parindent = 0cm
De même, toute modification des données RDF nécessite de refaire la saturation (car du moment où l'ensemble de données est saturé, aucun nouveau triplet ne peut être inféré avec une ou plusieurs règles d'inférence). Afin de résoudre ce problème, en particulier si les données \textit{RDF} sont dynamiques, quelques BD-RDF proposent la technique de reformulation de requêtes. 

\subsubsection{Reformulation des requêtes}

\parindent =0.5cm
C'est une alternative à la technique de saturation de données et consiste à réécrire la requête (au moment de son exécution) en prenant en considération un ensemble de règles d'inférence. Le langage d'interrogation \textit{SPARQL} (section \ref{sparql}) dispose de beaucoup de régimes d'inférence sous lesquels une requête peut être exécutée. Parmi les régimes d'inférence de \textit{SPARQL}, nous pouvons citer le régime d'inférence \textit{RDFS}.


\parindent=0cm
De même, pour les utilisateurs et les systèmes qui désirent définir eux-mêmes leurs règles d'inférences, le langage d'interrogation SPARQL propose un régime d'inférence \textit{RIF (\textbf{R}ules \textbf{I}nterchanged \textbf{F}ormat)} permettant de représenter et prendre en considération des règles d'inférence non prédéfinies dans \textit{SPARQL}. \\

\parindent =0.5cm
$\ominus$ Le principal inconvénient de la reformulation réside dans le fait que les requêtes reformulées ont tendance à être syntaxiquement complexes. Cela augmente généralement et de manière significative, le coût de l'évaluation. \\

\parindent = 0.5cm
Afin d'exploiter les données \textit{RDF} stockées dans les \textit{BD-RDF}, qu'elles sont saturées ou non, des langages d'interrogation ont été proposés. Ces derniers sont de différents types. Nous pouvons citer, par exemple, la catégorie de langages qui traitent les données sous forme de triplets sans prendre en compte la sémantique (qui est associée aux éléments des triplets), \textit{e.g.} le langage \textit{SPARQL, RDQL, SquishQL}, \textit{etc.} Dans une autre catégorie de langages, la séparation entre le schéma de données et de l'ontologie est bien mentionnée au moment de l'interrogation (\textit{e.g.} le langage\textit{ RQL, SeRQL} et \textit{OntoQL}). Il existe également une autre famille de langages qui sont inspirés de \textit{XPath} ou \textit{XQuery} afin d'interroger des données et des ontologies (\textit{e.g.} \textit{RDF-Path }et \textit{Versa}). 

\subsection{Le langage d'interrogation de graphes RDF : SPARQL }
\label{sparql}

\parindent =0.5cm 
Stocker une quantité immense de données de façon structurée n'aurait aucun intérêt s'il était irréalisable et impossible de pouvoir accéder à ces données. Alors, il est indispensable d'avoir un langage d'interrogation, à l'instar des autres langages tels que \textit{SQL}. Pour le Web sémantique, c'est \textit{SPARQL} qui représente le langage standard d'interrogation de graphes \textit{RDF}. \\ 

\textbf{SPARQL}, est l'acronyme de \textit{\textbf{S}imple \textbf{P}rotocol \textbf{A}nd \textbf{R}DF \textbf{Q}uery \textbf{L}anguage}, est une recommandation du \textit{W3C} depuis 15 janvier 2008. Il joue le rôle d'un pont entre les technologies du Web sémantique (dont \textit{RDF}), et les plateformes Web déjà existantes. Il est une \textit{API} universelle d'accès aux données. \\

\parindent=0cm
\textit{SPARQL} est, d'une part, un protocole et un langage de requête permettant l'accès aux données RDF. Il est aussi un protocole d'accès comme un service Web \textit{(SOAP : \textbf{S}imple \textbf{O}bject \textbf{A}ccess \textbf{P}rotocol)}, et également, un langage de présentation des résultats (\textit{XML}). \\

\parindent=0cm
Il ne prend pas en charge l'inférence en elle-même. Il ne fait rien de plus que de prendre les descriptions de ce que l'application veut, sous la forme d'une requête, et renvoie le résultat sous forme de graphe \textit{RDF}. Par ailleurs, \textit{SPARQL} peut être utiliser afin d'exprimer des requêtes sur différentes sources de données. \\

\parindent=0.5cm
 La requête \textit{SPARQL} officielle adopte quatre formes diverses: \\  
\parindent=0cm
- Requête de la forme \textbf{\textit{SELECT}}, renvoie la valeur de la variable, qui peut être attachée par un modèle de requête équivalent/correspondant; \\
- Requête ayant la forme \textbf{\textit{ASK}}, renvoie vrai si la requête correspond aux données et faux sinon; \\
- Requête de la forme \textbf{\textit{CONSTRUCT}}, retourne les réponses qui satisfont un ensemble de contraintes, sous forme de graphe \textit{RDF}. La structure du graphe retourné est décrite par un patron (ou \textit{template}) dans la requête. Elle est comparable à une vue matérialisée dans les \textit{SGBDR}; \\
- Requête de la forme \textbf{\textit{DESCRIBE}}, renvoie un graphe RDF décrivant une ressource RDF particulière.\\

\parindent=0.5cm
 Parmi les fonctionnalités du langage \textit{SPARQL}, nous citons : \\  
- \textbf{\textit{FILTERS}} : contraint les résultats de la requête à uniquement ceux où l'expression du filtre est évaluée à \textit{TRUE}; \\
- \textbf{\textit{OPTIONAL}} : puisque les données RDF sont des données semi-structurées, lorsqu'une requête est exécutée, elle n'échoue jamais même si les données n'existent pas. Ceci est réalisé grâce à la clause \textit{OPTIONAL}; \\
- \textbf{\textit{LIMIT}} : met une limite au nombre de résultats de requête retournés; \\
- \textbf{\textit{ORDER BY}} : cette clause est utilisée pour classer (dans l'ordre croissant ou décroissant) les résultats de la requête; \\
- \textbf{\textit{DISTINCT}} : est utilisée pour éliminer les doublons présents dans le résultat de la requête; \\
- \textbf{\textit{REGEX}} : cet opérateur appelle la fonction de correspondance pour faire correspondre le texte avec un patron d'expression régulière; \\
- \textbf{\textit{UNION}} : combine des patrons graphiques. \\ 

\parindent =0.5cm
\textbf{Modification} : La norme \textit{SPARQL 1.1} introduit, entre autres, les requêtes \textit{SPARQL} d'ajout ou de suppression de triplets dans une base de données RDF grâce aux clauses \textit{\textbf{INSERT}} et \textit{\textbf{DELETE}}. Ces requêtes permettent de créer de nouveaux triplets suivant un patron de graphe (\textit{i.e. TP}, section \ref{tp}). Elles peuvent utiliser une clause \textit{WHERE} afin d'insérer des données qui sont générées à partir des données existantes.  

\subsubsection{Patron de triplet (TP) et patron de graphe (BGP)}
\label{tp}

\parindent =0.5cm
 L'élément de base d'une requête \textit{SPARQL} est le \textit{\textbf{BGP}} \textit{(\textbf{B}asic \textbf{G}raph \textbf{P}attern)} (\textit{i.e. patron de graphe}), qui peut être considéré comme un ensemble de fragments de requête élémentaires, appelés \textit{\textbf{TP} (\textbf{T}riple \textbf{P}attern)}. Un \textit{TP} est représenté sous la forme d'un triplet RDF, pouvant comporter une ou plusieurs variables (\textit{i.e.} un identifiant alphanumérique commençant par \textit{"?"}) en tant que sujet, prédicat ou objet. Les sujets et les prédicats peuvent être des \textit{URIs}, des nœuds blancs ou des variables, alors que les objets peuvent, en plus, être des littéraux. Le patron de triplet (\textit{TP}) permet de trouver les triplets \textit{RDF}, à partir d'un graphe, qui satisfont le modèle qu'il définit, ce processus est appelé \textit{\textbf{matching}}. 
 
\parindent =0.5cm
 
Dans un patron de graphe (\textit{BGP}), les variables qui sont partagées par au moins deux patrons de triplets (\textit{TP}), sont nommées \textit{\textbf{variables de jointures}}. Ainsi, interroger un graphe \textit{RDF} avec un \textit{BGP}, revient à chercher des valeurs pour chacune des variables du patron de graphe pour que le graphe \textit{RDF} obtenu (en remplaçant les variables par ces valeurs) soit un sous-graphe du graphe \textit{RDF} global. 

\subsubsubsection{Définition formelle d'un patron de triplet et d'un patron de graphe}

\parindent =0.5cm
Dans \textit{SPARQL}, un graphe \textit{RDF} est vu comme un ensemble de triplets \textit{RDF}. Pour un graphe \textbf{G}, nous supposons que: \\
- \emph{U}, l'ensemble des \textit{URIs} associés à \textbf{G} ; \\
- \emph{L}, l'ensemble des littéraux dans \textbf{G} ; \\
- \emph{B}, l'ensemble des nœuds blancs / anonymes de \textbf{G}. \\

\parindent = 0.5cm
Un triplet \textit{RDF} est, formellement, un élément de l'ensemble $(U \cup B) \times U \times (U \cup B \cup L)$. Comme nous venons de le dire, un patron de triplet \textit{(TP)} est un triplet \textit{RDF} qui peut contenir des variables. Ainsi, un patron de triplet \textit{(TP)} peut être défini comme un élément de l'ensemble $(U \cup V ) \times (U \cup V ) \times (U \cup V \cup L)$ ; où \textit{V} est un ensemble de variables disjoint de \textit{U, L} et \textit{B}. On supposera que \textit{var(t)} est l'ensemble des variables d'un patron de triplet \textit{(TP)}. Un patron de graphe \textit{(BGP)} est interprété comme une conjonction de patron de triplets, \textit{i.e.} si \textit{Q} est un patron de graphe, ainsi, $Q = t_1 \wedge t_2 \wedge ... \wedge t_n$ ; où chaque $t_i$ (tel que $1 \leq i \leq n$) est un patron de triplet. On notera alors $\mid Q \mid$ le nombre de patrons de triplets \textit{(TP)} d'un patron de graphe \textit{(BGP)}. Tout comme pour \textit{var(t)}, on note \textit{var(Q)}, l'ensemble des variables distinctes des patrons de triplet de \textit{Q}.  

\subsubsection{Avantages de SPARQL} 
 Avec les services Web 2.0 existants, les Web services sont bien évidemment disponibles, alors qu'ils ne sont pas normalisés. Il faut donc savoir les méthodes du Web services et la structure des données afin de pouvoir les interroger. \\ 

\textbf{$\oplus$} Avec \textit{SPARQL}, nous n'avons pas besoin de connaître à l'avance la structure et le contenu des informations afin de pouvoir les récupérer. En fait, \textit{SPARQL} permet d'interroger n'importe quel constituant d'un triplet admettant la forme \emph{Sujet\textbf{-}Prédicat\textbf{->}Objet}. \\ 

\parindent =0.5cm
\textbf{$\oplus$} Une base de données graphique, qui peut être interrogée par des requêtes \textit{SPARQL} via \textit{Internet}, propose un point d'accès, appelé \textit{endpoint SPARQL}, où l'envoie de requêtes se fait via \textit{HTTP (i.e. \textbf{H}yper \textbf{T}ext \textbf{T}ransfer \textbf{P}rotocol)}. \\

\subsubsection{Limitations de SPARQL} 
$\ominus$ \textit{SPARQL} est une technologie jeune, la spécification du \textit{W3C} en cours possède encore quelques limites. Ces limitations deviennent claires lorsqu'on compare \textit{SPARQL} à des langages d'interrogation stables et établis, à l'instar de \textit{SQL} ou \textit{XQuery}. La liste suivante présente les caractéristiques manquantes dans \textit{SPARQL}. \\

\textbf{ Agrégation :} la spécification courante ne prend pas en charge les fonctions d'agrégation, comme l'addition des valeurs  numériques, le calcul de la moyenne, \textit{etc} ; \\

\textbf{ Négation :} \textit{SPARQL} ne supporte pas bien la "négation", \textit{e.g.} trouver toutes les personnes qui sont les amis de Bob mais qui ne connaissent pas Alice ; \\

\textbf{ Les prédicats ne peuvent pas avoir des propriétés :} Cela peut être une limitation \textit{RDF} héritée par \textit{SPARQL}, puisque \textit{RDF} représente tout dans des triplets. Il est facile d'implémenter les propriétés d'un nœud en utilisant des triplets supplémentaires. En revanche, il est très difficile d'implémenter des propriétés dans des arcs.
Dans \textit{SPARQL}, il n'existe aucun moyen d'attacher une propriété à un prédicat, \textit{e.g.} Bob connaît Peter depuis 5 ans ; \\

\textbf{ Support des correspondances floues avec des résultats classés :} \textit{SPARQL} est basé sur un modèle de requête booléen conçu pour une correspondance exacte. Exprimer une correspondance floue avec un résultat classé, est très difficile.
Par exemple, retrouvez les 20 premiers postes "similaires" à ce poste et classés par degré de similarité ; \\

\textbf{Expressions de chemin :} \textit{SPARQL} ne traite pas la spécification des expressions de chemin, \textit{e.g.} à partir d'une seule requête \textit{SPARQL}, il est impossible de calculer la fermeture transitive d'un graphe ou bien d'extraire tous les nœuds qui sont accessibles à partir d'un nœud donné. Cette lacune a été plusieurs fois identifiée dans des publications précédentes. Ainsi, différentes propositions pour l'intégration des expressions de chemin dans le langage, ont été proposées par \textit{Pérez et al., (2009), Kochut et Janik, (2007) et Alkhateeb et al., (2009)} ; \\

\textbf{Vues :} dans les langages d'interrogation traditionnels tels que SQL, des vues logiques sur les données jouent un rôle considérable. Ils sont indispensables pour la conception de base de données et pour la gestion de l'accessibilité. \textit{SPARQL} ne prend en ce moment pas en charge la spécification de vues logiques sur les données ; \\

\textbf{Prise en charge des contraintes :} le mécanisme d'affirmation et de vérification des contraintes d'intégrité dans les bases de données \textit{RDF}, n'est pas géré au niveau de la spécification actuelle de \textit{SPARQL}. \\

\parindent =0.5cm
En somme, les ontologies et les technologies liées à ces dernières, telles que le langage \textit{SPARQL}, sont encore inconnues ou peu maîtrisées par des utilisateurs non-experts. Ces utilisateurs peuvent alors avoir besoin d'un appui ou d'une assistance lors de l'exploitation de ces données. Par ailleurs, les ontologies contiennent des données hétérogènes et dynamiques, pouvant rapidement évoluer avec le temps et rendre, par la suite, les connaissances des usagers obsolètes pour l'exploitation de ces données \textit{RDF}. Ce manque de connaissance des utilisateurs sur les données des \textit{RDF}, est l'un des causes des formulations incomplètes ou incorrectes des requêtes qui renverront, par conséquent, des réponses insatisfaisantes aux utilisateurs. Pour éviter cela, les usagers ont besoin d'être accompagnés et soutenus dans la construction des requêtes et/ou dans le traitement de ces dernières. Pour palier ce problème, plusieurs approches coopératives ont ainsi été développées afin de proposer des solutions alternatives aux utilisateurs.

\section{Revue sur les différents travaux de comparaison des BD-RDF}

\parindent = 0.5cm
\textit{Rohloff et al. (2007)} dans \cite{rohloff2007evaluation}, ont évalué les nouvelles technologies des \textit{Triples Stores} les plus prometteurs du Web sémantique. Cette étude se concentre sur les préoccupations particulières de l'application client telles que, le volume de données, les mesures de performance fondamentales (temps du chargement et vitesse de l'interrogation), et les inférences de base. Cette étude tire parti du benchmark \textit{LUBM  (i.e. \textbf{L}ehigh \textit{U}niversity \textbf{B}ench\textbf{M}ark}). 
Les triples stores évalués dans cette étude sont: \textit{MySQL, DAML BL, SwiftOWLIM, et BigOWLIM}, et qui supportent différents langages d'interrogation (\textit{e.g. SPARQL, SeRQL}). Idéalement, un Triple Store doit avoir un temps de réponse qui n'augmente pas (ou à la limite augmente faiblement), quand le nombre de triplets à charger accroît. Ceci n'était pas le cas avec les bases de données graphiques testées. \\

\parindent=0cm
Les métriques prises en considération sont : \\
- Temps du chargement accumulé : le temps mesuré en heure, pour charger les fichiers OWL décrivant les départements universitaires, dans des triples stores, pour un nombre donné de triplets ; 

- Temps de réponse à la requête : est calculé comme la moyenne des temps d'exécution pour chacune des quatre requêtes identifiées ; 

- Les exigences d'espace disque : la quantité d'espace disque demandée pour charger les données d'évaluation\textit{, etc}. 
 \\ 
 
\parindent=0.5cm
\textit{Stegmaier et al.} en 2009 ont évalué les bases de données \textit{RDF} existantes à ce moment-là, et qui supportent le langage d'interrogation \textit{SPARQL}, tout en se basant sur quelques critères \cite{stegmaier2009evaluation} : \\
- L'extensibilité : ce qui permet l'intégration de nouvelles fonctionnalités ; \\
- Aperçu architectural ; \\
- Disponibilité de langages d'interrogation (\textit{i.e.} si elles supportent d'autres langages d'interrogation) ;  \\
- Interprétation des formats de données \textit{RDF} ; \\
- L'expressivité de \textit{SPARQL} et la performance des \textit{frameworks}/applications. \\

\parindent =0.5cm
En 2011, \textit{Morsey et al.} ont proposé une méthodologie de création de \textit{Benchmark SPARQL} générique appelée \textit{DBPSB (i.e. \textbf{DBP}edia \textbf{S}PARQL \textbf{B}enchmark)}, afin d'évaluer la performance et la capacité des \textit{Triples Stores} les plus populaires : \textit{Virtuoso, Sesame, Jena-TDB }et \textit{BigOWLIM} \cite{morsey2011dbpedia}. Pour ce faire, ils ont utilisé 24 modèles de requête \textit{SPARQL}, couvrant les caractéristiques de \textit{SPARQL} les plus couramment utilisées, et contenant toutes les fonctionnalités de ce dernier (\textit{e.g. FILTER, REGEX, etc.})
\\ Leur méthodologie suit les quatre exigences clés pour les \textit{Benchmarks} spécifiques au domaine : \\
-	Pertinence : le test des opérations typiques dans un domaine spécifique ; \\
-	Portabilité : le \textit{Benchmark} doit être exécutable sur différentes plateformes ; \\
-	Évolution : il est possible d'évaluer ce Benchmark sur de petits et de très grands ensembles de données ; \\
-	Compréhensible. \\
\textit{Virtuoso} était clairement le plus rapide, suivi par \textit{BigOWLIM}, \textit{Sesame} et \textit{JenaTDB}. Il y avait en moyenne une baisse linéaire de la performance de la requête avec l'augmentation de la taille de l'ensemble de données. \textit{Virtuoso} a été le seul qui a répondu à toutes les requêtes à temps, et celui qui a pu évoluer sur le \textit{DBPSB}. \\

\parindent =0.5cm
D'après \emph{Angles (2012)} dans \cite{angles2012comparison}, une base de données graphique doit fournir : \\
- Des interfaces externes (interfaces utilisateur ou \textit{API}) ; \\
- Des langages de base de données (pour la définition, la manipulation et l'interrogation de données) ; \\
- Un optimiseur de requêtes ; \\
- Un moteur de base de données (modèle de niveau intermédiaire) ; \\
- Un moteur de stockage (modèle de bas niveau) ; \\
- Un moteur de transaction ; \\
- Des fonctionnalités d'opération et de gestion (pour le réglage, la sauvegarde, la récupération, \textit{etc}). \\
Parmi les BD graphiques qui respectent ces conditions, il a cité, \textit{AllegroGraph, DEX, HyperGraphDB, InifiniteGraph, Neo4j} et \textit{Sones}. \\

\begin{figure}[H]
\begin{center}

\includegraphics[scale=0.9]{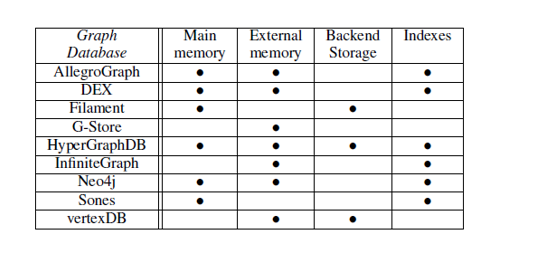}
\end{center}
\caption{Fonctionnalités du stockage des données RDF (Angles (2012)) \label{figure 2.50}}

\end{figure}

\parindent=0cm
Par la suite, il a examiné le support de trois schémas/manières de stockage (à savoir, la mémoire principale, la mémoire externe et le stockage principal) des différentes \textit{BD} graphiques citées auparavant, ainsi que la mise en œuvre des indexs (figure \ref{figure 2.50}). Il est important de souligner que la gestion d'une énorme quantité de données est une exigence importante dans les applications de la vie réelle pour les bases de données graphiques. Par conséquent, la prise en charge du stockage en mémoire externe est une exigence principale. De plus, les indexs sont importants pour améliorer les opérations de récupération de données.

\begin{figure}[H]
\begin{center}

\includegraphics[scale=0.9]{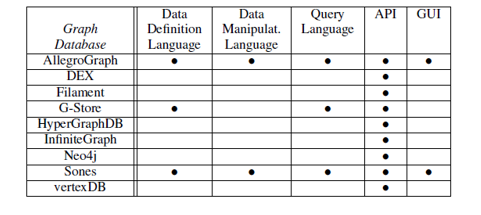}
\end{center}
\caption{Fonctionnalités de manipulation et d'opération (Angles (2012)) \label{figure 2.51}}

\end{figure}

\parindent=0cm
Du point de vue fonctionnement et manipulation des données (figure \ref{figure 2.51}), il a évalué si une base de données graphique implémente des langages de base de données, des interfaces de programmation d'application (API) et des interfaces graphiques utilisateur (GUI). Il a considéré trois langages de base de données : \\ 
- Le langage de définition de données \textit{(\textbf{D}ata \textbf{D}efinition \textbf{L}anguage)}, qui permet de modifier le schéma de la base de données en ajoutant, modifiant ou supprimant ses objets ; \\
- Le langage de manipulation de données \textit{(\textbf{D}ata \textbf{M}anipulation \textbf{L}anguage)}, qui permet d'insérer, de supprimer et de mettre à jour des données dans la BD ; \\
- Le langage d'interrogation \textit{(\textbf{Q}uery \textbf{L}anguage)}, qui permet d'extraire des données.

Il a constaté, également, que les contraintes d'intégrité sont mal étudiées par les \textit{BD} graphiques. Il a, entre autres, considéré plusieurs contraintes d'intégrité : 

\parindent = 0cm
- La vérification des types, pour tester la cohérence d'une instance par rapport aux définitions dans le schéma; \\
- L'identité de nœud / de relation, pour vérifier qu'une entité ou une relation peut être identifiée par une valeur (\textit{e.g.} un nom ou un \textit{ID}) ou par les valeurs de ses attributs (\textit{e.g.} une identification de voisinage); \\
- L'intégrité référentielle, pour vérifier que seules les entités existantes sont référencées; \\
- Vérification de cardinalité, pour vérifier l'unicité des propriétés ou des relations; \\
- Dépendance fonctionnelle, pour tester qu'un élément du graphe détermine la valeur d'un autre; \\
- Contraintes du patron graphique pour vérifier une restriction structurelle (\textit{e.g.} des contraintes de chemin). \\ 

Il a finalement conclu que la prise en charge des schémas évolutifs est la caractéristique des bases de données graphiques couramment utilisées pour justifier l'absence de contraintes d'intégrité. Cela n'est pas un argument valide, sachant que que la cohérence des données, dans une base de données, est égale ou même plus importante qu'un schéma flexible. \\

\begin{figure}[H]

\begin{center}
\includegraphics[scale=0.9]{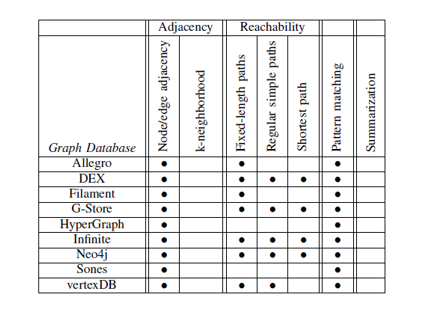}
\end{center}
\caption{BD graphiques actuelles et leur support pour des requêtes essentielles (Angles (2012)) \label{figure 2.52}}
\end{figure}

Il a testé la puissance des bases de données graphiques à résoudre plusieurs types requêtes, qui peuvent être considérées essentielles (figure \ref{figure 2.52}). \\

\textbf{1. Requêtes d'adjacence :} \\
Deux nœuds sont adjacents (voisins), s'il existe un arc qui les relie. De même, deux arcs sont dits adjacents s'ils relient un nœud en commun.
Quelques requêtes typiques dans ce groupe sont, l'adjacence de nœud/arc de base, afin de tester si deux nœuds sont adjacents, de calculer les k-voisins d'un nœud, de lister tous les voisins d'un nœud\textit{, etc.} \\

\textbf{2. Requêtes d'accessibilité :} \\
Ce type de requêtes est caractérisé par les problèmes de chemin et de parcours. Il a considéré deux types de chemins : \\

\parindent=0.5cm
-	Les chemins à longueur fixe : contiennent un nombre fixe de nœuds et d'arcs. 

\parindent=0.5cm
-	Les chemins simples réguliers : qui fournissent certaines restrictions de nœud et d'arc.

Dans ce contexte, un problème plus complexe consiste à trouver le plus court chemin entre deux nœuds. \\

\parindent=0cm
\textbf{3. Requêtes de correspondance de patrons (pattern matching) :} \\
Elles permettent de trouver tous les sous-graphes d'un graphe de données, qui sont isomorphes au patron graphique.
Elles traitent deux problèmes : \\

\parindent=0.5cm
- Le problème d'isomorphisme graphique qui a une complexité de calcul inconnue.

\parindent=0.5cm
- Le problème d'isomorphisme sous-graphique qui est un problème NP-complet. \\

\parindent=0cm
\textbf{4. Requêtes de synthèse :} \\
Ces requêtes ne sont pas censées consulter la structure du graphe. Cependant, elles sont basées sur des fonctions spéciales qui permettent de résumer ou d'opérer sur les résultats de la requête, renvoyant normalement une seule valeur. Les fonctions agrégées (\textit{e.g.} moyenne, count, maximum\textit{, etc}) sont incluses dans ce groupe. \\

De plus, \textit{Angles} a considéré des fonctions pour calculer certaines propriétés d'un graphe ainsi que ses éléments. Par exemple: \\

\parindent=0.5cm
-	L'ordre du graphe (\textit{i.e.} le nombre de sommets) ; 

\parindent=0.5cm
-	Le degré d'un nœud (\textit{i.e.} le nombre de voisins d'un nœud) ; 

\parindent=0.5cm
-	Le minimum, le maximum et le degré moyen dans le graphe ; 

\parindent=0.5cm
-	La longueur d'un chemin (\textit{i.e.} le nombre d'arêtes dans le chemin) ; 

\parindent=0.5cm
-	La distance entre les nœuds (la longueur du plus court chemin entre les nœuds) ;

\parindent=0.5cm
-	Le diamètre du graphe (la plus grande distance entre deux nœuds quelconques)\textit{, etc.} \\

\parindent=0cm
Sachant que la plupart des bases de données graphiques implémentent une API au lieu d'un langage d'interrogation, il a, ainsi, évalué si une API peut répondre à une requête essentielle. \\

\parindent = 0.5cm
D'un autre côté, et puisque il n'existe pas de langages standards pour l'interrogation de graphes RDF (l'absence d'un langage d'interrogation standard est un inconvénient des bases de données graphiques actuelles), en 2012 \textit{Mike Buerli} a divisé les modèles de bases de données graphiques par langage \cite{buerli2012current} : \\
- Bases de données graphiques; \\
- BD graphiques distribuées (\textit{e.g.} \textit{Horton (2010)} créée par \textit{Microsoft, InfiniteGraph (2010))}; \\
- Bases de données graphiques à clé-valeur (\textit{e.g.} \textit{VertexDB (2009), CloudGraph (2010))} ; \\ 
-  Bases de données de documents  (\textit{e.g.} \textit{Orientdb (2009), CouchDB)}; \\
- Bases de données graphiques SQL (\textit{e.g. Filament (2010), G-Store (2010))} ; \\
- Bases de données graphiques Map/Reduce \textit{(e.g. Pregel (2009), Phoebus (2010), Giraph (2011))}. \\

\parindent = 0.5cm
Dans un graphe, les arcs, définis entre les paires de sommets, forment la topologie de ce dernier. Ainsi, une base de données graphique doit être efficace pour exploiter les informations topologiques. Elle doit fournir un moyen rapide et efficace pour traverser les graphes. Lorsqu'on traite des problèmes de parcours du graphe, l'approche habituelle consiste à charger les données du graphe dans la mémoire principale, et à effectuer la (les) opération (s) de traversée, tout en gardant la structure entière dans la mémoire. \\

\parindent = 0cm
Cependant, cette approche impose une limite à la taille du graphe. Les solutions, pour parcourir le graphe sur les supports persistants, ne sont pas courantes, car le coût des accès aléatoires à un support persistant est problématique. C'est un grand défi pour les bases de données graphiques, où un utilisateur s'attend à la persistance des données et à la prise en charge des opérations de parcours rapides. C'est la caractéristique principale qui distingue les bases de données graphiques des autres systèmes de gestion de données. \\

\parindent = 0cm
Du point de vue performance et capacité d'effectuer des opérations de parcours sur la structure graphique, \textit{Ciglan et al. (2012)} dans \cite{ciglan2012benchmarking}, ont développé un \textit{benchmark} de parcours afin de tester la capacité des bases de données graphiques, \textit{Neo4J, DEX, OrientDB}, le référentiel RDF natif (\textit{NativeSail}) et le prototype de recherche \textit{SGDB8}, à:  \\
-	Effectuer des traversées locales à partir d'un ou plusieurs sommets, et explorer leur voisinage \textit{k-hop} ; \\
-	Effectuer des parcours de tout le graphe ; \\
-	Tester les cas où le graphe entier ne peut pas être mis en cache dans la mémoire principale. \\

\parindent=0cm
La conclusion des résultats préliminaires est que les traversées locales dans un grand réseau, sont plus adaptées aux systèmes testés, que ceux nécessitant des traversées de toute la structure du graphe. \\

\parindent =0.5cm 
\textit{Jouili et al. (2013)} ont évalué, dans leur travail \cite{jouili2013empirical}, quatre \textit{BD} graphiques (\textit{Neo4j, Titan, OrientDB et DEX}). Ils ont mis l'accent sur l'utilisation des \textit{Blueprints} (une \textit{API} Java générique pour les bases de données graphiques) dans les applications, ce qui permet de changer de base de données graphique sans des efforts d'adaptation. \\
Ils ont également utilisé, pour leur comparaison, \textit{\textit{GDB} (i.e. \textbf{G}raph \textbf{D}atabase \textbf{B}enchmark)} qui est un cadre d'analyse comparative distribué, open-source sous Java. Son principal objectif est de comparer les bases de données graphiques en utilisant des opérations graphiques habituelles telles que, l'exploration des voisins d'un nœud, la découverte du plus court chemin entre deux nœuds, ou simplement l'obtention de tous les sommets qui partagent une propriété spécifique. \\

\parindent=0cm
Ils ont voulu aussi analyser l'impact de la taille d'un graphe sur les performances d'une base de données, les parcours et les opérations intensives. Ils ont alors utilisé deux graphes : l'un ayant 250 000 sommets (1 250 000 arcs), et l'autre ayant 500 000 sommets (2 500 000 arcs). \\

\parindent =0.5cm
\textit{Lissandriniet et al.} en 2017 \cite{lissandrini2017evaluation} ont fourni une évaluation systématique et complète des bases de données graphiques les plus modernes \textit{ArangoDB et BlazeGraph} (qui sont les BD graphiques les plus récentes), \textit{Neo4J, OrientDB, Sparksee (DEX)} et \textit{Titan}. 

\parindent=0cm
Leurs expérimentations ont été réalisées sur 28 millions nœuds et 31 millions arcs, par opposition aux travaux existants (250 000 nœuds et 2.2 millions arcs). Ils ont utilisé plusieurs bases de test telles que \textit{yeast}, \textit{Frb-S}, \textit{Mico} et \textit{ldbc}. Ils ont évalué pour chaque BD : \\

\textbf{- Le chargement de données :} 

\parindent =1cm
Le temps du chargement : pour tous les ensembles de données, ArangoBD avait le temps le plus court, suivi par Neo4j. Sur l'ensemble de données le plus grand (\textit{Frb-L}), ArangoDB était encore une fois le plus rapide. Neo4j (v.3.0) est juste quelques minutes plus lent, suivi par Neo4j (v.1.9). 

\parindent=1cm
L'espace du chargement : ils ont mesuré la taille du disque occupée par les données pour chaque système. 

\parindent=0cm
\textbf{- Taux d'achèvement :} puisque les bases de données graphiques sont souvent utilisées pour les applications en ligne, il est important de s'assurer que les requêtes s'exécutent dans un délai raisonnable. 

\textbf{- Insertion, mis à jour et suppression :} les tests montrent des performances extrêmement rapides pour Sparksee, Neo4j (v.1.9) et ArangoDB. 

\textbf{- Les sélections générales} : BlazeGraph est le plus lent, avec des performances autour de 500 ms pour la recherche des nœuds, et 4 secondes ou plus pour la recherche des arêtes. Neo4j (v.1.9), OrientDB et Sparksee retournent un résultat dans environ 10 ms. 

\textbf{- Les performances globales :} ils ont comparé le temps cumulé pris par chaque système pour compléter l'ensemble des requêtes. 

\section{Revue sur les méthodes existantes pour l'interrogation par des requêtes SPARQL des graphes RDF}

\parindent =0.5cm
Dans la littérature, plusieurs approches ont essayé d'exploiter et de mettre en œuvre les notions précédemment introduites. \\ 

\parindent = 0.5cm
Ainsi, avoir des données sémantiques sur le Web et compréhensibles par les machines, ne suffit pas. Il est nécessaire d'avoir des outils pour traiter les informations, les transformer et faire des raisonnements au-dessus. \\

\parindent = 0.5cm
Le travail de \textit{Hartig et Heese (2007)} dans \cite{hartig2007sparql}, est fondé sur un modèle graphique de requête \textit{SPARQL}, appelé SQGM \textit{(\textbf{S}PARQL \textbf{Q}uery \textbf{G}raph \textbf{M}odel)}. Au-dessus du \textit{SQGM}, ils ont défini des règles de transformation afin de faciliter et de réécrire la requête en une autre sémantiquement équivalente. Le \textit{SQGM} est utilisé pour stocker les informations sur la requête en cours de traitement. \\

\parindent=0cm
Les éléments de base d'un SQGM sont les opérateurs et les flux de données. Un opérateur traite les données, et un flux de données connecte l'entrée et la sortie de deux opérateurs. Un SQGM peut être interprété comme étant un graphe étiqueté orienté avec des sommets et des arcs représentant des opérateurs et des flux de données, respectivement. Les opérateurs ont une tête, un corps et des annotations supplémentaires. Un arc symbolise le flux de données entre deux opérateurs indiquant qu'un opérateur consomme la sortie d'un autre opérateur. Un flux de données permet également de transférer les données fournies par l'un des opérateurs et consommées par un autre.  \\
Ils ont défini un ensemble de types d'opérateurs afin de pouvoir couvrir les structures du langage SPARQL. Parmi ces opérateurs, il existe l'opérateur du patron graphique (BGP) qui est définit dans la spécification SPARQL. Dans la représentation graphique d'un opérateur du patron graphique, la tête liste les noms des variables liées par l'opérateur, alors que le corps contient les propriétés de l'opérateur.
\\

\parindent =0.5cm
Une autre idée pour l'interrogation en utilisant \textit{SPARQL} a été présentée dans le travail de \textit{Quilitz et Leser (2008)}, où ils ont développé \textit{\textbf{DARQ}} \cite{quilitz2008querying}. Ce système fournit un accès de requête transparent à plusieurs services \textit{SPARQL}, c'est-à-dire qu'il donne à l'utilisateur l'impression d'interroger un seul graphe \textit{RDF} malgré la distribution de données sur le Web. \\
Le moteur de recherche décompose une requête en sous-requêtes, auxquelles chaque service peut répondre. Un optimiseur de requêtes prend les sous-requêtes et construit un plan d'exécution optimisé. Enfin, les sous-requêtes sont envoyées aux sources de données et les résultats sont intégrés. \\

\parindent=0.5cm
Les données RDF sont considérées comme un graphe orienté étiqueté, et le résumé du graphe est représenté comme une structure d'index, où certains nœuds sont fusionnés tout en conservant tous les arcs. Plus une requête est optimisée, moins il faut du temps pour trouver une réponse correspondante. L'efficacité de la requête joue un rôle important lorsqu'il s'agit de données à grande échelle.

\parindent=0cm
Dans ce cadre, \textit{Nguyen et al.,} ont examiné les approches récentes pour l'évaluation des requêtes \cite{nguyensparql}. La plupart de ces approches récupèrent des sous-graphes (\textit{i.e.} des données RDF) pour chaque patron de triplet dans la requête. Puis, les données RDF sont jointes (fusionnées) pour trouver une réponse correspondante. Ainsi, le nombre des opérations de jointure augmente avec le nombre de patrons de triplets. Cette approche entraîne un grand nombre de données intermédiaires inutiles pour chaque requête, et nécessite un temps considérable pour générer et traiter des données qui ne contribueront pas aux résultats de la requête. Lorsque l'ensemble de données RDF évolue, le volume de résultats intermédiaires peut avoir un effet significatif sur les performances d'une requête.

Pour cette raison, ils ont adopté un nouvel index structurel permettant d'améliorer le stockage et la récupération des données pour le traitement des requêtes SPARQL. \\

Le moteur d'interrogation proposé, possède trois sous-composants : \\
- Query Parser : reçoit les requêtes d'entrée des utilisateurs, extrait leurs BGPs pour les envoyer par la suite à l'optimiseur de requête. Il crée, également, une liste de variables pour l'étape du traitement des requêtes ; \\
- Query Optimizer : génère un plan d'exécution pour la requête. Les patrons de triplets sont disposés dans un ordre, tel que le résultat de correspondance d'un patron sert d'entrée pour le patron suivant dans le plan. Comme le résultat de chaque patron est vérifié pour sa validité à chaque étape de traitement, le nombre de résultats intermédiaires est sensiblement réduit ; \\
- Query Processor : consiste à trouver les points de correspondance avec les variables de la requête, à vérifier ces derniers et ensuite à les combiner pour retrouver l'ensemble de la réponse pour toute la requête. \\

\parindent =0.5cm
En 2009, \textit{Tzoganis et al.} \cite{tzoganis2009querying} ont déployé une application orientée utilisateur pour l'interrogation des ontologies. L'utilisateur peut facilement charger une ontologie existante et proposer des requêtes écrites en langage d'interrogation \textit{SPARQL-DL}. Leur application Web peut s'exécuter à partir de n'importe quel navigateur Web majeur. Elle a été développée avec \textit{Google Web Toolkit (GWT)}. \\
Google Web Toolkit permet aux développeurs d'écrire une interface AJAX avec le langage de programmation JAVA, puis de compiler ce code source JAVA en JavaScript. Il peut facilement être combiné avec plusieurs outils de développement JAVA comme Eclipse. Ainsi, il offre un environnement de développement facile à gérer qui facilite également de déboguer le code JAVA. \\
AJAX est un ensemble de technologies existantes (\textit{e.g.} HTML, CSS, JavaScript, XML\textit{, etc.}) qui viennent ensembles afin de construire des applications Web modernes. \\

\parindent =0cm
La fonctionnalité de leur application devient disponible via son interface. Elle est très simple et minimaliste. Elle se compose de deux parties principales, dont la $1\up{ère}$ prend l'entrée de l'utilisateur (\textit{i.e.} l'ontologie et la requête), ainsi que la $2\up{ème}$ (nommée « résultats ») affiche la réponse à l'utilisateur. \\
Le moteur d'interrogation \textit{SPARQL-DL} de \textit{Pellet} (le raisonneur utilisé) effectue le pré-traitement de la requête. Cela inclut la suppression des atomes redondants et l'élimination des atomes « \textit{sameAs} » avec les b-nœuds :\textit{ « bnodes »} (ils s'agissent des nœuds vides ou de ressources anonymes ou bien ils peuvent être des nœuds dans un graphe RDF qui ne sont ni identifiés par un \textit{URI} ni ils sont des littéraux). 
L'étape du pré-traitement suivante consiste à supprimer les atomes triviaux \textit{e.g. SubClassOf (C, C)} et \textit{SubClassOf (C, owl:Thing)}. 
La dernière étape du pré-traitement consiste à diviser la requête en composants connectés afin d'éviter le calcul croisé de leurs résultats. \\ 

\parindent =0.5cm
Dans la même année, \textit{Hartig et al.} ont proposé une approche qui, durant l'exécution d'une requête, permet de découvrir les sources de données qui peuvent être pertinentes pour répondre à cette requête \cite{hartig2009executing}. Parce que quelques requêtes ne peuvent pas être répondues depuis un seul ensemble de données, le moteur d'interrogation exécute chaque requête à partir d'un ensemble croissant de données pertinentes retrouvées du Web. \\ 
Cette approche a plusieurs limitations : \\
- Besoin des \textit{URIs} initiaux dans les requêtes afin de pouvoir traverser les liens ; \\
- Découverte de liens infinis ; \\
- Dé-référencement d'\textit{URI} qui est une tâche longue ; \\
- Les serveurs du \textit{Linked Data} mettent des restrictions aux clients, comme ne servant qu'un nombre limité de requêtes par seconde. \\

\parindent = 0.5cm
Une méthode qui a été déployée par \textit{Huang et al.} en \textit{2011} dans \cite{huang2011scalable}, fournit un système de \textit{BD-RDF} en cluster, parce que vue la quantité de données \textit{RDF} est en augmentation permanente, il n'est plus possible de stocker ni d'accéder à toutes ces données sur une seule machine tout en assurant des performances raisonnables. Par conséquent, au lieu d'affecter aléatoirement les triplets aux partitions, ils ont utilisé  un algorithme de partitionnement de graphes qui permet aux triplets (qui sont proches dans le graphe \textit{RDF}) d'être stockés sur la même partition, pour enfin paralléliser l'accès à ces machines au moment de l'interrogation. Ils ont introduit une méthode automatique permettant la décomposition des requêtes à des morceaux qui peuvent être exécutés indépendamment sans aucune communication réseau.  \\

\parindent = 0.5cm
\textit{Kollia et al.,} en 2011, ont pensé à développer un nouvel algorithme \cite{kollia2011sparql} qui permet de : \\
- Répondre aux requêtes SPARQL sous OWL 2 (désormais SPARQL-OWL) ; \\
- Décrire une implémentation basée sur le raisonnement HermiT ; \\
- Utiliser cette implémentation pour étudier une gamme de techniques d'optimisation, améliorant les performances de réponse aux requêtes SPARQL-OWL. \\

\begin{figure} [H]
\begin{center}
\includegraphics[scale=1]{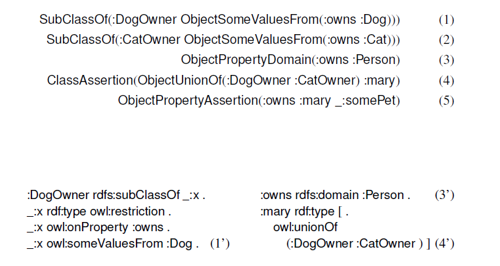}
\end{center}
\caption{Mapping des objets OWL en des triplets RDF (Kollia et al. (2011)) \label{figure 2.58}}
\end{figure}

\parindent=0cm
Ainsi, comme SPARQL est un langage d'interrogation RDF basé sur des triplets, ils ont montré brièvement comment les objets OWL peuvent être mappés en des triplets RDF. La figure \ref{figure 2.58} montre une représentation RDF des axiomes (1), (3) et (4) dans la syntaxe de Turtle.

Les axiomes OWL qui utilisent uniquement l'expressivité du schéma RDF, par exemple les restrictions de domaine et d'image, aboutissent généralement à une traduction directe. Par exemple, l'axiome (3) est mappé au triplet (3').
 
Les expressions complexes de classe, telles que la super-classe dans l'axiome (1), nécessitent généralement des nœuds vides auxiliaires, \textit{e.g.} le nœud vide auxiliaire \textit{-: x} (axiome (1')), pour l'expression de la super-classe, qui est ensuite utilisé comme sujet des triplets suivants.
  
Dans la traduction de l'axiome (4), ils ont utilisé le constructeur de nœud vide de Turtle [ ] et ( ) comme raccourci pour les listes dans RDF (axiome (4')). \\

\begin{figure} [H]
\begin{center}
\includegraphics[scale=0.9]{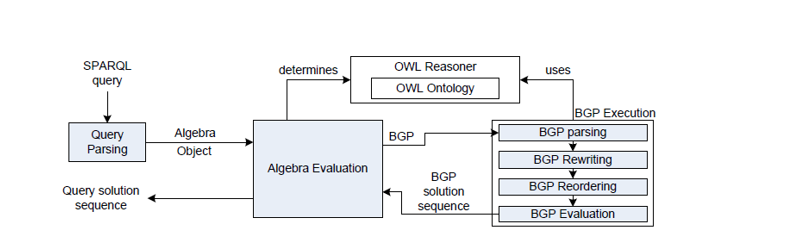}
\end{center}
\caption{Les principales phases de traitement de requêtes (Kollia et al. (2011)) \label{figure 2.59}}
\end{figure}

Ils ont utilisé la bibliothèque ARQ3 du Jena Semantic Web Toolkit pour analyser la requête et pour les opérations d'algèbre SPARQL. 
Le BGP (basic graph pattern) est analysé et mappé dans des modèles d'axiomes en utilisant l'extension OWL API. Les modèles d'axiomes résultants sont ensuite transmis à un optimiseur de requête. Ce dernier applique la réécriture du modèle d'axiome, puis recherche un bon plan d'exécution pour la requête, en se basant sur les statistiques fournies par le raisonneur (figure \ref{figure 2.59}). \\

\parindent = 0.5cm
En effet, le défi principal est de développer des interfaces permettant d'exploiter l'expressivité du modèle de données sous-jacent et du langage d'interrogation, tout en diminuant leur complexité. Ainsi, il est indispensable d'avoir une approche pour la réponse aux questions permettant aux utilisateurs d'exprimer, d'une manière arbitraire, leurs besoins en information complexe en langage naturel sans les obliger à connaître le schéma sous-jacent, le vocabulaire ou le langage d'interrogation. \\

Plusieurs systèmes de réponse aux questions pour les données \textit{RDF} ont été proposés, \textit{e.g.} \textit{Aqualog, Power-Aqua, NLP-Reduce} et \textit{FREyA}. La plupart de ces systèmes permettent de transformer la requête du langage naturel en une représentation à base de triplets. Ensuite, avec l'application des métriques de similarité et des heuristiques de recherche, ils vont retrouver les correspondances des sous-graphes à partir des référentiels \textit{RDF}.
Cependant, parfois la structure sémantique originale de la question ne peut pas être fidèlement capturée en utilisant des triplets. \\

\begin{figure}[H]
\begin{center}
\includegraphics[scale=0.9]{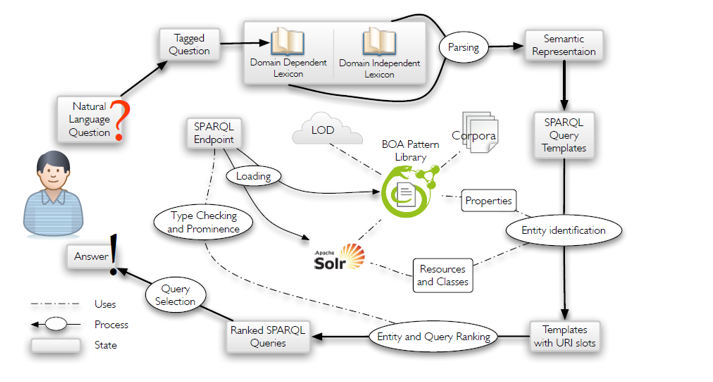}
\end{center}
\caption{Le modèle basé sur la génération de requêtes SPARQL (Unger et al. (2012)) \label{figure 2.61}}
\end{figure}

\parindent=0cm
C'est pour cette raison que \textit{Unger et al. (2012)} dans \cite{unger2012template}, ont proposé une nouvelle approche, qui permet de générer tous les modèles de requête \textit{SPARQL}. Cela se fait en capturant la structure sémantique du besoin en information de l'utilisateur ainsi qu'en ne laissant ouverts que des emplacements spécifiques pour les ressources, les classes et les propriétés. Ces emplacements doivent être déterminés par rapport à l'ensemble de données sous-jacent (\textit{i.e.} doivent être traités avec des \textit{URIs}), en utilisant l'extension \textit{WordNet} (figure \ref{figure 2.61}). \\

 Ils ont essayé de générer tous les modèles possibles. Par la suite, ils ont sélectionné celui qui capture le mieux la structure des données considérées. Pour ce faire, ils ont adopté le mécanisme de l'analyse (parsing) et de la construction du sens du système de réponse aux questions \textit{Pythia}, car il s'adapte tout simplement avec leurs buts.
Le processus du parsing de \textit{Pythia} repose sur un lexique qui spécifie, pour chaque expression, une représentation syntaxique et une sémantique. \\

\parindent =0cm
Une fois une liste de requêtes \textit{SPARQL} est disponible, ils doivent décider laquelle doit être exécutée par rapport au \textit{Triple Store (TP)}. Cependant, cela ne garantit pas que la combinaison des patrons de triplets, dans une requête, est significative et mène à un résultat non vide. Par conséquent, il est nécessaire d'exécuter et de tester les requêtes avant de retourner un résultat à l'utilisateur. Leur système retourne la requête ayant le score le plus élevé avec un résultat non vide. \\

\parindent =0.5cm
Le principal obstacle que l'utilisateur rencontre, est quand il essaie de formuler une requête, pourtant il n'a aucune information sur la structure sous-jacente et le vocabulaire de données. 
Dans ce cadre, une méthode a été présentée dans le travail de \textit{Campinas et al.} en 2012 \cite{campinas2012introducing}, basée sur un résumé du graphe de données et assiste l'utilisateur pendant la formulation de sa requête, par la recommandation des éléments de requêtes structurels possibles, tout en se basant sur l'état actuel de la requête. Ce modèle assure la synthèse du graphe et contient trois couches : \\
- Une couche d'entité et une autre d'ensemble de données : elles représentent les ensembles de données, les entités et leurs relations ; \\
- Une couche de collection de nœuds : représente les collections de nœuds et leurs relations.  \\

\parindent =0cm
Ils ont développé un éditeur SPARQL, qui est une application tirant partie du résumé du graphe de données afin d'aider l'utilisateur dans la formulation efficace de requêtes SPARQL, même sans une connaissance préalable de la structure et le vocabulaire des sources de données. Cet éditeur supporte quatre types de recommandations : la classe, le prédicat, les relations entre les variables, et les graphes nommées. \\

Pendant la formulation de la requête, cet éditeur fournit à l'utilisateur un parmi ces différents types de recommandations en se basant sur l'état de la requête modifiée. L'état de la requête modifiée est composé d'un patron graphique incomplet, ainsi que de la position du curseur. Ils ont également développé un arbre de syntaxe abstrait \textit{(\textbf{A}bstract \textbf{S}yntax \textbf{T}ree : AST)}, qui est une structure de données utilisée par le système, afin de traduire les besoins courants de l'utilisateur en des recommandations possibles. \\

\parindent= 0.5cm
\textit{Makris et al. (2012)} ont déployé \textit{SPARQL-RW} \cite{makris2012sparql}, qui est un système supportant l'interrogation transparente des ensembles de données, gérés par différentes organisations. En utilisant cette infrastructure, les utilisateurs peuvent exprimer des requêtes \textit{SPARQL} en se basant sur leur propre schéma \textit{RDF/S} et accéder automatiquement aux données grâce à une fédération de ressources \textit{RDF} dans le Web. 
Formellement, étant donné une ontologie source $(O_{s})$, une autre cible $(O_{r})$, et un ensemble de correspondances \textit{M} entre $(O_{s})$ et $(O_{r})$; \textit{SPARQL-RW}, prend en entrée une requête \textit{SPARQL} exprimée sur $(O_{s})$ et la réécrit en une autre requête correspondante à elle sémantiquement exprimée sur $(O_{r})$, tout en respectant \textit{M}.   \\

\parindent= 0cm
Le système est composé des éléments de base : \\
- L'analyseur et le composeur de la requête : analysent la requête d'entrée et composent une autre réécrite ; \\
- L'analyseur de correspondance (mapping parser) : analyse les correspondances prédéfinies ; \\
-	Élément qui détermine le type de correspondances : identifie le type de chaque correspondance afin d'être exploitée par le processus de la réécriture ; \\
-	Logiciel de réécriture : réécrit le patron graphique de la requête SPARQL d'entrée en se basant sur les correspondances prédéfinies.
\\

\parindent =0.5cm
\textit{Yahya et al. (2012)} \cite{yahya2012natural} ont développé \textit{\textbf{DEANNA}} \textit{(\textbf{DE}ep \textbf{A}nswers for ma\textbf{N}y \textbf{N}atural \textbf{A}sked questions)} qui est une méthode basée sur un programme linéaire entier (\textit{ILP}), afin de résoudre plusieurs tâches de désambiguïsation. Cette méthode assure la segmentation des questions du langage naturel en éléments, le mapping de ces éléments en des entités, classes et relations sémantiques. Enfin, la construction des patrons de triplet \textit{(TP) SPARQL}. \\
L'\textit{ILP} utilise la richesse des bases de connaissances larges comme \textit{Yago2}, qui possède des informations non seulement sur les entités et les relations, mais également sur les noms de surface et les patrons textuels par lesquels les sources Web se réfèrent à eux (\textit{e.g.} \textit{Yago2} sait que \textit{« Casablanca »} peut se référer à une ville ou à un film, et \textit{« joué dans »} est un patron qui peut désigner une relation). \\

\parindent = 0.5cm
Les travaux existants de question/réponse \textit{RDF}, prennent en considération deux approches : la compréhension de la question (afin de traiter la désambiguïsation des phrases du langage naturel) et l'évaluation de la requête. \\
La difficulté dans la tâche de question/réponse en RDF, est dans l'ambiguïté du langage naturel. Afin de traduire \textit{N} en SPARQL, chaque terme doit être transformé en un élément sémantique (\textit{i.e.} une entité, une classe ou un prédicat) dans le graphe RDF (G).
La désambiguïsation d'un seul terme dans \textit{N}, peut influencer le mapping du reste des termes. Les méthodes existantes de désambiguïsation considèrent que la sémantique de la question, ce qui engendre alors un coût élevé dans l'étape de la compréhension de la requête. \\

\parindent = 0cm
Dans ce contexte, \textit{Zou et al. (2014)} ont proposé de traiter la désambiguïsation, plutôt au niveau de l'étape d'évaluation de la requête (\textit{i.e.} $2\up{ème}$ étape) au lieu de la $1\up{ère}$ étape \cite{zou2014natural}. Cela permet d'éviter le processus de désambiguïsation coûteux de l'étape de la compréhension de la question, ainsi d'accélérer la totalité de la performance (\textit{i.e.} la précision et le temps de réponse). 
Plus précisément, ils ont permis que les termes (dans \textit{N}) puissent correspondre à plusieurs éléments sémantiques (\textit{i.e. }sujets, objets et prédicats) dans le graphe RDF G, au niveau de l'étape de la compréhension de la question. \\

Ils ont proposé un graphe de requête sémantique $Q_{s}$ (représente l'intention de la requête utilisateur), afin de décrire la sémantique de la question \textit{N}. Il faut tout d'abord extraire les relations sémantiques impliquées de la question \textit{N}. Les relations sémantiques sont la base sur lesquelles ils ont construit le graphe de requête sémantique $Q_{s}$.

Intuitivement, une correspondance d'une question N à partir d'un graphe \textit{RDF}, \textit{G}, est une correspondance du sous-graphe sémantique $Q_{s}$ avec \textit{G}. \\

\parindent = 0.5cm
Plusieurs recherches et outils commerciaux offrent des gestionnaires de données RDF centralisés, tandis que l'accroissement des données exige des stockages distribués et de l'indexation. Dans ce cadre, \textit{Papailiou et al. (2014)} ont développé \textit{H2RDF+} \cite{papailiou2014h}, qui est un système de gestion de données \textit{RDF} (s'applique sur une infrastructure Cloud), open source, assurant une forte efficacité. Il utilise HBase pour l'indexation des triplets. Ils ont permis aux utilisateurs de choisir un ensemble de données à charger et à indexer. Également, ils ont permis aux utilisateurs d'écrire leur propre requête \textit{SPARQL} ou bien de sélectionner une prédéfinie.  \\

\parindent=0.5cm
En 2015,\textit{ Hammoud et al.,} ont présenté \textit{\textbf{DREAM} (\textbf{D}istributed \textbf{R}DF \textbf{E}ngine with \textbf{A}daptive Query Planner and \textbf{M}inimal Communication}) qui est un moteur RDF distribué avec un planificateur de requêtes adaptatif et une communication minimale. Il conserve les avantages des systèmes RDF centralisés et distribués, et évite leurs inconvénients \cite{hammoud2015dream}. 

\parindent=0cm
DREAM stocke un ensemble de données intacte sur chaque machine du cluster, et utilise un planificateur de requêtes qui partitionne efficacement n'importe quelle requête SPARQL. En effet, le planificateur de requêtes transforme \textit{Q} en un graphe \textit{G}. Il décompose \textit{G} en plusieurs ensembles de sous-graphes, et mappe chaque ensemble sur une machine séparée. Ensuite, toutes les machines traitent leurs ensembles de sous-graphes en parallèle et se coordonnent les unes avec les autres pour produire le résultat final de la requête. \\

DREAM sélectionne, de manière adaptative, différents nombres de machines pour différentes requêtes SPARQL (en fonction de leurs complexités). \\

\parindent =0.5cm
\textit{Allocca et al. (2016)} ont développé un outil de recommandation de requête \textit{SPARQL}, appelé \textit{\textbf{SQUIRE}} \cite{allocca2016sparql} qui, sur la base d'un ensemble de critères permettant de mesurer la similarité entre la requête initiale et les recommandations, démontre la faisabilité du processus de reformulation des requêtes sous-jacentes. \textit{SQUIRE} classe les requêtes recommandées et offre un support précieux pour les recommandations de requêtes sur un ensemble de données \textit{RDF} cible, inconnu et non mappé. Alors, non seulement il aide l'utilisateur à apprendre le modèle de données et le contenu d'un ensemble de données \textit{RDF}, mais aussi il n'exige pas de l'utilisateur une connaissance intrinsèque des données. \\
Étant donné une requête \textit{SPARQL} $Q_{o}$, qui est satisfaisante avec un ensemble de données \textit{RDF} source $(D_{s})$, \textit{SQUIRE} fournit des recommandations de requête en reformulant automatiquement $Q_{o}$ en d'autres $Q_{r}$ qui sont satisfaisantes avec un ensemble de données \textit{RDF} cible $(D_{t})$.  \\

\parindent= 0cm
SQUIRE permet aux utilisateurs de se référer à un ensemble de données RDF source, écrire la requête \textit{Q}, et spécifier l'ensemble de données RDF cible. Une fois l'utilisateur clique sur le bouton \textit{« Recommend »}, SQUIRE retourne une liste de requêtes recommandées, classées de haut en bas. Ces dernières, non seulement sont exprimées en termes de l'ensemble de données cible, mais sont également garanties être satisfaisantes (\textit{i.e.} l'ensemble de résultats n'est pas vide). \\

\parindent= 0cm
Pour résumer, \textit{SQUIRE} vise à recommander des requêtes dont les reformulations: \\
- Reflètent autant que possible la même signification, la même structure, le même type de résultats et la même taille de résultat que la requête initiale et ; \\
- Ne nécessitent pas de faire la correspondance entre les deux ensembles de données.  \\

\parindent =0.5cm
La recherche facettée est devenue une approche de recherche exploratoire dans de nombreuses applications en ligne. Elle a été dernièrement proposée comme un paradigme approprié pour l'interrogation des données \textit{RDF}.  \\
Pour cette raison, \textit{Arenas et al.(2016)} ont proposé \textit{\textbf{SemFacet}} qui est un cadre théorique rigoureux pour la recherche facettée dans le contexte des graphes de connaissances basés sur \textit{RDF} \cite{arenas2016faceted}. Ils ont développé une interface de requête simple et puissante pour les non-experts. Cette interface permet la navigation à travers les collections inter-connectées des entités, auxquelles ils ont défini une ontologie comme un ensemble fini de règles et de faits et ils ont capturé les facettes qui sont pertinentes pour cette ontologie. Ils ont formalisé leurs notions de « interface facettée » et « requête facettée », dont les requêtes à facettes sont décrites par des « termes » de la logique du premier ordre et par des fragments du langage de requête \textit{SPARQL}. Cette requête est également guidée par les valeurs sélectionnées dans l'interface. \\

\parindent =0.5cm
Les méthodes existantes d'interrogation exigent, non seulement que l'utilisateur ait une connaissance approfondie sur le schéma du graphe \textit{RDF}, mais aussi se concentrent sur la similarité structurelle du graphe et ne considèrent pas la similarité sémantique. Cependant, nous pouvons trouver deux patrons de graphe (\textit{BGP}) qui possèdent une large différence structurelle, mais qui ont une signification sémantique identique. C'est dans ce cadre que \textit{Zheng et al. (2016)} étaient les premiers à proposer une nouvelle mesure de similarité appelée \textbf{la distance d'édition du graphe sémantique}. Cette mesure permet de mesurer la similarité entre les graphes \textit{RDF} (réellement entre le graphe \textit{G} et la requête \textit{Q} pour enfin renvoyer les \textit{"top-k"} sous-graphes  de \textit{G} comme des candidats à \textit{Q}). Ils ont également conçu un nouvel index \textit{« résumé sémantique du graphe »} permettant de résumer le graphe de connaissances afin de réduire l'espace de recherche, ce qui améliore les performances de l'interrogation \cite{zheng2016semantic}. 

\subsection{Quelques approches existantes pour l'interrogation par mots-clés}

\parindent=0.5cm
\textit{Wang et al.,} en 2008 \cite{wang2008q2semantic}, ont implémenté une interface pour les mots-clés, appelée \textit{Q2Semantic}. Ils ont considéré des termes extraits de Wikipédia pour enrichir les littéraux décrits dans les données RDF d'origine. De cette manière, les utilisateurs n'ont pas besoin d'utiliser des mots-clés qui correspondent exactement aux données RDF. Lorsque le premier mot-clé soit entré, Q2Semantic génère automatiquement une liste d'expressions contenant ces mots-clés, parmi lesquels l'utilisateur peut choisir. Q2Semantic prend en charge la traduction des mots-clés utilisateurs en des requêtes formelles. \\

\parindent=0cm
Ils ont également adopté plusieurs mécanismes pour le classement des requêtes, car le processus de construction de la requête peut aboutir à de nombreuses requêtes, \textit{i.e.} les interprétations possibles des mots-clés. Alors, un schéma de classement est bien nécessaire pour envoyer les requêtes qui correspondent le plus à la signification voulue par l'utilisateur. Ce classement prend en compte de nombreux facteurs pertinents, tels que la longueur de la requête, la pertinence des éléments des ontologies. \\

Ils ont proposé un algorithme d'exploration et une nouvelle structure de données graphique appelée RACK, qui ne représente qu'un résumé de l'ontologie ou des données RDF d'origine. Cela réduit l'espace de données pertinent pour l'exploration. \\

Les résultats expérimentaux montrent de bonnes performances sur de nombreux ensembles de données de différentes tailles. Cependant, ce travail supporte que les mots-clés qui correspondent aux littéraux et aux concepts contenus dans les données RDF. Il ne prend pas en charge les mots-clés sous la forme de relations et d'attributs. \\

\parindent=0.5cm
La recherche par mot-clé offre l'avantage de la facilité d'utilisation, mais présente des défis en raison de leur nature souvent laconique et ambigüe. Dans ce cas, chaque mot-clé doit être interprété, et la requête entière doit être mappée à un ensemble d'expressions conditionnelles, \textit{i.e.} clause WHERE et clause RETURN. Cependant, il n'est pas toujours facile de trouver un mappage unique. Alors, le problème devient celui d'identifier les \textit{K} interprétations souhaitées par l'utilisateur. \\

\parindent=0cm
C'est dans ce cadre que \textit{Anyanwu et al.,} en 2011, ont abordé le problème de la génération d'interprétations de requêtes (mots-clés) sur les BD-RDF, en utilisant des informations provenant de l'historique des requêtes de l'utilisateur \cite{fu2011effectively}. 
Les utilisateurs posent, souvent, une série de questions connexes, en particulier dans des scénarios exploratoires. Dans ces scénarios, les informations sur les requêtes précédentes peuvent être utilisées pour influencer l'interprétation d'une requête plus récente. \\
Ils ont fait face à deux principaux défis : \\
- Capturer et représenter efficacement l'historique des requêtes ; \\
- Exploiter efficacement l'historique des requêtes pendant l'interprétation de cette dernière. \\

Ils ont proposé et mis en œuvre : \\
- Un modèle de graphe de synthèse pondéré dynamique, qui est utilisé pour capturer, de façon concise, les caractéristiques essentielles de l'historique des requêtes d'un utilisateur. \\
- Une fonction de pondération dynamique qui assigne des pondérations aux éléments du graphe récapitulatif de manière à capturer leur pertinence par rapport au contexte d'interrogation actuel. \\
$\rightarrow$ Il ont finalement présenté une évaluation complète de leur approche en utilisant un sous-ensemble de \textit{DBpedia}. \\

\parindent=0.5cm
En 2011, \textit{Elbassuoni et al.,} ont proposé un modèle d'interrogation utilisant des requêtes par mot-clé sur les graphes RDF \cite{elbassuoni2011keyword}. Ce modèle récupère un ensemble de sous-graphes correspondants aux mots-clés de la requête, ainsi qu'il les classe en se basant sur le modèle du langage statistique (LMs). 
Leur base de connaissances est constituée d'un ensemble de triplets S-P-O. Pour pouvoir traiter les requêtes par mot-clé, ils ont construit un document virtuel pour chaque triplet $t_i$, qu'ils l'ont appelé $D_i$. \\
$D_i$ contient un ensemble de mots-clés qui sont extraits du sujet et de l'objet des triplets, et des mots-clés représentatifs pour les prédicats. \\

\parindent=0cm
Ils ont stocké tous les termes du document $D_i$ dans un index inversé. Alors, étant donnée une requête par mot-clé, ils ont utilisé cet index inversé pour récupérer, pour chaque mot-clé d'une requête, une liste de triplets correspondants. Ils ont rejoint ensuite les triplets à partir des listes différentes en fonction de leurs sujets et objets pour récupérer des sous-graphes avec un ou plusieurs triplets. 

Afin d'éviter de récupérer des sous-graphes arbitrairement longs, ils ont limité les sous-graphes récupérés pour avoir les deux propriétés suivantes : \\
- Les sous-graphes doivent être uniques et maximaux, \textit{i.e.} chaque sous-graphe récupéré ne doit pas être un sous-ensemble de tout autre sous-graphe récupéré ; \\
- Les sous-graphes doivent contenir des triplets correspondants à différents ensembles de mots-clés. Autrement dit, aucun triplet, dans le même sous-graphe, ne correspond exactement au même ensemble de mots-clés. Si deux triplets correspondent au même ensemble de mots-clés, alors ils font partie de deux différents résultats possibles de la requête utilisateur, et doivent être considérés comme des parties de deux sous-graphes distincts. \\

\parindent=0.5cm
Selon \textit{Shekarpour et al (2015)} \cite{shekarpour2015sina}, le développement de systèmes de recherche par mot-clé est difficile pour trois raisons : 

\parindent=0cm
- Les ressources proviennent de différents ensembles de données ; 

\parindent=0cm
- Différents ensembles de données utilisent des schémas hétérogènes et chacun d'entre eux ne peut contenir qu'une partie de la réponse pour une certaine requête utilisateur ; 

\parindent=0cm
- La création d'une requête formelle fédérée à partir de mots-clés sur différents ensembles de données, nécessitent l'exploitation de liens entre différents ensembles de données sur les niveaux de schéma et d'instance. \\

C'est dans ce cadre qu'ils ont présenté \textit{SINA}. Il est un système de recherche par mot-clé (ou requête en LN) fournis par l'utilisateur, et traduits en requête SPARQL conjonctives, qui sont exécutées sur un ensemble de sources de données inter-connectées. SINA utilise un modèle de Markov caché pour déterminer les ressources les plus appropriées pour une requête, à partir de différents ensembles de données. \\
SINA est capable également de construire des requêtes fédérées en exploitant les liens entre les ressources. \\
Ce système a été évalué sur trois ensembles de données différents. Il a pu répondre exactement à 25 requêtes du QALD-1 \textit{(i.e. \textbf{Q}uestion \textbf{A}nswering over \textbf{L}inked \textbf{D}ata)}. De même, ils ont étudié l'exécution de Sina dans ses implémentations mono-cœurs et parallèles. \\

\parindent =0.5cm
La plupart des approches disponibles, exploitent le parcours du graphe en temps réel afin d'identifier les sous-graphes appropriés. Certaines d'entre elles fournissent des résultats approximatifs parce que le parcours des graphes RDF avec des millions ou des milliards de données est très coûteux et prend beaucoup du temps. Par conséquent, il est nécessaire de fournir des méthodes pouvant réduire la traversée du graphe, dans le temps de génération de requête. \\

\parindent=0cm
\textit{Dudhani et al. (2016)} \cite{spa} ont proposé une technique de \textit{« segmentation profonde »} pour l'interrogation par mots-clés, traitée sur des bases de données \textit{RDF}. Cette approche réalise un élagage agressif des interprétations non pertinentes à partir de l'espace des interprétations considéré. \\
Ils ont conçu les premiers algorithmes efficaces permettant d'énumérer toutes les réponses dans l'ordre, afin de calculer les \textit{"top-k"} réponses (selon certains critères de classement), où chaque réponse est une sous-structure du graphe contenant tous les mots-clés saisis. \\
Ils ont présenté également un schéma d'indexation pour \textit{RDF} qui décrit les caractéristiques structurelles du graphe, ses chemins et les informations sur l'accessibilité des nœuds, afin d'accélérer la récupération des sous-structures représentant les résultats de la requête. En outre, l'index permet de faciliter les opérations de maintenance au cas où l'ensemble de données évolue.

\section{Discussion et Synthèse}

\parindent =0.5cm
La considération de RDF comme étant un modèle de base de données à part entière, la conception ainsi que les primitives d'un langage d'interrogation, sont un sujet moins développé. Il constitue un parmi les besoins si nous voulons profiter de l'intégralité des potentialités du modèle de données RDF (\textit{i.e.} l'optimisation de requêtes, la réécriture de requêtes, les vues, les mises à jour). \\

\parindent =0.5cm
Dans ce contexte, certains travaux précédents de stockage, de gestion et d'interrogation de graphes RDF volumineux, se limitaient sur le processus de question/réponses à partir des données \textit{RDF} inter-connectées sur le Web. Ils fournissent des interfaces, avec lesquelles l'utilisateur n'a qu'à saisir ses mots-clés en langage naturel décrivant son besoin en informations. La plupart de ces systèmes permettent de transformer la question du langage naturel en une représentation à base de triplets. Ensuite, avec l'application des métriques de similarité et des heuristiques de recherche, ils vont retrouver les correspondances des sous-graphes à partir des graphes RDF.\\

\parindent = 0.5cm
D'autres travaux ont laissé la tâche d'écriture des requêtes \textit{SPARQL} à l'utilisateur (même pour les non-experts), mais en lui fournissant un appui et une assistance au moment de l'écriture de ces dernières.
Parmi les autres méthodes développées, nous trouvons celles qui permettent d'optimiser les requêtes \textit{SPARQL} (\textit{e.g.} en éliminant les triplets redondants, en réorganisant les clauses de la requête, en proposant un résumé du graphe \textit{RDF}, \textit{etc}). \\

\parindent = 0.5cm
D'autres approches se sont concentrées sur le stockage des données RDF dans des \textit{BD-RDF}, ou des \textit{Triples Stores}, et ensuite ils ont fourni des méthodes et des algorithmes permettant d'assurer une interrogation rapide et efficace de ces données, en utilisant des requêtes \textit{SPARQL}. Entre autres, la plupart des méthodes existantes s'appuient sur des applications Web (interface depuis un point d'accès: endpoint : permet d'effectuer des requêtes \textit{SPARQL} sur une base de connaissances. Une manière de transfert des requêtes \textit{SPARQL} des clients à un service Web capable de les exécuter et de renvoyer les résultats via HTTP). Alors, ils n'assurent pas une gestion persistante des données RDF, permettant de stocker et d'accéder aux données graphiques sur un support persistant. \\

\parindent=0cm
Les solutions pour parcourir un graphe RDF sur les supports persistants, ne sont pas courantes car le coût des accès aléatoires à un support persistant est problématique. C'est un grand défi pour les BD graphiques, où l'utilisateur s'attend à la persistance des données et à la prise en charge des opérations de parcours rapides. \\

\parindent = 0.5cm
Certaines méthodes existantes, non seulement elles exigent que l'utilisateur ait une connaissance approfondie sur le schéma du graphe \textit{RDF}, mais aussi, l'utilisateur ne peut interroger qu'une seule ontologie, à la fois, stockée dans une \textit{BD-RDF}. \\

\parindent =0.5cm
D'autres approches se sont concentrées sur l'indexation des données \textit{RDF}, et comme le stockage et l'interrogation des données fortement indexées peuvent être efficaces et faisables pour des requêtes simples, ils sont, par contre, pas conseillés et difficiles à manipuler en exécutant des requêtes compliquées. \\

\parindent=0.5cm
En outre, le problème de passage à l'échelle est parmi les causes des limites de performance des systèmes réalisant les travaux d'interrogation. En effet, les solutions offertes par les \textit{Triples Stores} dépassent leur capacité de stockage. Ceci donne l'opportunité d'améliorer les capacités des algorithmes d'interrogation et de recherche.   \\

\parindent =0.5cm
Dans ce contexte, et afin de simplifier le processus d'interrogation, une solution envisageable est la génération automatique de requêtes \textit{SPARQL} à partir de graphes \textit{RDF}, basée sur des mots-clés fournis par l'utilisateur.

\section{Conclusion}

\parindent =0.5cm
RDF et RDFS utilisent, pour décrire des connaissances, une structure particulière (\textit{i.e.} graphe), dans laquelle les nœuds sont des concepts représentés par des propriétés et les arcs sont des prédicats décrivant les relations entre ces concepts. Ils permettent également de formuler des règles de raisonnement qui peuvent être employées afin d'inférer de nouvelles connaissances. Dans ce chapitre nous avons expliqué le stockage et l'interrogation des connaissances à partir des \textit{BD-RDF}. Certaines sont dites non-natives (\textit{i.e.} s'appuient sur les \textit{SGBDR}) alors que d'autres sont natives et elles utilisent la structure en graphe pour la représentation des connaissances. Certaines \textit{BD-RDF} se chargent, dans leur système, d'un moteur de raisonnement qui est capable de fonctionner sous deux modes différents, \textit{la saturation des données} et \textit{la reformulation des requêtes}. Les requêtes exprimées sur ces connaissances sont représentées grâce aux langages d'interrogation adaptés, dont le plus utilisé est \textit{SPARQL}. 
Nous avons effectué une étude d'art sur quelques méthodes de stockage des données \textit{RDF} existantes, ainsi que sur les différentes approches qui visent à générer, d'une manière automatique ou semi-automatique, des requêtes formelles à partir de mots-clés ou de phrases exprimés par l'utilisateur en langage naturel.

\chapter{Proposition d'une nouvelle approche de génération de requêtes SPARQL à partir de graphes RDF}

\section{Introduction}

\parindent =0.5cm
La récupération des données de la gigantesque masse disponible, devient de plus en plus difficile dans le cadre du Web sémantique, ainsi que l'accès aux données représente, de plus en plus, un réel besoin pour les utilisateurs. Comme il est très difficile pour un utilisateur final de surmonter et maîtriser la complexité des langages et des schémas des connaissances, la mise à disposition des interfaces des moteurs \textit{SPARQL}, est devenue une nécessité : afin de pouvoir former une requête valide sur les bases de connaissances du Web sémantique, l'utilisateur doit avoir une idée sur le langage \textit{SPARQL}, les ontologies employées, ainsi que, la forme ou la structure des graphes \textit{RDF} considérés. \\
Pour simplifier le processus d'interrogation, une solution envisageable est la génération automatique de requêtes \textit{SPARQL} basée sur les mots-clés fournis par l'utilisateur. Notre travail se place dans ce contexte de recherche, \textit{i.e.} comment pouvons-nous interpréter et élucider une requête en langage naturel et la traduire en une requête \textit{SPARQL} ?

\section{Problématique et motivations dans le contexte d'interrogation des graphes RDF}

\parindent= 0.5cm
Dans la littérature, la pertinence et l'efficacité des interfaces en langage naturel ont été largement étudiés, sans que l'on atteigne une conclusion finale. Quelques travaux \cite{pradel2013passage}, évaluent l'utilisation des interfaces qui permettent d'assurer, aux utilisateurs, un accès aux connaissances en utilisant la langage naturel. \\ 

\parindent=0cm
De ce fait, ils déterminent trois problèmes majeurs dans ces interfaces : \\
- Les \textit{ambiguïtés} du langage naturel sont le principal problème, et qui est dû à la variabilité linguistique, \textit{i.e.} les diverses façons d'exprimer une même requête ; \\
- La \textit{barrière adaptative} et qui fait que les interfaces assurant de bonnes performances de recherche, ne sont malheureusement pas très portables : la plupart du temps, ces systèmes sont dépendants du domaine et par conséquent, difficiles à adapter à d'autres contextes ; \\
- Le problème de \textit{l'habitabilité} : lorsqu'on donne trop de liberté à l'utilisateur pour exprimer ses requêtes, il peut les exprimer au-delà des capacités du système. Par conséquent, une interface doit exiger des structures à l'utilisateur pour l'aider pendant le processus de formulation de sa requête ; \\

Un autre problème qui nous intéresse est celui du passage à l'échelle des méthodes de stockage et d'interrogation des données RDF.

\section{La base de données graphique Neo4j}

\parindent = 0.5cm
La version stable actuelle de \textit{Neo4j} est 3.2.5 (Septembre 2017), et elle est sous licence \textit{GPLv3}. Elle est écrite en java et son objectif principal est de stocker les données d'une manière consistante, permanente et efficace. Elle fournit des interfaces permettant aux applications et aux utilisateurs d'accéder, d'évaluer, de modifier et de gérer des ensembles de données RDF. Les requêtes peuvent être envoyées via le protocole \textit{HTTP / REST} ou directement en Java. 
\textit{Neo4j} est une BD "intégrable" (\textit{i.e. embeddable}) car elle peut être ajoutée à une application et utilisée comme toute autre bibliothèque. \\ 

\textbf{Avantages du mode intégré : }
 
\parindent =0cm
\textbf{$\oplus$} Faible latence : parce que l'application communique directement avec la BD, il n'y a pas de surcharge réseau. Afin de réduire la latence, Neo4j fournit deux niveaux de mise en cache : \\

\parindent=0.5cm 
\textbf{Système de fichiers} : est une zone de mémoire libre, ou RAM, qui n'a été allouée à aucun processus. Lorsqu'un processus demande un fichier ou une partie d'un fichier, ce fichier est chargé en mémoire dans la cache dy système de fichiers. 

\parindent=0.5cm
\textbf{Cache d'objet} : parce que Neo4j s'exécute dans une JVM, il gère les nœuds et les relations en tant que objets Java en mémoire. \\

\parindent=0cm
\textbf{$\oplus$} Choix des APIs : nous avons accès à toute la gamme d'\textit{API}s pour créer et interroger des données ; \\
\textbf{$\oplus$} Transactions explicites : grâce au \textit{Core API} (section \ref{core}, page \pageref{core}), nous pouvons contrôler le cycle de vie transactionnel en exécutant une séquence de commandes arbitrairement complexes sur la BD dans le contexte d'une transaction unique, \textit{etc.} \\

\parindent = 0.5cm
\textbf{$\Leftrightarrow$} Lors de l'exécution en mode intégré, nous devons garder à l'esprit ce qui suit : \\ 
- \textit{JVM (Java Virtuel machine)} : uniquement \textit{Neo4j} est une BD JVM. Beaucoup de ses \textit{APIs} sont, alors, accessibles uniquement à partir d'un langage basé sur \textit{JVM}. Puisque Neo4j est écrit en Java, et comme toutes autres applications écrites en Java, nous devons configurer la JVM pour avoir des performances optimales ; \\
- Le comportement du \textit{GC (Garbage Collector)} : lors de l'exécution en mode intégré, \textit{Neo4j} est soumis au comportement de récupération de place (GC) de l'application hôte. Les longues pauses \textit{GC} peuvent affecter le temps d'interrogation ; \\
$\ast$ Cycle de vie de la BD : l'application est responsable du contrôle du cycle de vie de la BD, ce qui inclut le démarrage et la fermeture en toute sécurité. \\

\parindent = 0.5cm
\textbf{$\bullet$ \textit{Neo4j embedded} peut être mis en cluster pour une haute disponibilité ainsi qu'une mise à l'échelle horizontale. }

\subsection{Architecture de Neo4j}

\begin{figure}[H]
\begin{center}

\includegraphics[scale=0.7]{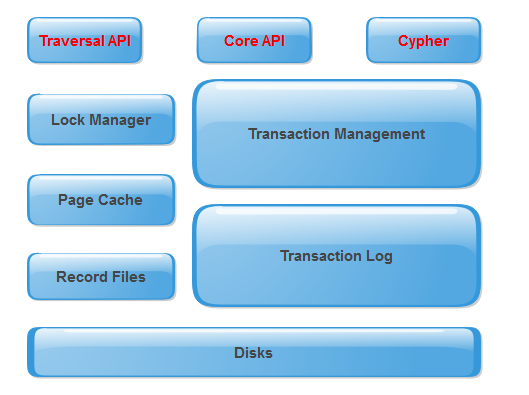}

\end{center}
\caption{Arichtecture de Neo4j \label{figure 3.1}}
\end{figure}

\parindent = 0.5cm
Comme le montre la figure \ref{figure 3.1}, \textit{Neo4j} stocke les données graphiques dans un certain nombre de différents fichiers de stockage. Chaque fichier de stockage contient les données d'une partie spécifique d'un graphe. La répartition des responsabilités de stockage facilite les parcours performants de graphes. Cela signifie que la vue de l'utilisateur sur le graphe et les enregistrements sur le disque, sont structurellement dissemblables.

\subsubsection{Nœuds et relations}

\parindent = 0.5 cm
\textit{Neo4j} utilise le modèle de données de graphe de propriété. Ce modèle garantit que les arcs soient dirigés, les sommets et les arcs soient étiquetés, les sommets et les arcs possèdent des données de paires clé/valeurs associés (\textit{i.e.} des propriétés) où la clé est une chaîne, ainsi que les valeurs de propriété peuvent être une primitive Java ou un tableau (\textit{e.g.} les valeurs \textit{String, int et int[]} sont valides). Il peut y avoir plusieurs arêtes entre deux sommets. Dans \textit{Neo4j}, les sommets attribués sont appelés "nœuds" et les arêtes attribuées dirigées sont appelées "relations". les valeurs attribuées elles-mêmes sont nommées "propriétés". La combinaison des nœuds, les relations entre eux et les propriétés, forment un espace de nœuds (\textit{i.e.} un réseau cohérent représentant les données du graphe (\textit{Neo Technology 2006})). 

\begin{figure}[H]
\begin{center}

\includegraphics[scale=0.7]{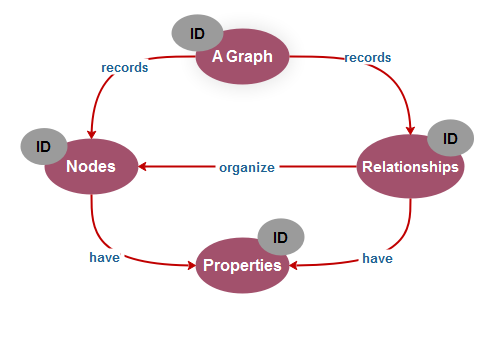}

\end{center}
\caption{Un exemple d'un graphe Neo4j \label{figure 3.2}}
\end{figure}

\parindent = 0.5cm
La figure \ref{figure 3.2} montre les composants d'un graphe de propriété simple. \textit{Neo4j} utilise, également, ce modèle pour le stockage des données, dont tous les nœuds ont des identifiants \textit{(ID)} et toutes les relations sont dirigées et ont un type \textit{(e.g. records, have, etc)}.

\begin{figure}[H]
\begin{center}

\includegraphics[scale=0.6]{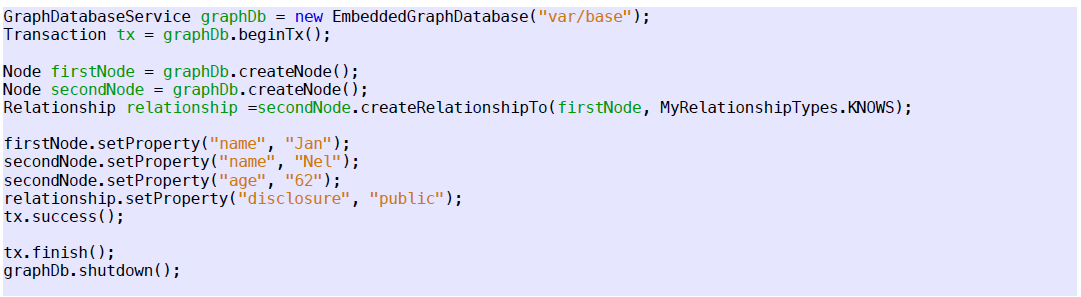}

\end{center}
\caption{Création d'un petit graphe Neo4j \label{figure 3.3}}
\end{figure}

La figure \ref{figure 3.3} décrit comment créer un petit graphe, sous Java, composé de deux nœuds connectés avec une relation et quelques propriétés.

\subsubsection{APIs de Neo4j et le langage Cypher}

\begin{figure}[!ht]
\begin{center}

\includegraphics[scale=0.7]{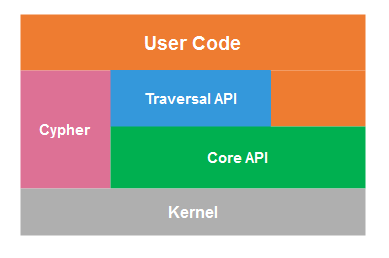}

\end{center}
\caption{Vue logique des APIs dans Neo4j \label{figure 3.4}}
\end{figure}

\parindent = 0.5cm
Bien que le système de fichiers et les infrastructures de mise en cache soient fascinants, les développeurs interagissent rarement directement avec eux. Cependant, ils manipulent une BD graphique via un langage d'interrogation, qui peut être impératif ou déclaratif, à noter le langage \textit{Cypher} (propre à \textit{Neo4j}). Toutefois, d'autres \textit{APIs} existent (figure \ref{figure 3.4}). Toutes les BD graphiques n'ont pas forcément le même nombre de couches, ni nécessairement les couches qui se comportent et interagissent exactement de la même manière.

\subsubsubsection{API kernel}
\parindent = 0.5cm
Au niveau le plus bas se trouve les gestionnaires d'événements de transactions de \textit{Kernel}. Ceux-ci permettent au code utilisateur d'écouter les transactions, et ensuite de réagir (ou non) en fonction du contenu des données et de l'étape du cycle de vie de la transaction.
\subsubsubsection{Core API}
\label{core}
\parindent = 0.5cm 
Une \textit{API Java} impérative qui expose les primitives de graphes de nœuds, de relations, de propriétés et des étiquettes, à l'utilisateur. Pour l'écriture, le \textit{Core API} fournit des fonctionnalités de gestion de transactions afin d'assurer une persistance atomique, cohérente, isolée et durable.
\subsubsubsection{Le traversal Framework}
\parindent = 0.5cm
Une \textit{API Java} déclarative. Elle permet à l'utilisateur de spécifier un ensemble de contraintes qui limitent les parties du graphe que la traversée est autorisée à visiter. Nous pouvons spécifier les types de relations à suivre et dans quelle direction (en spécifiant efficacement les filtres de relation). Nous pouvons également indiquer si nous voulons que la traversée soit effectuée en largeur ou en profondeur, \textit{etc.}

\subsubsubsection{Cypher} 
\parindent = 0.5cm
Il est un langage d'interrogation déclaratif, inspiré de \textit{SQL} et \textit{SPARQL}. Il assure une interrogation et une mise à jour efficaces du \textit{Triples Store}. Neo4j prend en charge aussi le langage d'interrogation impératif \textit{\textbf{Gremlin}} (il était au départ le langage par défaut de Neo4j, mais \textit{Neo Technology} a décidé de créer \textit{Cypher} pour faciliter la mise en main, la lecture et l'écriture des requêtes). \\

\parindent = 0cm
Les requêtes Cypher suivent une certaine syntaxe, et les clauses les plus importantes sont :  \\

\parindent = 0cm
- \textbf{MATCH :} retourne toutes les données ayant la forme du sous-graphe décrit ; \\
- \textbf{CREATE :}  crée des nœuds ou des relations ; \\
- \textbf{DELETE :} supprime des nœuds, des relations et des propriétés ; \\
- \textbf{DETACH DELETE :} supprime un nœud et tous les arcs connectés à lui ; \\
- \textbf{SET :} modifie ou met à jour une pair clé/valeurs ; \\
- \textbf{REMOVE :} supprime une pair clé/valeurs ; \\
- \textbf{WHERE :} définit les propriétés d'un graphe et le résultat, fournit des critères pour filtrer les résultats de la correspondance du patron graphique ; \\
- \textbf{RETURN :} définit les valeurs retournées ; \\
- \textbf{START :} spécifie un ou plusieurs points de départ (nœuds ou relations) dans le graphe ; \\
- \textbf{UNION :} fusionne les résultats de deux ou plusieurs requêtes ; \\
- \textbf{ORDER BY :} dans la commande MATCH pour ordonner les résultats renvoyés ; \\
- \textbf{LIMIT et SKIP :} pour limiter ou filtrer le nombre de lignes retournées. \\

\parindent = 0.5cm
En outre, Neo4j prend en charge des \textbf{indexs} en utilisant \textit{Apache Lucene}. Un index porte sur l'attribut d'un nœud et permet de parcourir aisément les nœuds d'un graphe afin d'appliquer des requêtes efficaces avec le langage \textit{Cypher}.

\subsection{Avantages de Neo4j}

\parindent = 0cm
Parmi les avantages de \textit{Neo4j}, nous citons : \\
\textbf{$\oplus$} \textit{Neo4j} fournit un modèle de données flexible, simple mais puissant, qui peut être facilement modifié en fonction des applications et des industries ; \\
\textbf{$\oplus$} \textit{Neo4j} est hautement disponible pour les applications en temps réel de grandes entreprises avec des garanties transactionnelles ; \\
\textbf{$\oplus$} Avec \textit{Neo4j}, nous pouvons facilement représenter des données connectées et semi-structurées ; \\
\textbf{$\oplus$} Elle fournit son propre langage d'interrogation \textit{Cypher} ; \\
\textbf{$\oplus$} \textit{Neo4j} n'exige pas de jointures complexes pour récupérer les données connectées, car il est très facile de récupérer les détails d'un nœud ou d'une relation sans jointures ou index ; \\
\textbf{$\oplus$} \textit{Neo4j} prend en charge les contraintes / règles \textit{ACID} ; \\ 
\textbf{$\oplus$} Nous pouvons mettre à l'échelle la BD en augmentant le nombre de lectures / écritures et le volume des informations, sans affecter la vitesse du traitement de requêtes et l'intégrité de données ; \\
\textbf{$\oplus$} \textit{Neo4j} prend également en charge la réplication de données pour assurer la sécurité et la fiabilité de ces dernières.

\section{Description de la nouvelle méthode d'interrogation \textbf{OntoNeo} }

\parindent = 0.5cm
 Notre processus de question/réponse à des mots-clés utilisateurs à partir du graphe RDF stocké dans Neo4j (dont ce graphe est déduit à partir de l'ontologie de départ), est décrit dans l'\textit{algorithme} \ref{Algorithme 1}. Il prend en entrée une ontologie \textit{\underline{ou plusieurs}} (section \ref{plsonto}, page \pageref{plsonto}), et retourne une réponse à une requête utilisateur écrite en langage naturel. Dans ce qui suit, nous allons décrire chaque phase séparément. \\ \\
 
\begin{algorithm}
\caption{Algorithme récapitulatif des différentes phases de notre processus \label{Algorithme 1}} 

\textbf{ \\ Entrée : } 
Une ontologie, \textit{O}. \\
\textbf{ Sortie : } 
Une réponse à une requête (mots-clés) utilisateur en LN, \textit{Q}.

\begin{algorithmic} [1]
 
\State Analyse de l'ontologie O en des ensembles d'entités ;
\State Extraction des triplets RDF \textit{(S,P,O)} à partir de ces ensembles d'entités ;
\State Chargement des triplets RDF dans Neo4j ; 
\State Interrogation de la base de données graphique Neo4j ;

\end{algorithmic}
\end{algorithm}

\subsection{Phase 1 : Analyse (Parsing) de l'ontologie }

\parindent = 0.5cm
Notre système admet en entrée l'ontologie du domaine qui englobe divers types d'informations : taxonomie de concepts de même type modélisés par la relation \textit{is-a}, thésaurus de concepts de différents types liés par les relations \textit{Object-Property} et \textit{Data-Type-Property}, \textit{etc}.  \\

\parindent=0.5cm
- Il retourne comme sorties les éléments suivants  : 

\parindent =0cm
$\triangleright$ Les entités de l'ontologie : celles-là représentent la partie la plus importante ;  \\
$\triangleright$ Les types de ces entités, et qui peuvent être des classes, des instances, des propriétés d'objet ou des propriétés de type de données.  \\

Le parsing en détail : 

\parindent = 1cm
$\bullet$ On parse les classes de(des) l'ontologie(s) d'entrée et on extrait pour chaque classe ses annotations (ses labels, ses commentaires), ses sous-classes, ses classes équivalentes, et disjointes ; 

$\bullet$ On parse les propriétés d'objet/ de type de données de(s) l'ontologie(s) d'entrée et on extrait pour chaque propriété ses annotations, ses sous-propriétés, ses super-propriétés, ses domaines, ses images, ses propriétés inverses (seulement pour les propriétés d'objet), équivalentes, disjointes, et son type (fonctionnelle, symétrique, \textit{etc}); 

$\bullet$ On parse les instances et on extrait pour chacune ses annotations, ses classes qui l'instancient, ses "\textit{Same}" et "\textit{Different}" instances.  \\

\textbf{$\Leftrightarrow$} Pour le parsing, nous avons choisi d'utiliser \textit{OWL API} qui est une API performante, reconnue et professionnelle. Nous avons utilisé également le raisonneur \textit{Hermit}, qui est un raisonneur d'ontologies écrites en OWL, et dont son utilisation n'est pas obligatoire, car \textit{OWL API} a aussi un raisonneur par défaut. \textit{Hermit} permet de déterminer si l'ontologie est cohérente, identifier les relations de subsomption entre les classes, déduire des inférences, \textit{etc}. \\ 

\parindent=0.5cm
Nous avons choisi d'effectuer le raisonnement pendant le parsing de l'ontologie, et donc avant l'exécution des requêtes, en appliquant la technique de la saturation de données.

\subsection{Phase 2 : Génération des triplets RDF}

\parindent = 0.5cm
Dans cette phase se fait la création des triplets RDF (à partir de l'ensemble d'entités déjà généré dans la première phase), afin de traduire la séquence précédente vers un ensemble de triplets RDF sous la forme \textit{\textit{<} Sujet, Prédicat, Objet \textbf{>}}. Cela aboutit à deux types de triplets : \\ \\ 

\parindent = 0cm
 \textbf{\emph{Triplets de type primitives OWL :} } \\ 

\begin{figure}[H]
\begin{center}

\includegraphics[scale=0.6]{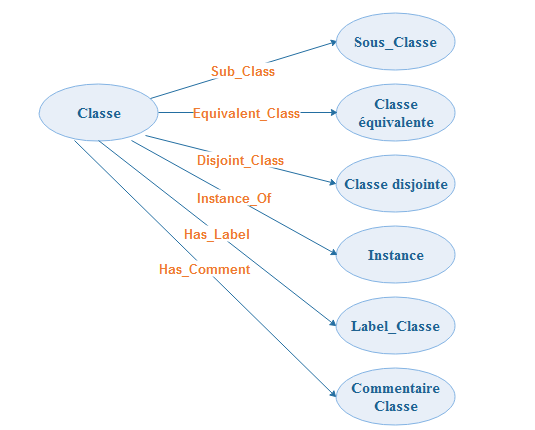}

\end{center}
\caption{Triplets des classes \label{figure 3.5}}
\end{figure}

\parindent =0cm
\textbf{Pour les classes}, figure \ref{figure 3.5} :  \\
 
 \parindent=0.5cm
- Le sujet est une classe ; 

 \parindent=0.5cm
- L'objet peut être une classe, une instance, un label, un commentaire ; 

 \parindent=0.5cm
  - Le prédicat existe de différents types : \textit{Sub-Class-Of, Equivalent-Class, Disjoint-Class, Instance-Of, Has-Label, Has-Comment}. \\ 

\parindent=0.5cm
Les concepts sont organisés hiérarchiquement à travers la relation conceptuelle \textit{« Sub-Class-Of »} ou \textit{« is-a »} d'héritage ou de spécialisation, utilisée pour construire une taxonomie / hiérarchie de concepts. D'autres relations prédéfinies telles que l'équivalence et la disjonction peuvent également lier les concepts pour véhiculer plus de sémantique. \\

\begin{figure}[H]
\begin{center}

\includegraphics[scale=0.6]{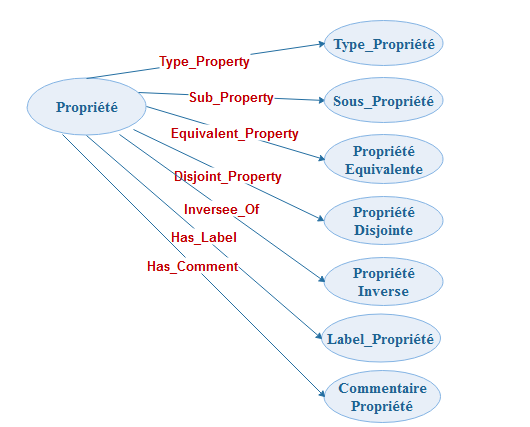}

\end{center}
\caption{Triplets des propriétés \label{figure 3.6}}
\end{figure}

\textbf{Pour les propriétés d'objet/de type de données} (figure \ref{figure 3.6}) : \\

\parindent =0.5cm  
- Le sujet est une propriété ;

\parindent =0.5cm
- L'objet est une propriété, un label ou un commentaire ;

\parindent =0.5cm
- Le prédicat peut être de divers types : \textit{Type-Property, Sub-Property, Equivalent-Property, Disjoint-Property, Inverse-Of, Has-Label, Has-Comment}.   \\

\parindent=0cm
Il n'y a pas de propriétés inverses pour les propriétés de types de données. \\

\parindent=0.5cm
Il existe en \textit{OWL} deux types de propriétés : les propriétés d'objet et celles de types de données. \\
Une propriété d'objet qui définit une relation entre deux individus d'une classe ou de plusieurs classes, et aussi, une propriété de type de données qui est une relation entre un individu d'une classe et une valeur ou une donnée. Les propriétés d'objet et de type de données, définies par l'utilisateur, sont toutes (respectivement) des enfants de \textit{”owl:TopObjectProperty”} et \textit{”owl:TopDataProperty”}, et des parents des sous-propriétés\textit{ ”owl:BottomObjectProperty”} et \textit{”owl:BottomDataProperty”}. Les propriétés d'annotation telles que \textit{« owl:versionInfo »}, \textit{« rdfs:label »}, \textit{« rdfs:comment »}, \textit{« rdfs:seeAlso »}, \textit{« owl:priorVersion »,} \textit{etc}. sont des constructeurs intégrés dans OWL. \\
Les propriétés peuvent aussi être organisées hiérarchiquement et liées par des relations conceptuelles prédéfinies, telles que l'équivalence, la disjonction et beaucoup d'autres. \\ 

\begin{figure}[H]
\begin{center}

\includegraphics[scale=0.6]{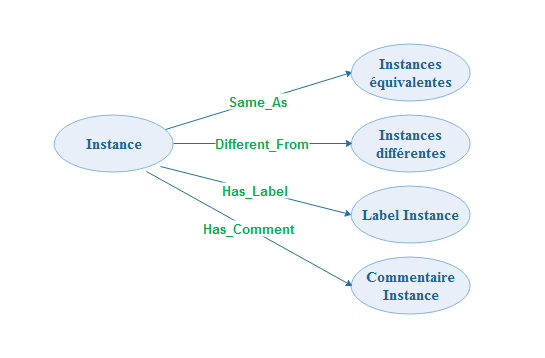}

\end{center}
\caption{Triplets des instances \label{figure 3.7}}
\end{figure}

 \textbf{Pour les instances}, figure \ref{figure 3.7}: \\
 
\parindent =0.5cm
- Le sujet est obligatoirement une instance ; 

\parindent =0.5cm
- L'objet est une instance, label ou commentaire ; 
 
\parindent =0.5cm 
- Le prédicat existe de quatre types : \textit{Same-As, Different-From, Has-Label, Has-Comment}. \\

\parindent = 0.5cm
Une instance est un objet particulier instancié par les classes à l'aide de la relation prédéfinie \textit{"Instance-Of"}. \\
Un individu peut ne pas avoir de classe(s) qui l'instancie(nt), dans ce cas, il sera implicitement une instance de la classe \textit{« owl:Thing »}.

\parindent = 0cm
 \textbf{\emph{Triplets dont les prédicats sont les propriétés elles-mêmes :} } 

\parindent = 0.5cm
Dans ce cas, le sujet est le domaine (qui représente la classe source ou bien l'individu d'une classe), ainsi que l'objet est le co-domaine et qui est la classe cible (ou bien il peut être une valeur de données quand le prédicat s'agit d'une propriété de type de données). Le domaine et l'image sont reliés par l'intermédiaire des propriétés (d'objet ou de type de données) extraites déjà à partir de l'ontologie du départ.  \\

\parindent=0cm
- Nous avons déterminé les propriétés d'objet définies pour chaque classe, puis les images de chacune de ces propriétés (si elles existent). \\

\parindent=0cm
- Pour les propriétés de type de données, nous avons extrait pour chaque instance (\textit{i.e.} domaine), sa valeur de données qui est un littéral (\textit{i.e.} co-domaine). \\ \\ 
Tout cela nous a permis d'avoir des triplets de type :
\begin{center}
 \textbf{\textit{\textbf{<<} Domaine \textbf{ -} Propriété(s) \textbf{->} Co-domaine(s) \textbf{>>}}}.
 \end{center}

\begin{figure}[H]
\begin{center}

\includegraphics[scale=0.7]{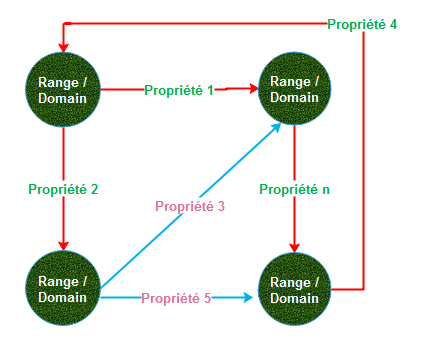}

\end{center}
\caption{Domaines et Images \label{figure 3.50}}
\end{figure}

 \parindent=0.5cm
Nous pouvons trouver un domaine relié à plusieurs images à travers différentes propriétés (figure \ref{figure 3.50}). En outre, un domaine dans un triplet peut être une image dans un autre triplet.  \\

\parindent=0.5cm
\textbf{Remarque :} la technique de saturation de données permet de produire quasiment tous les triplets possibles en appliquant toutes les règles d'inférence sur les données RDF. C'est le rôle du moteur de raisonnement (Hermit dans notre cas) de les produire. Ces triplets dérivés (inférés) sont rajoutés à l'ensemble de données générés. Pendant la saturation, chaque donnée inférée peut participer à l'inférence d'autres données, jusqu'à ce qu'un point fixe soit atteint (\textit{i.e.} pas de nouveaux faits peuvent être inférés). Ensuite, la réponse à la requête est réduite à une simple évaluation de la requête sur cet ensemble de données saturé.

\subsection{Phase 3 : Chargement des triplets RDF dans Neo4j }

\parindent = 0.5cm
Cette phase consiste à stocker les triplets RDF dans la BD graphique \textit{Neo4j}, où le sujet et l'objet sont des nœuds, ainsi que les prédicats sont les relations qui les relient (elles peuvent être de type primitive OWL;\textit{ e.g. Sub-Class-Of}, \textit{Equivalent-Object-Property, Same-As, etc,} ou bien des propriétés de l'ontologie; telles que \textit{hasBase, UNDEFINED-part-of, etc}). Les nœuds et les relations, dans Neo4j, peuvent être étiquetés et attribués. \\

\begin{figure}[H]
\begin{center}

\includegraphics[scale=0.6]{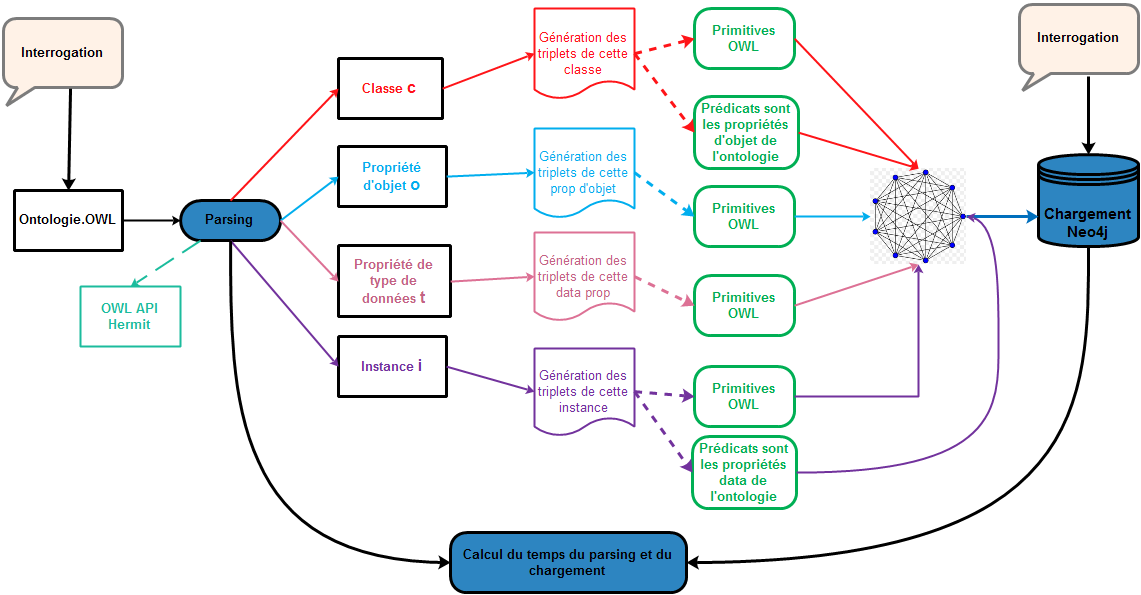}

\caption{Parsing de l'ontologie, génération et chargement du graphe RDF dans Neo4j \label{figure 3.57}}
\end{center}
\end{figure}

Cette phase s'effectue au fur et à mesure avec le parsing de l'ontologie (figure \ref{figure 3.57}). 

\subsection{Phase 4 : Interrogation de la BD Neo4j}

\begin{figure}[H]
\begin{center}

\includegraphics[scale=0.7]{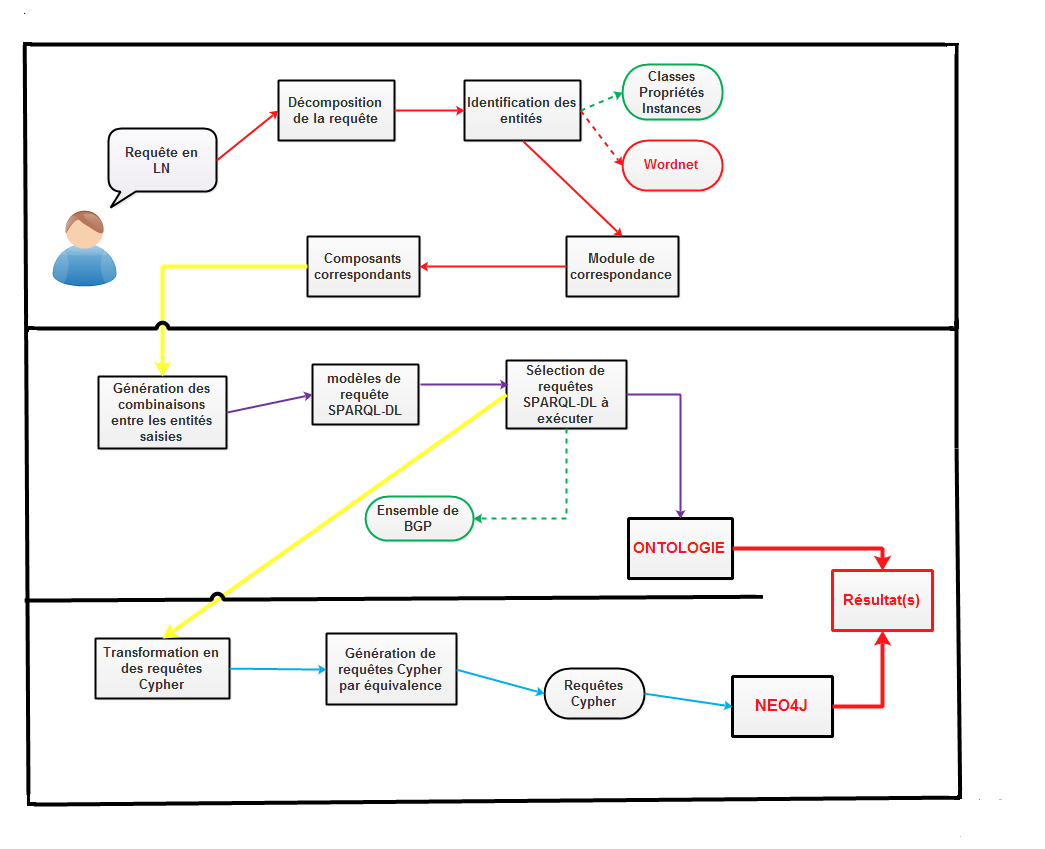}

\end{center}
\caption{Processus d'interrogation de la BD Neo4j \label{figure 3.9}}
\end{figure}

Afin d'interroger la BD graphique Neo4j, la requête utilisateur, initialement formulée en langage naturel, doit passer par différentes étapes de traitement (Figure \ref{figure 3.9}). \\ 

Ces étapes comportent : \\
- L'analyse de la requête utilisateur ; \\
- L'identification du type de chaque entité saisie ; \\
- La génération des différentes liaisons sémantiques possibles entre les composants de la requête ; \\
- La génération des requêtes SPARQL-DL afin de les exécuter sur l'ontologie ;  \\
- Équivalence de langages afin d'aboutir à des requêtes Cypher ; \\
- Exécution des requêtes Cypher (à partir de la BD graphique Neo4j) ;  \\ 
- Comparaison des résultats de ces derniers. \\

\subsubsection { Étape de l'analyse de la requête utilisateur }

\parindent = 0.5cm
Cette étape consiste à segmenter / décomposer la requête saisie par l'utilisateur afin d'identifier ses différents composants (\textit{i.e.} entités nommées). \\

\parindent=0.5cm
Il arrive que l'utilisateur ne saisit pas le nom de l'entité mais plutôt son synonyme, son hyperonyme et/ou son hyponyme. Par conséquent, nous avons eu recours à la base de données lexicale \textit{WordNet}, avec laquelle, et dans le cadre de ce travail, nous avons utilisé juste les synonymes, car ils sont au même niveau, dans l'ontologie, que l'entité saisie. Cependant les hyperonymes et les hyponymes sont de niveau supérieur et inférieur, respectivement. Cela veut dire que l'hyperonyme est considéré comme la super-entité de l'entité saisie, ainsi que l'hyponyme est sa sous-entité. Ce qui signifie qu'ils ne possèdent pas forcément les mêmes informations (cette étape reste à améliorer en prenant également en considération les hyperonymes et les hyponymes). 

\parindent=0cm
La requête d'entrée passe par trois étapes de pré-traitement, et c'est WordNet qui assure ces étapes : 

\parindent=0.5cm
- Tokenization ; 

\parindent=0.5cm
- Suppression des mots vides ;

\parindent=0.5cm
- Lemmatisation du mot : déterminer le lemme des mots-clés restants. \\

\parindent=0.5cm
Alors, après avoir terminé le pré-traitement de la requête en LN, il faut préciser et déterminer si les mots-clés saisis sont les noms prédéfinis des entités, ou bien ils se référent à leurs synonymes. Pour ce faire, et au moment du chargement dans la phase précédente, des triplets RDF (dans lesquels le sujet est soit une classe, une propriété ou une instance), nous avons stocké également tous les synonymes de chaque sujet. Nous avons, au même temps, relié chaque synonyme d'un sujet à tous les objets de ce sujet par le biais des mêmes relations (\textit{e.g.} \textit{Sub-Class-Of}, \textit{hasIngredientOf}, \textit{etc}), afin de former un sorte de graphe de synonymes. Réellement, nous avons considéré les synonymes comme étant les noms des entités eux-mêmes. \\
Par exemple, nous supposons que l'utilisateur a saisi \textit{" What are FishTopping and thermal ?"}, tel que "FishTopping" est le nom d'une classe, ainsi que thermal est le synonyme de la classe "Hot" ; ces deux classes font parties de l'ontologie \emph{Pizza}. Notre système va devoir répondre à cette requête.  \\

\parindent=0.5cm
Cependant, la majorité des noms des entités dans les ontologies que nous avons utilisées, se composent de deux parties ou voire plus (\textit{e.g. VegetableTopping, ThinandCrispyPizza, etc}). Dans ce cas, le moteur de WordNet segmente ces noms (\textit{e.g.} il va segmenter \textit{"VegetableTopping"} en Vegetable et Topping), puis il détermine les synonymes de chaque partie du nom, à part. Cependant, nous pouvons trouver deux noms de deux entités différentes qui se composent de quelques parties similaires, tels que \textit{"OliveTopping"} et \textit{"VegetableTopping"}. Ainsi, nous allons stocker les synonymes de Vegetable reliés à tous les objets de \textit{VegetableTopping}, et de même pour Topping, nous allons charger tous ses synonymes associés à tous les objets de \textit{VegetableToping}. \\
Dans ce cas, si jamais l'utilisateur saisit un synonyme de Topping, le système va lui retourner des informations sur les deux classes OliveTopping et VegetableTopping.  Cela veut dire que notre système va comprendre que ces deux entités partagent des informations en commun et par conséquent, il va stocker pour chaque nœud (correspondant à une entité) quelques informations de l'autre entité.

En somme, ces informations stockées ne sont pas forcément correctes sémantiquement.\\

\parindent=0.5cm
D'autre part, nous pouvons trouver des ontologies dans lesquelles les noms des entités peuvent contenir des chiffres et des symboles. Ces noms n'ont pas de synonymes dans \textit{WordNet}, par exemple, \textit{MA-0001480, MA-0001435, etc}, de l'ontologie \emph{Mouse}. \\
Dans ce cas, nous avons proposé d'utiliser les labels des entités au lieu de leurs noms (\textit{i.e.} l'utilisateur pourra saisir le synonyme du label d'une entité). cependant, les entités ne possèdent pas toutes des labels. En  gros, ce deux points restent à améliorer. \\

Enfin, un module de correspondance permet essentiellement de correspondre chaque composant avec les entités de l'ontologie déjà parsée, afin d'identifier et de distinguer la nature de chaque entité de la requête en LN. Pour ce faire, le système de stockage comprend quatre premières listes, une stocke les classes, et les trois autres stockent les propriétés d'objet, de type de données et les instances, respectivement. Les quatre autres listes comprennent les synonymes de chaque type d'entités (\textit{i.e.} classes, propriétés d'objet, propriétés de type de données et instances). Il suffit alors de comparer chaque composant de ces mots-clés avec chaque liste afin d'identifier sa nature / son type.

\subsubsection { Étape de la génération de requêtes SPARQL-DL }

\parindent =0.5cm
\textbf{SPARQL-DL} est un sous-ensemble important de SPARQL. Il est un langage expressif qui permet de mélanger des requêtes T-Box (\textit{i.e.} axiomes décrivant les concepts et le relations entre eux), A-Box (\textit{i.e.} représentent les instances) et R-Box. \\\textit{SPARQL-DL} s'avère être un langage d'interrogation qui peut soutenir l'expressivité avancée d'\textit{OWL 2}. Il se diffère de SPARQL au niveau de la syntaxe, en outre, SPARQL fournit plus de fonctionnalités lors de l'interrogation d'ontologies (\textit{e.g.} la possibilité d'interroger un graphe via le Web).

\parindent = 0.5cm
La cause pour laquelle nous avons utilisé \textit{SPARQL-DL} au lieu de SPARQL, est parce que \textit{OWL API} n'a aucun support natif pour SPARQL. En effet, \textit{OWL API} ne fonctionne pas au niveau RDF où SPARQL le fait, mais plutôt au niveau OWL abstrait. SPARQL est un langage standardisé pour interroger des données d'un graphe exprimé en RDF et non pas des ontologies écrites en OWL. \\

\parindent = 0.5cm
Les mots-clés décomposés auparavant permettent de générer plusieurs sous-requêtes \textit{SPARQL-DL}. Il reste enfin à traduire / mapper les mots-clés saisis et décomposés au cours de l'étape précédente, en des requêtes en langage d'interrogation \textit{SPARQL-DL} (\textit{i.e.} en un ensemble de \textit{Triple Pattern TP}, formant un \textit{BGP : Basic Graph Pattern}). Une correspondance d'une requête utilisateur N à partir d'une ontologie, est finalement une correspondance du BGP avec l'ontologie O. \\
Alors, pour chaque type d'entité, nous allons exécuter un ensemble de requêtes \textit{SPARQL-DL} permettant de retourner les informations qui sont liées à elle. Nous utilisons un ensemble de modèles de patrons graphiques (\textit{BGP}) de base prédéfinis (pour chaque type d'entité), permettant de déterminer les interprétations possibles et adéquates des requêtes utilisateurs. Ceci est réalisé en obtenant les listes des identifiants des entités (\textit{i.e.} des mots-clés fournis) et par la suite en injectant ces identifiants dans les positions appropriées dans les \textit{TP}, de telle sorte que les TP obtenus suite à cette injection, soient des sous-ensembles de l'ontologie globale. \\

Ainsi, si nous exécutons directement les requêtes \textit{SPARQL-DL }générées, les résultats seraient retournés à partir de l'ontologie. Cependant, notre but est d'interroger la BD \textit{Neo4j}.

\subsubsection { Étape de transformation/équivalence des requêtes SPARQL-DL en des requêtes Cypher }

\parindent = 0.5cm
En effet, l'utilisateur va fournir une requête en langage naturel et il n'a pas à se soucier et ne lui intéresse pas réellement d'où vient le résultat à sa requête. Il saisit ses mots-clés décrivant son besoin en information et il attend un résultat qui lui satisfait en un temps court. Alors pourquoi faire une équivalence (une sorte de transformation) entre les deux langages ? \\ 

Tout d'abord, parce que notre but dès le départ est d'interroger un graphe RDF stocké dans une BD graphique, alors que nous pouvons pas accéder aux graphes RDF avec le langage \textit{SPARQL-DL}. En outre et ce qui est le plus important, cela nous permettrait de changer de BD-RDF tout en gardant le travail de la génération des requêtes \textit{SPARQL-DL} à partir des mots-clés,  intacte. Cela veut dire que l'avantage de cette équivalence est qu'elle nous permet d'implémenter l'architecture de notre système au-dessus d'une grande variété de BD-RDF implémentant leurs langages d'interrogation, sans changer aucun des autres composants du système. Alors, en changeant de BD-RDF, il suffirait juste de stocker les triplets existants (déjà). Ensuite, il faut changer les requêtes du langage Cypher par des requêtes écrites en langage propre à la nouvelle BD-RDF utilisée. Ainsi, l'équivalence reste applicable entre SPARQL-DL et le nouveau langage. \\

\begin{figure}[H]
\begin{center}

\includegraphics[scale=0.98]{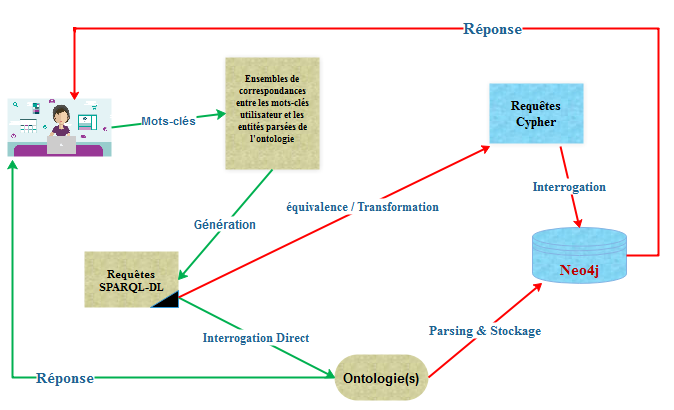}

\end{center}
\caption{Architecture générale de notre système \label{figure 3.11}}
\end{figure}

\parindent = 0.5cm
Le but également (comme le montre la figure \ref{figure 3.11}) est : \\

\parindent=0cm
- Interroger en premier lieu l'ontologie elle-même directement en utilisant \textit{SPARQL-DL} ; \\
- Interroger, par la suite, les données stockées dans la BD Neo4j en utilisant Cypher, plus précisément, en passant par \textit{SPARQL-DL} et en faisant l'équivalence. \\ 

\parindent =0.5cm
$\Rightarrow$ Comparer les résultats de ces derniers : \textbf{il faut absolument trouver les mêmes résultats}. Cela signifie que le passage d'une ontologie à un graphe RDF n'influence pas la précision des résultats, et que les données d'une ontologie sont converties convenablement et proprement vers un graphe RDF. Ces données reflètent fidèlement la sémantique décrite dans l'ontologie.  \\

\parindent = 0.5cm
Alors, après avoir modélisé les requêtes SPARQL-DL d'une requête initialement formulée par l'utilisateur, il faut maintenant passer à la phase de réécriture afin d'avoir des requêtes Cypher. Les requêtes \textit{SPARQL-DL} seront transformées en d'autres équivalentes à elles sémantiquement afin de pouvoir accéder à \textit{Neo4j}. Il est nécessaire, pour la réécriture de la requête, que l'application de la transformation n'influence pas le résultat final de la requête. \\

Notre système vise à réécrire les requêtes \textit{SPARQL-DL} (en d'autres \textit{Cypher}) en s'assurant que les reformulations reflètent autant que possible la même signification, la même structure, le même type de résultats et la même taille de résultat que les requêtes \textit{SPARQL-DL} initiales. \\

\textbf{Remarque :} véritablement, il n'y a pas une manière universelle permettant de mesurer la distance ou/et la similarité entre deux requêtes formelles. \\ 

\begin{figure}[!ht]
\begin{center}

\includegraphics[scale=0.78]{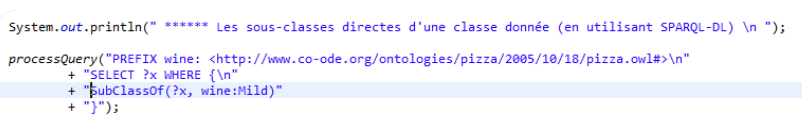}

\end{center}
\caption{Exemple de requête SPARQL-DL \label{figure 3.12}}
\end{figure}

\begin{figure}[!ht]
\begin{center}

\includegraphics[scale=0.74]{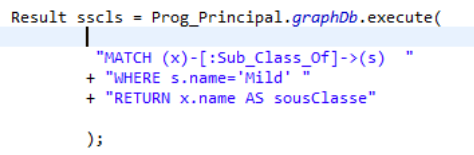}

\end{center}
\caption{Exemple de requête Cypher \label{figure 3.13}}
\end{figure}

Par conséquent, afin d'assurer cette équivalence : \\
- Nous avons pris chaque deux requêtes (\textit{SPARQL-DL} et \textit{Cypher}) renvoyant, normalement, les mêmes résultats ; \\
- Nous avons essayé d'identifier de chaque côté, les parties similaires (qui se répètent) entre chaque deux requêtes\textit{ SPARQL-DL }et \textit{Cypher} ; \\
- Afin de générer des requêtes \textit{Cypher} à partir de celles SPARQL-DL, nous avons gardé les parties similaires syntaxiquement des deux requêtes et nous avons écrasé les autres parties (différentes) des requêtes \textit{SPARQL-DL }et les fait remplacer par leurs correspondantes du langage \textit{Cypher}. Les figures \ref{figure 3.12} et \ref{figure 3.13} montrent des exemples de requête \textit{SPARQL-DL} et \textit{Cypher} ayant pour but de déterminer les sous-classes d'une classe donnée. En analysant ces deux requêtes, on constate que les deux langages sont presque totalement différents syntaxiquement, et par conséquent, il n'y avait pas de grandes équivalences dégagées. \\ \\

\parindent = 0.5cm
Toutes les variables qui apparaissent dans les \textit{TP} de la requête SPARQL-DL, apparaissent également dans la requête Cypher.  \\

\parindent =0.5cm
Dans les requêtes SPARQL et SPARQL-DL, nous utilisons le mécanisme des espaces de noms pour définir les préfixes afin de rendre les requêtes plus lisibles. Un \textit{IRI (\textbf{I}nternationalised \textbf{R}esource \textit{I}dentifier)} d'une entité, est composé d'un "IRI de préfixe" suivi du symbole \textbf{"\#"} puis du "nom" de l'entité. En général, la partie "IRI de préfixe" de l'entité, est exactement l'IRI de l'ontologie.

\section{Interface de recherche et les différentes approches de réponse aux requêtes}
\subsection{Interface de recherche}

\begin{figure}[!ht]
\begin{center}

\includegraphics[scale=0.95]{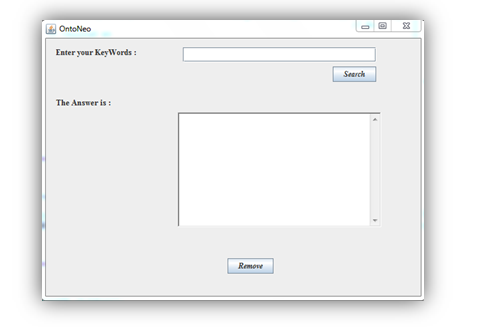}

\end{center}
\caption{Interface utilisateur \label{figure 3.14}}
\end{figure}

\parindent =0.5cm
C'est une interface simple pour l'utilisateur donnée par figure \ref{figure 3.14}. La $1\up{ère}$ partie intitulée \textit{"Enter Your KeyWords"}, consiste à une zone du texte où l'utilisateur tape le nom des entités appartenant aux ontologies considérées. L'utilisateur doit saisir des noms valides et la zone du texte ne doit pas rester vide, sinon un message d'erreur s'affiche. Une fois le bouton \textit{"Search"} soit cliqué, le résultat s'affiche dans la zone du texte en lecture seule \textit{"The Answer is"}.  \\

\subsection{Les approches de réponse aux requêtes}

\subsubsection{Affichage de toutes les informations à l'utilisateur}

\parindent =0.5cm
Pour ce faire, il faut tout d'abord effectuer des combinaisons entre les mots-clés décomposés de la requête en LN de l'utilisateur, afin d'établir les liaisons sémantiques possibles entre eux. On doit alors pouvoir extraire une représentation sémantique du contenu de la requête (N) en LN, qui nous permettra de générer la (ou les) requête(s) \textit{SPARQL-DL}, et par la suite les requêtes Cypher correspondantes. Chaque requête Cypher  représente un sous-graphe du graphe RDF global. \\

\parindent=0.5cm Parce que N est une donnée non structurée et G est un graphe contenant des données structurées, nous avons dû combler l'écart entre deux types de données pour avoir une représentation graphique de N. Pour ce faire, il faut tout d'abord extraire les relations sémantiques (\textit{i.e.} les propriétés) impliquées de la question N; dont chaque relation sémantique est un triplet $({arg_1}, rel, {arg_2})$ où \textit{rel} est la relation, ${arg_1}$, ${arg_2}$ sont les éléments reliés par \textit{rel}. \\

\parindent = 0.5cm
En effet, nous avons étudié le cas de un à trois mots-clés maximums, sinon ça va être beaucoup de combinaisons à déterminer. Au delà de 3 mots-clés, notre système retourne les informations de chaque entité à part sans effectuer des combinaisons sémantiques entre elles. \\

\parindent =0.5cm
Au moment où l'utilisateur saisit un (ou plusieurs) mot-clé, toutes les informations primitives de l'entité saisie (Sub-Class-Of, Equivalent-Property, \textit{etc}.), en plus, les résultats des combinaisons des mots-clés vont être affichés. \\

\subsubsubsection {Le cas d'un seul mot-clé}

C'est le cas le plus facile à traiter, dont le mot-clé peut être une classe, une propriété (d'objet ou de type de donnés) ou bien une instance (tableau \ref{Tableau 3.1}).    \\

\parindent=0.5cm
L'utilisateur saisit le nom (ou bien le synonyme du nom) d'une entité appartenant à l'ontologie en question (ou bien au graphe RDF de cette ontologie et qui est stocké dans Neo4j) et il ne précise pas l'information qu'il veut avoir en retour. Ce mot-clé entré va être comparer à toutes les structures existantes afin de déterminer sa nature (\textit{i.e.} classe, propriété ou instance). Ensuite, le système lui retourne toutes les informations liées à l'entité fournie. \\

Le saisi d'une seule entité signifie que deux composants du triplet RDF sont vides. Les informations qui s'afficheront à l'utilisateur sont les résultats  des requêtes SPARQL-DL (sur l'ontologie) et ceux des requêtes Cypher par équivalence (sur le graphe RDF). \\

Un mot-clé est valide et pertinent, s'il peut être mis dans un patron de triplet (respectivement, dans une requête Cypher) et que ce dernier soit un sous-ensemble de l'ontologie considérée (respectivement, un sous-graphe du graphe RDF généré), ainsi qu'il retourne un résultat. Même une valeur de 0 est prise en considération car cela peut signifier que la clause \textit{WHERE} de la requête \textit{SPARQL-DL} et de la requête Cypher, ne produit aucune correspondance ni dans l'ontologie ni dans le graphe RDF.

\begin{table}[H]
\begin{center}
\renewcommand {\arraystretch }{1.2}

\begin{tabular}{|l||l|}

\hline
\textbf{Mot-clé} & \textbf{Résultats attendus} \\
\hline \hline
\multirow {3} {*} {\textit{\textbf{Classe}}} & $\ast$ Ses classes équivalentes ; \\ \cline{2-2}
                          & $\ast$ Ses classes disjointes ; \\ \cline{2-2}
                          & $\ast$ Ses sous-classes ; \\ \cline{2-2}
                          & $\ast$ Ses instances (s'il y en a) ; \\ \cline{2-2}
                          & $\ast$ Son label et son commentaire. \\  \hline \hline \hline
                          
\multirow {3} {*} {\textit{\textbf{Propriété}}}  & $\ast$ Ses propriétés équivalentes ; \\ \cline{2-2}
                          & $\ast$ Ses propriétés disjointes ; \\ \cline{2-2} 
                           & $\ast$ Ses sous-propriétés ; \\ \cline{2-2}
                           & $\ast$ Ses propriétés inverses (que pour les propriétés d'objet) ; \\ \cline{2-2} 
                            & $\ast$ Son label et son commentaire ; \\ 
 \cline{2-2}
                          & $\ast$ Son domaine (classe source) ; \\ 
                          & $\ast$ Son image (classe cible). \\ 
                          & \textbf{$\Rightarrow$ }qui sont reliés par la propriété donnée : \\
                         & \textcolor{red}{\textbf{?}} $\rightarrow$ \textcolor{green}{propriété \textbf{X}} $\rightarrow$ \textcolor{red}{\textbf{?}}  \\
                         \hline \hline \hline
\multirow {3} {*} {\textit{\textbf{Instance}}} & $\ast$ Ses instances équivalentes \\ \cline{2-2}
                          & $\ast$ Ses instances différentes \\ \cline{2-2}
                          & $\ast$ Son label et son commentaire \\ \cline{2-2}              
                           & $\ast$ Déterminer la classe à laquelle l'instance (saisie) appartient, \\
                           & via la relation \textit{Instance-Of} : \\     
                          & Instance $\rightarrow$ Instance-Of $\rightarrow$ \textcolor{red}{\textbf{?}} \\ 
                          
                          \hline \hline \hline

\end{tabular}

\end{center}
\caption{Cas de recherche d'un seul mot-clé \label{Tableau 3.1}}
\end{table}

\subsubsubsection {Le cas de deux mots-clés}

Dans ce cas, il y en a six combinaisons à déterminer (tableau \ref{Tableau 3.2}). \\

Par exemple, si les deux entités saisies sont des classes, alors le problème devient celui de déterminer une relation qui peut relier ces deux classes. Cette relation est le prédicat, qui peut être de type primitive OWL (\textit{i.e.} Sub-Class-Of, Equivalent-Class-Of, etc), ou bien une propriété d'objet.\\ Le système retourne également toutes les informations (citées dans le cas précédent) de ces deux classes. \\

\parindent=0.5cm
Le cas le plus compliqué est quand l'utilisateur fournit deux arguments dont l'un est une classe et l'autre est une propriété. Tout d'abord, le système considère la classe saisie comme étant le domaine (le sujet du triplet) et il détermine alors son (ses) co-domaine(s) tout en considérant la propriété fournie comme un prédicat. Ensuite, cette fois le système met la classe à la place de l'objet du triplet RDF (\textit{i.e.} comme étant une image), et alors il doit déterminer son (ses) domaine(s) par le biais de la propriété initialement fournie par l'utilisateur. \\
Les informations de chacune de ces entités à part, s'afficheront également à l'utilisateur.

\begin{table}[H]
\begin{center}

\resizebox{\textwidth}{!}{%
\begin{tabular}{||l||l||}

\hline
\textbf{Mots-clés} & \textbf{Résultats attendus} \\
\hline \hline
\textit{\textbf{Deux classes}} & $\ast$ Retourner les informations (déjà mentionnées dans \\
               & le cas précédent) de chacune de ces deux classes. \\ \cline{2-2}
               & La (les) relation(s) qui relie(ent) ces deux classes (s'il y en a). \\
                  \hline \hline \hline
                          
\multirow {3} {*} {\textit{\textbf{Deux propriétés}}}  & $\ast$ Retourner les informations (déjà mentionnées dans \\
                 & le cas précédent) de chacune de ces deux \\
                 & propriétés \textbf{+} \\ \cline{2-2}
                               & $\ast$ Leurs domaines et leurs images. \\
                         \hline \hline \hline
                         
\multirow {3} {*} {\textit{\textbf{Deux instances}}} & $\ast$ Les informations (déjà mentionnées dans le $1\up{er}$ \\
         & cas) de chacune de ces deux instances \textbf{+} \\ \cline{2-2}
                         & $\ast$ La (les) classe(s) qui les instancie(ent). \\ 
                          
                          \hline \hline \hline    
                        
\multirow {3} {*}{\textit{\textbf{Une classe-Une propriété}}} & $\ast$ Les informations de chaque entité saisie ; \\ \cline{2-2}
                          & $\ast$ Au début, on considère la classe (donnée) comme\\
                          & un domaine (classe source) et donc on va déterminer le\\
                          & co-domaine (la classe cible) et qui sont reliées à\\
                          & elle par l'intermédiaire de la propriété saisie :\\
                          & \textcolor{blue}{Domaine (classe saisie)} $\rightarrow$ \textcolor{green}{propriété saisie} $\rightarrow$ \textcolor{red}{\textbf{?}} \\ \cline{2-2}
                          & $\ast$ Dans un $2\up{ème}$ lieu, soit la classe saisie une image (classe\\
                          & cible), et donc on va rechercher sa (ses) classe(s) source(s)\\
                          & (domaines), sachant que le prédicat est la propriété saisie.\\
                        &  \textcolor{red}{\textbf{?}} $\rightarrow$ \textcolor{green}{propriété saisie} $\rightarrow$ \textcolor{blue}{Image (classe saisie)} 
                         \\  \hline \hline \hline
                          
\multirow {3} {*} {\textit{\textbf{Une classe-Une instance}}} & $\ast$ En plus de déterminer les informations de chacune,\\
                          & le système va aussi retourner si la classe saisie\\
                          & par l'utilisateur contient l'instance donnée. \\
                         \hline \hline \hline
                         
\textit{\textbf{Une propriété-Une instance}} & $\ast$ Dans ce cas, on va déterminer juste les informations de \\ 
                         & la propriété et de l'instance données.\\ 
                              \hline \hline \hline

\end{tabular}
}

\end{center}
\caption{Cas de recherche de deux mots-clés \label{Tableau 3.2}}
\end{table}

\subsubsubsection {Le cas de trois mots-clés}

Cela impliquera dix combinaisons à produire (tableau \ref{Tableau 3.3}). \\

\parindent = 0.5cm
Les mots-clés saisis doivent être regroupés ensembles en des segments. Pour chaque segment, des informations appropriées sont alors déterminées.  Un segment valide est un segment pour lequel au moins une information peut être trouvée dans le graphe RDF sous-jacent. Notre algorithme montre une approche naïve pour trouver tous les segments valides, en considérant l'ordre des mots-clés. Il commence par le $1\up{er}$ mot-clé dans la requête donnée en tant que $1\up{er}$ segment, puis ajoute le mot-clé suivant au segment en cours et vérifie si cet ajout rendrait le nouveau segment invalide. Ce processus est répété jusqu'à ce qu'il atteigne la fin de la requête (\textit{i.e.} toutes les combinaisons possibles).

\begin{table}[H]
\begin{center}
\renewcommand {\arraystretch }{1.1}
   
\begin{tabular}{||l||l||}

\hline
\textbf{Mots-clés} & \textbf{Résultats attendus} \\
\hline \hline
\textit{\textbf{Trois classes}} & $\ast$ Déterminer leurs informations respectivement.  \\
                  \hline \hline \hline
                          
\textit{\textbf{Trois propriétés}}  & $\ast$	Déterminer leurs informations, ainsi que leurs\\
& domaines et leurs images : \\ & \textcolor{red}{\textbf{?}} $\rightarrow$ propriété \textbf{X} $\rightarrow$ \textcolor{red}{\textbf{?}}. \\
                     \hline \hline \hline
                         
\textit{\textbf{Trois instances}} & $\ast$ Déterminer leurs informations et à quelle(s)\\ 
              & classe(s) elles appartiennent. \\ 
                        \hline \hline \hline    
                        
\multirow {3} {*}{\textit{\textbf{Deux classes-Une propriété}}} & $\ast$ Les informations des entités saisies  \textbf{+}\\ \cline{2-2}
                          & $\ast$ Domaine(s) et image(s) de la propriété saisie \textbf{+}\\ \cline{2-2}
                          & $\ast$ Image(s) de chacune des classes via la propriété \\
                          & saisie \textbf{+}\\ \cline{2-2}
                          & $\ast$ Domaine(s) de chacune des classes via \\ 
                          & la propriété saisie.  
                         \\  \hline \hline \hline
                         
\multirow {3} {*}{\textit{\textbf{Deux classes-Une instance}}} & $\ast$ Les informations de chacune des entités saisies \textbf{+}\\ \cline{2-2}
                          & $\ast$ A quelle classe appartient l'instance saisie \textbf{+}\\ \cline{2-2}
                          & $\ast$ Déterminer si l'une des classes (entrées) contient \\
                          & l'instance donnée. 
                         \\  \hline \hline \hline
                          
\multirow {3} {*} {\textit{\textbf{Deux propriétés-Une classe}}} & $\ast$ L'ensemble des informations des entités saisies \textbf{+}\\ \cline{2-2} 
                  & $\ast$ Domaine(s) et image(s) de chacune des propriétés \textbf{+}\\ \cline{2-2}
                  & $\ast$ Domaine(s) de la classe saisie via les deux propriétés \\
                  & données (une par une)\\ 
                  & \textcolor{red}{\textbf{?}} $\rightarrow$ \textcolor{green}{\emph{P1}} $\rightarrow$ C \\
                  & \textcolor{red}{\textbf{?}} $\rightarrow$ \textcolor{blue}{\emph{P2}} $\rightarrow$ C \\
                  & $\ast$ Image(s) de la classe saisie via les \\
                  & deux propriétés données. \\
                  
                  & C $\rightarrow$ \textcolor{green}{\emph{P1}} $\rightarrow$ \textcolor{red}{\textbf{?}} \\
                  & C $\rightarrow$  \textcolor{blue}{\emph{P2}} $\rightarrow$ \textcolor{red}{\textbf{?}} \\
                 
                   \hline \hline \hline
                                                                   
\textit{\textbf{2 propriétés-Une instance}} & $\ast$ Juste affichage des informations de chaque entité \\
                       & entrée.
                         \\  \hline \hline \hline     
                         
\multirow {3} {*}{\textit{\textbf{Deux instances-Une classe}}} & $\ast$ Informations de ces entités  \textbf{+}\\ \cline{2-2}
                          & $\ast$ Déterminer la(les) classe(s) auxquelles les instances \\
                          & saisies appartiennent.
                         \\  \hline \hline \hline
                         
\textit{\textbf{2 instances-Une propriété}} & $\ast$ Détermination des informations de chaque entité \\
                          & entrée.
                         \\  \hline \hline \hline
                         
\textit{\textbf{1 instance- 1 propriété- 1 classe}} & $\ast$ Détermination des informations de chaque entité.

                         \\  \hline \hline \hline
\end{tabular}

\end{center}
\caption{Cas de recherche de trois mots-clés \label{Tableau 3.3}}
\end{table}

\newpage
\parindent = 2cm
\textit{\textbf{Exemple d'exécution}} \\

\begin{figure}[H]
\begin{center}

\includegraphics[scale=0.85]{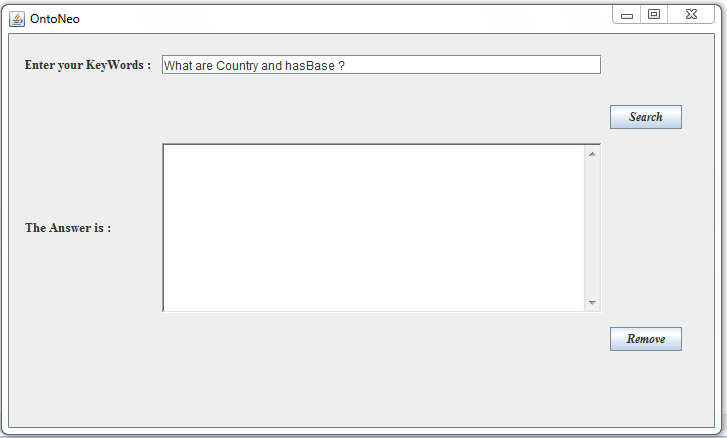}

\end{center}
\caption{Requête utilisateur en LN \label{figure 3.15}}
\end{figure}

\parindent = 0.5cm
On suppose que l'utilisateur a saisi la requête en LN (dans la figure \ref{figure 3.15}) qui comporte un nom d'une classe et un nom d'une propriété $\Rightarrow$ le résultat va contenir les informations de type primitives OWL de ces entités, de plus, les résultats des combinaisons de ces deux mots-clés, \textit{i.e.} : \\
- Les informations de la classe et de la propriété saisies \textbf{+} \\ 
- Le domaine et l'image de la propriété saisie + \\ 
- L'image (respectivement le domaine) de la classe saisie tout en considérant à chaque fois la propriété fournie comme étant le prédicat. \\

Dans cet exemple, nous travaillions sur l'ontologie \textit{Pizza} (nous allons définir cette ontologie dans le chapitre suivant). Après la décomposition de la requête, nous saisissons que  l'utilisateur a fourni deux noms de deux entités : \textit{« Country »} qui est une classe et \textit{« hasBase »} qui représente une propriété d'objet. Nous supposons que l'utilisateur ne connait ni la nature des entités saisies ni la structure du graphe interrogé. \\ 

(\textcolor{red}{Remarque} : nous avons utilisé l'ontologie Pizza dans cet exemple parce qu'elle est riche en informations. Elle comporte des classes, des propriétés d'objet et de type de données ainsi que des instances. Alors, elle nous permet d'évaluer l'exhaustivité de notre travail.) \\

\begin{figure}[H]
\begin{center}

\includegraphics[scale=1]{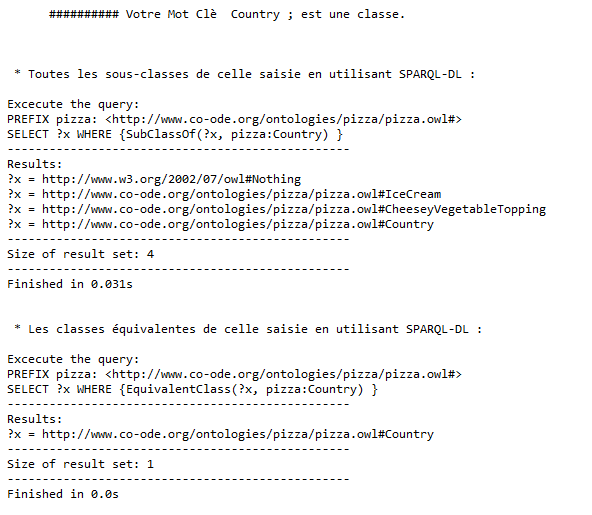}

\end{center}
\caption{Résultats de la requête en LN en utilisant SPARQL-DL (1) \label{figure 3.16}}
\end{figure}

\begin{figure}[H]
\begin{center}
\includegraphics[scale=0.8]{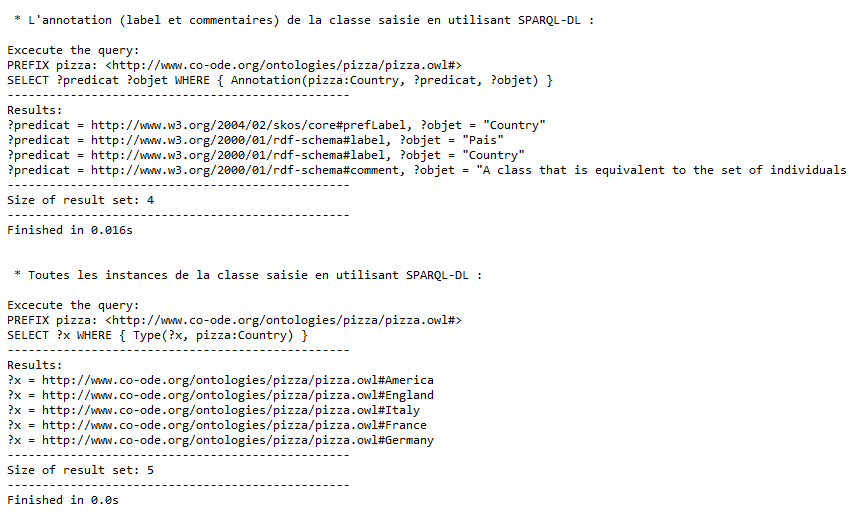}

\end{center}
\caption{Résultats de la requête en LN en utiliant SPARQL-DL (2) \label{figure 3.17}}
\end{figure}

\begin{figure}[H]
\begin{center}

\includegraphics[scale=0.95]{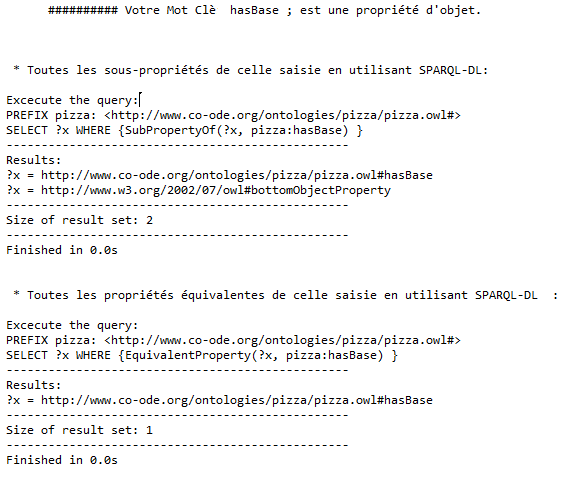}

\end{center}
\caption{Résultats de la requête en LN en utilisant SPARQL-DL (3) \label{figure 3.18}}
\end{figure}

\begin{figure}[H]
\begin{center}

\includegraphics[scale=1.12]{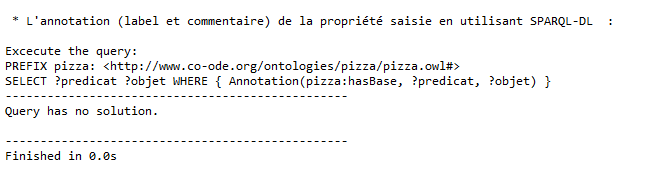}

\end{center}
\caption{Résultats de la requête en LN en utilisant SPARQL-DL (4) \label{figure 3.19}}
\end{figure}

Les figures \ref{figure 3.15}, \ref{figure 3.16}, \ref{figure 3.17}, \ref{figure 3.18} et \ref{figure 3.19} présentent les résultats de cette requête utilisateur en exécutant (tout d'abord) des requêtes \textit{SPARQL-DL} sur l'ontologie.  

\parindent = 0.5cm
La figure \ref{figure 3.16} montre les sous-classes et les classes équivalentes de celle saisie par l'utilisateur, tout en précisant le type de l'entité \textit{Country}. Nous constatons la présence du préfixe qu'on en a déjà parlé (pour les requêtes SPARQL-DL, le préfixe est une chose indispensable dans l'écriture de la requête puisqu'il permet de déterminer l'IRI de l'ontologie cible à interroger et ce qui est logique vu qu'on interroge l'ontologie directement, ainsi qu'il s'affiche également dans les résultats, mais cela ne change rien). \\

\parindent = 0.5cm
La classe \textit{"Nothing"} est la sous-classe de toutes les classes dans l'ontologie, c'est pour cette raison qu'on remarque sa présence à chaque fois qu'on demande les sous-classes d'une certaine classe. \\
De même, nous constatons que le moteur d'interrogation de \textit{SPARQL-DL} suppose toujours qu'une classe est équivalente sémantiquement à elle-même (\textit{i.e.} la classe \textit{Country} possède qu'une seule classe équivalente, qui est \textit{Country} elle-même). Dans ce contexte, la règle dit qu'un axiome \textit{EquivalentClasses ($C_1$ $C_2$)} entre 2 entités est formellement équivalent à 2 axiomes de subsomption réciproques : \\

\parindent = 0cm
- \textit{SubClassOf ($C_1$ $C_2$)} ;  \\
- \textit{SubClassOf ($C_2$ $C_1$)}. \\

\begin{figure}[H]
\begin{center}

\includegraphics[scale=0.9]{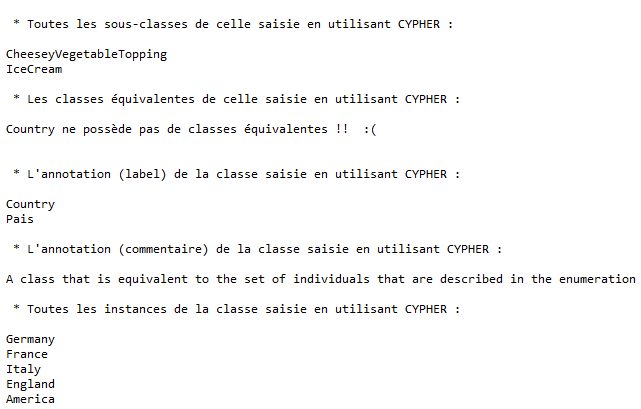}

\end{center}
\caption{Résultats de la requête en utilisant Cypher (par équivalence) (1) \label{figure 3.20}}
\end{figure}

\parindent = 0.5cm
Cela veut dire que, puisque \textit{Country} est équivalente à elle-même alors elle est, suivant la règle, une sous-classe à elle-même. Ces informations sont inférées par le raisonneur et elles me semblaient triviales que je ne les ai pas stocké dans \textit{Neo4j}, c'est pour cette raison qu'elles ne sont pas affichées en appliquant les requêtes Cypher par équivalence (figure \ref{figure 3.20}, page \pageref{figure 3.20}). \\

\parindent = 0.5cm
Ensuite, l'annotation (le label ainsi que le commentaire) de la classe saisie et ses instances, sont présentés dans la figure \ref{figure 3.17}. \\ \\

\parindent = 0.5cm
Après avoir affiché toutes les informations de la classe saisie, il faut maintenant afficher celles de la propriété. La figure \ref{figure 3.18} montre les sous-propriétés et les propriétés équivalentes de \textit{« hasBase »}. Finalement, cette propriété ne possède pas d'annotation comme le montre la figure \ref{figure 3.19}. \\

\parindent=0.5cm
De même, la propriété d'objet hasBase ne possède qu'une seule propriété équivalente, qui est hasBase elle-même. Par conséquent, hasBase est une sous-propriété à elle-même (suivant la règle précédemment définie). \\
La sous-propriété \textit{BottomObjectProperty} est la sous-propriété d'objet de toutes les propriétés d'objet de l'ontologie.

\newpage
Maintenant, nous devons afficher à l'utilisateur, également, les résultats de sa requête (en LN) en utilisant Cypher (en passant par l'équivalence définie auparavant). La figure \ref{figure 3.20} présente les sous-classes, classes équivalentes, annotation et instances de la classe \textit{« Country »} saisie. On constate l'absence du préfixe dans les résultats et c'est parce que nous les avons enlevé lors du chargement dans la BD. \\ \\

\begin{figure}[H]
\begin{center}

\includegraphics[scale=0.72]{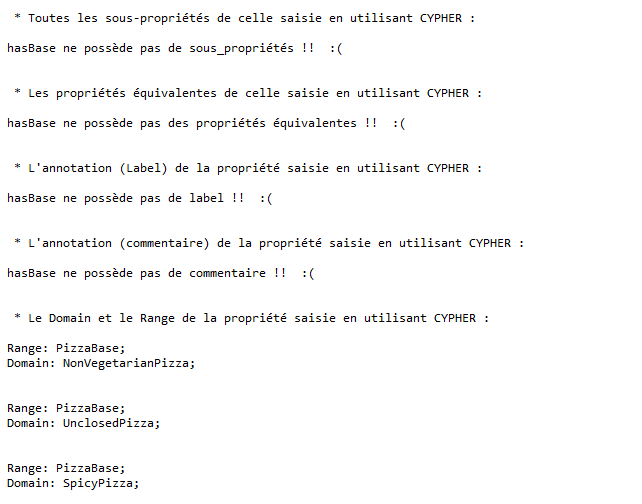}

\end{center}
\caption{Résultats de la requête en utilisant Cypher (par équivalence) (2) \label{figure 3.21}}
\end{figure}

\begin{figure}[H]
\begin{center}

\includegraphics[scale=0.72]{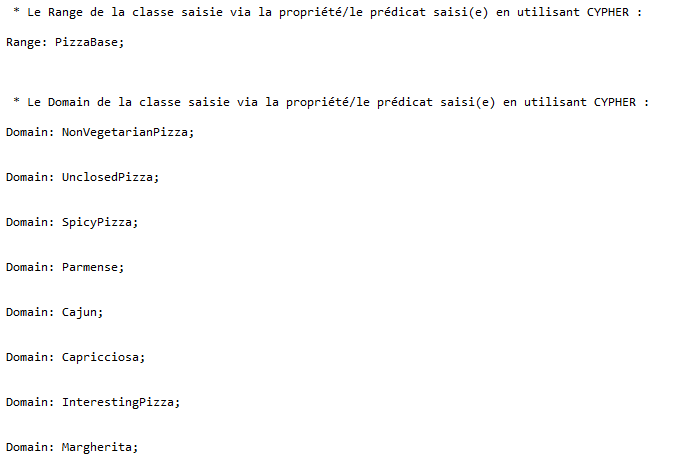}

\end{center}
\caption{Résultats de la requête en utilisant Cypher (par équivalence) (3) \label{figure 3.22}}
\end{figure}

La figure \ref{figure 3.21} indique les domaines et les co-domaines de la propriété \textit{« hasBase »}, ainsi que ses propriétés équivalentes, ses sous-propriétés et ses annotations (qu'elle ne les possède pas). \\
Dans la figure \ref{figure 3.22}, nous constatons que notre système a affiché les résultats des combinaisons entre les entités saisies. Dans notre exemple, il nous a renvoyé l'image (respectivement, le domaine) de la classe \textit{« Country »} via la propriété \textit{« hasBase »} (\textit{i.e.} il a considéré hasBase comme un prédicat).

\parindent =0cm
Nous avons obtenu les mêmes résultats que ce soit en utilisant \textit{SPARQL-DL} tout court ou bien en appliquant \textit{Cypher} (tout en passant par \textit{SPARQL-DL}). Les mêmes résultats s'affichent également dans l'interface, cependant nous avons choisi de faire des imprimes d'écran du console pour réduire le nombre de captures. \\

\parindent=0.5cm
Cette technique s'avère bonne pour des requêtes simples mais ce n'est pas le cas pour des requêtes plus compliquées, car le système doit parcourir les données sur tout le graphe RDF (respectivement, les données de l'ontologie) au moment de l'interrogation, afin de le faire correspondre aux requêtes Cypher complexes (respectivement, aux patrons \textit{SPARQL-DL} complexes).

\subsubsection{La recherche à facettes}

\parindent =0.5cm
En appliquant la méthode précédente, et qui permet de déterminer toutes les informations sur le(s) mot(s)-clé(s) utilisateur, et de générer toutes les combinaisons sémantiques entre eux, ce qui aboutit à un nouveau ensemble de requêtes, afin de déterminer le maximum d'informations sur la question utilisateur,  on constate que le résultat est long et parfois très long, parce que on ne sait pas ce que l'utilisateur cherche exactement et du coup, on lui affiche tous sur sa requête. \\
Afin de palier ce problème, nous avons proposé d'appliquer ce qu'on appelle de la « \textbf{recherche facettée} ». \\ 
Ce sont des interfaces qui s'enchaînent, un dialogue avec l'utilisateur lui permettant de raffiner l'interprétation de son besoin en informations, afin de mieux le satisfaire.

\begin{figure}[H]
\begin{center}

\includegraphics[scale=0.85]{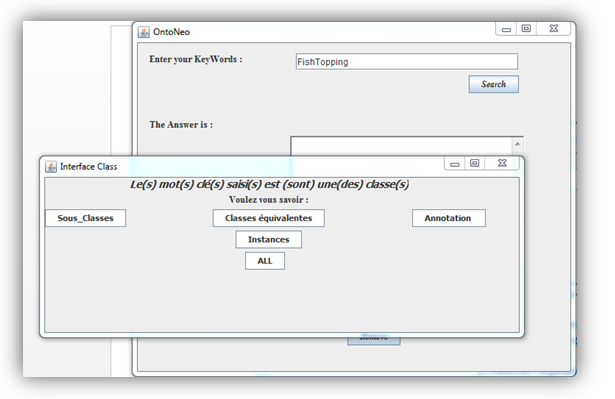}

\end{center}
\caption{Interface à facettes (1) \label{figure 3.23}}
\end{figure}

En saisissant ses mots-clés et en cliquant sur le bouton \textit{« Search »}, une $2\up{ème}$ interface (figure \ref{figure 3.23}) s'affiche à l'utilisateur contenant les différents choix possibles. Cette interface marque le début du processus de navigation. L'utilisateur peut sélectionner alors son besoin en information et le système réagit à ces actions en mettant à jour les résultats de la recherche. Nous avons capturé les facettes qui sont pertinentes pour chaque entité. \\

\begin{figure}[H]
\begin{center}

\includegraphics[scale=0.65]{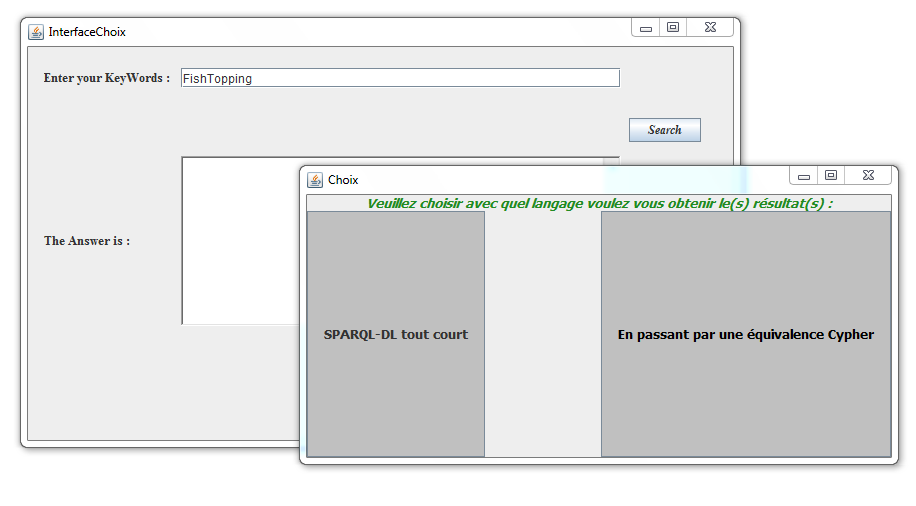}

\end{center}
\caption{Interface à facettes (2) \label{figure 3.24}}
\end{figure}

Une fois l'utilisateur fait son choix (parmi ces choix, nous trouvons le bouton \textit{« ALL »} qui nous permet d'afficher toutes les informations sur les entités saisies), une $3\up{ème}$ interface  (figure \ref{figure 3.24}) s'affiche indiquant à l'utilisateur qu'il peut choisir avec quel langage il préfère avoir les résultats. \\

Avec cette méthode qui assure la navigation à travers les collections inter-connectées des entités, les utilisateurs peuvent diminuer les résultats de recherche. Nous devons alors traduire les mots-clés et les choix de l'utilisateur en un ensemble de Triples Pattern (requête \textit{SPARQL-DL}). \\

\subsubsection{Interrogation de plus d'une ontologie}
\label{plsonto}

\parindent=0.5cm
L'utilisateur (en saisissant sa requête en LN), normalement il n'est pas obligé de savoir quelle entité est de quelle ontologie.
Par conséquent, il n'a pas à se soucier si son entré comporte des entités de différentes ontologies. \\

Alors, un défi, est que l'information permettant de répondre à une certaine question, puisse exister entre différents graphes RDF. Afin de répondre à ce défi, notre approche ressemble à une recherche horizontale, où chaque composant d'une requête d'entrée est comparé à tous les ensembles de données disponibles afin de déterminer à quelle ontologie il appartient. \\

\begin{figure}[H]
\begin{center}

\includegraphics[scale=0.75]{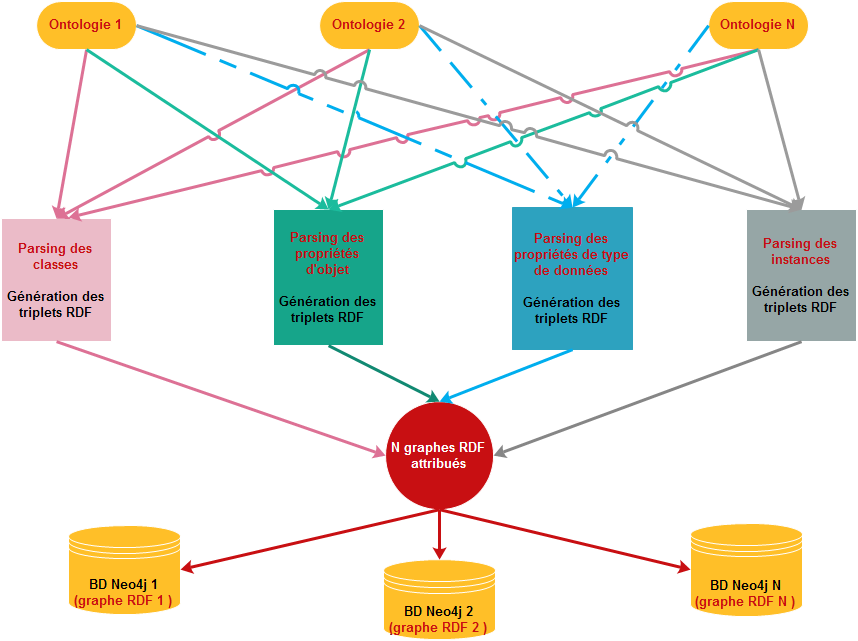}

\end{center}
\caption{Architecture générale du processus du traitement de plusieurs ontologies \label{figure 3.26}}
\end{figure}

Alors, après avoir parsé les ontologies, extrait les triplets RDF et les stocké dans \textit{Neo4j} (figure \ref{figure 3.26}), on aura la possibilité d'interroger, au même temps, soit ces ontologies directement (en utilisant \textit{SPARQL-DL}), soit leurs graphes RDF à partir de \textit{Neo4j} (en utilisant \textit{Cypher} par équivalence), tout en donnant, à l'utilisateur, l'impression d'interroger un seul graphe RDF (\textit{i.e.} l'utilisateur peut saisir, à la fois, N entités de N ontologies différentes). 

\begin{figure}[H]
\begin{center}

\includegraphics[scale=0.7]{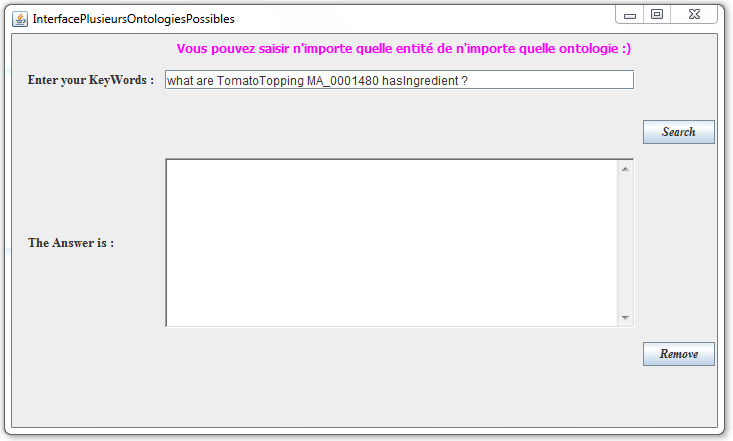}

\end{center}
\caption{Interrogation de plusieurs ontologies \label{figure 3.27}}
\end{figure}

\parindent=0cm
\textcolor {red}{\bf Exemple :} \\
La question d'entrée formulée par l'utilisateur en LN (\textit{e.g.} figure \ref{figure 3.27}), est d'abord traitée et analysée. L'étape suivante consiste à identifier la nature des entités décomposées, afin de les placer dans les partons de triplet (TP) \textit{SPARQL-DL} correspondants (respectivement, dans les requêtes Cypher correspondantes). Ces patrons de requêtes contiennent des emplacements vides (pour les préfixes, puisque nous ne pouvons interroger une ontologie, avec le langage SPARQL-DL, que si en indiquant son IRI, et vu que nous sommes en train d'interroger plusieurs ontologies en même temps, ces emplacements restent vides jusqu'à la détermination des ontologies qui contiennent les différentes entités saisies). \\
En somme, ces emplacements, qui sont des éléments manquants de la requête, doivent être traités par des \textit{IRIs} par rapport aux ontologies interrogées. \\ \\

Après que le système décompose la requête, nous obtenons trois noms d'entités qui sont : \\
- TomatoTopping ; \\
- MA-0001480 ; \\
- hasIngredient. \\

Le système doit déterminer d'abord à quelles ontologies appartienne chacune de ces entités. Cela se fait en comparant chacune d'elles avec les listes stockant les noms des classes, propriétés d'objet, propriétés de type de données et instances, respectivement. Ces listes stockent juste les noms des différentes entités de chaque ontologie, et ont pour intérêt d'aider à la découverte de l'appartenance de chaque entité fournie. Par exemple, dans notre cas, notre algorithme va déterminer que TomatoTopping et hasIngredient appartiennent à l'ontologie Pizza, ainsi que MA-0001480 appartient à l'ontologie Mouse. \\ \\

Une fois cette étape soit terminée, nous devons maintenant reconnaître la nature de chaque entité saisie pour savoir quelles requêtes devons nous exécuter. \\
Revenons à notre exemple, le système va reconnaître que TomatoTopping et MA-0001480 sont des classes, ainsi que hasIngredient est une propriété d'objet. \\
Enfin, les emplacements vides vont être remplis par les préfixes (IRIs) des deux ontologies en question, et les requêtes SPARQL-DL vont être exécutées, générant par la suite des requêtes Cypher. \\

\begin{figure}[H]
\begin{center}

\includegraphics[scale=1.3]{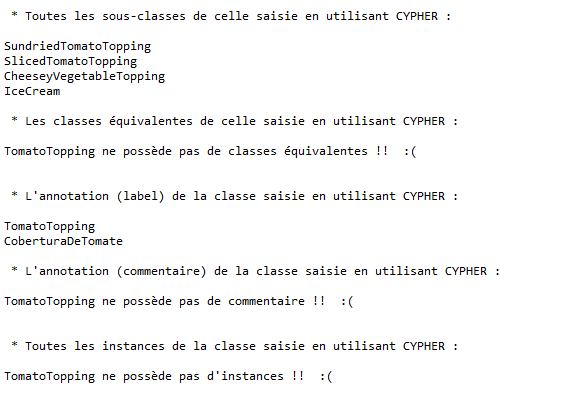}

\end{center}
\caption{Résultats des requêtes Cypher (1) \label{figure 3.28}}
\end{figure}

La figure \ref{figure 3.28} indique les résultats correspondants à la classe TomatoTopping. Ces résultats sont retournés suite à l'exécution des requêtes Cypher sur le graphe RDF de l'ontologie Pizza.

\begin{figure}[H]
\begin{center}

\includegraphics[scale=1.2]{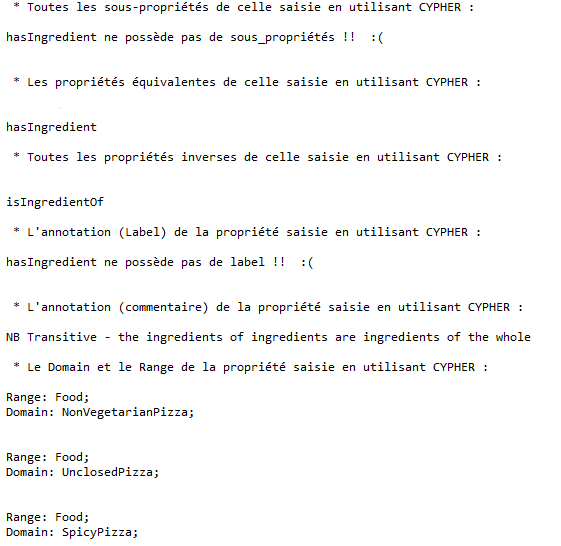}

\end{center}
\caption{Résultats des requêtes Cypher (2) \label{figure 3.29}}
\end{figure}

\parindent=0.5cm
La figure \ref{figure 3.29} affiche les informations de la propriété d'objet fournie (hasIngredient), suite à l'exécution des requêtes Cypher par équivalence sur le graphe RDF de l'ontologie Pizza. \\

\begin{figure}[H]
\begin{center}

\includegraphics[scale=0.9]{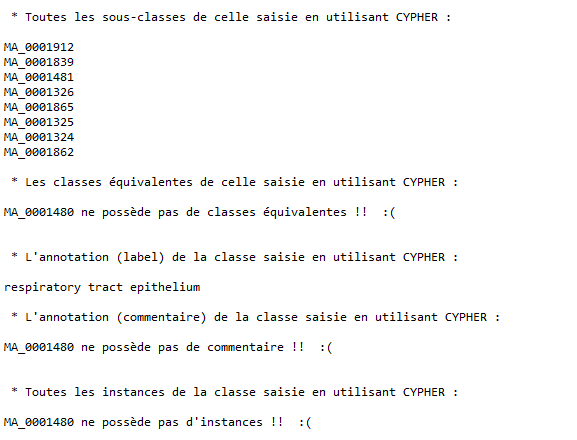}

\end{center}
\caption{Résultats des requêtes Cypher (3) \label{figure 3.30}}
\end{figure}

\parindent=0.5cm
Enfin, la figure \ref{figure 3.30} détermine la réponse correspondante à la classe MA-0001480. Cette réponse est le résultat des requêtes Cypher sur le graphe RDF de l'ontologie Mouse. \\

\section{Conclusion}

\parindent  =0.5cm
En somme, les \textit{Triples Stores} constituent de plus en plus la base de plusieurs applications. Il est alors très important que, pendant l'implémentation de n'importe quelle application, d'avoir une idée claire sur les forces et les faiblesses des implémentations des \textit{TS} actuels. L'efficacité de données joue un rôle important lorsqu'il s'agit de données à grande échelle. En outre, c'est évident qu'il est difficile pour les utilisateurs (même pour les professionnels) de concevoir des requêtes \textit{SPARQL} compliquées. \\

\parindent=0.5cm
Dans ce contexte, notre application fournit un outil pratique permettant de récupérer des données à partir des graphes RDF. Notre interface de recherche rend la procédure d'interrogation d'un graphe RDF suffisamment conviviale. Nous avons utilisé \textit{SPARQL-DL} comme langage d'interrogation, qui est conforme aux recommandations du \textit{W3C} et qui s'avère être un langage de requête pouvant soutenir l'expressivité avancée d'\textit{OWL 2}.  \textit{Hermit} lui-même s'avère être efficace pour soutenir cette expressivité. Sans oublier \textit{Neo4j} qui offre un accès et une gestion de données performants, ainsi qu'une fluidité de représentation guidée par le besoin. \\
Notre application effectue une récupération de données \textit{RDF} de manière efficace avec des temps d'exécution encourageants.















\chapter{Résultats expérimentaux et discussions}

\section{Introduction}

\parindent =0.5cm
Le chapitre précédent a fait l'objet d'une description détaillée de notre méthode d'interrogation de la BD \textit{Neo4j}. Dans ce chapitre, nous allons mettre l'accent sur l'évaluation et l'expérimentation de notre approche. Notre méthode est réalisée en utilisant un ensemble de métriques d'évaluation. Ces dernières permettent d'estimer la qualité des résultats de l'interrogation. L'expérimentation a été effectuée sur plusieurs types d'ontologies. \\

Ce chapitre est organisé comme suit : la première partie décrit l'environnement d'expérimentation et les outils utilisés afin d'implémenter notre méthode. La deuxième partie présente les métriques d'évaluation permettant d'estimer l'efficacité de notre approche. Par la suite, une évaluation sur différentes ontologies de différentes tailles, ainsi qu'une discussion, on été présentées.

\section{Environnement d'expérimentation et outils}

\parindent =0.5cm
Nous allons dans cette section, étudier les différents composants de l'environnement matériel et logiciel avec lesquels nous avons réalisé une étude expérimentale. \\

\parindent =0.5cm 
Notre algorithme a été implémenté en langage Java en utilisant l'\textit{OWL API 3.4.8}. Afin d'ajouter de l'intelligence aux données, il faut passer par une phase d'enrichissement de connaissances dans laquelle une ontologie joue un rôle fondamental. Grâce à ces données intelligentes, n'importe quel système peut réaliser des déductions, qui ne semblent pas évidentes pour un être humain (\textit{e.g.} extraction de résumés, recommandations, détection de conflits, \textit{etc}). C'est dans ce cadre que \textit{OWL API} a été développée. Il s'agit d'un ensemble d'interfaces riches en fonctionnalité, assurant une manipulation flexible des ontologies. Elle nous a permis de parser (analyser) les ontologies d'entrée ainsi d'identifier leurs différentes entités. \\

\parindent=0.5cm
Tout ce travail a été réalisé sous l'environnement de développement\textit{ Eclipse KELPER}. L'environnement \textit{Eclipse} garantie la réutilisation ou la modification de quelques modules faisant partie de notre méthode. \\

\parindent =0.5cm 
Pour le raisonnement sur les ontologies, nous avons utilisé la dernière version officielle, stable et la plus récente \textit{Hermit 1.3.8}. Concernant l'interrogation de l'ontologie directement, avons avons travaillé avec la version \textit{1.0.0} de \textit{SPARQL-DL}. \\
Nous avons également eu recours à la version \textit{3.1.2} de la BD graphique \textit{Neo4j}, afin de stocker les triplets RDF (graphe RDF). Il suffit de l'intégrer en \textit{Java} sous forme d'une bibliothèque (\textit{library}). \\ 

Afin de se connecter à la BD graphique Neo4j, nous devons : \\ 
- Définir un type énuméré implémentant la classe \textit{\textbf{RelationshipType}} (pour décrire, manuellement et à notre choix, toutes les relations possibles entre les nœuds du graphe RDF) ;  \\
- Créer une fonction \textit{\textbf{registerShutDownHook} }chargée de fermer la connexion en cas de problème ; \\
- Effectuer la connexion (\textit{i.e.} créer le graphe grâce à la fonction \textit{\textbf{GraphDatabaseFactory}}).  

\parindent =0.5cm 
\textbf{$\rightarrow$} Les opérations de mise à jour de la BD, se font dans un bloc \textit{« try ... finally »}. \\

L'ensemble de tests a été réalisé sur un \textit{PC} doté d'un système d'exploitation \textit{Windows 7 professionnel} muni d'un processeur \textit{Intel Core i3}, avec une horloge ayant une fréquence \textit{2.30 GHz} et \textit{6 Go} de \textit{RAM}. \\

Nous avons intégré la version \textit{2.0} de la base de données lexicale \textit{WordNet} pour le traitement des requêtes en langage naturel des utilisateurs.

\section{Métriques d'évaluation}

\parindent =0.5cm 
Dans le but de mesurer l'efficacité de notre méthode d'interrogation, un ensemble de métriques d'évaluation est bien nécessaire. Ces mesures d'évaluation incluent le temps du parsing, le temps du stockage et le temps de réponse. \\

\parindent=0cm
- \textbf{Le temps du parsing :} c'est le temps nécessaire pour que toutes les informations de l'ontologie soient analysées et parsées ; \\

- \textbf{Le temps du stockage :} c'est le temps nécessaire pour générer tous les triplets RDF à partir de l'ensemble d'entités résultant de l'étape du parsing, ainsi pour les sauvegarder dans la base de données graphique \textit{Neo4j}. \\

Le stockage se fait au fur et à mesure avec le parsing, alors nous allons les fusionner en une seule mesure : temps du parsing et du chargement ; \\ \\
- \textbf{Le temps de réponse :} c'est le temps nécessaire pour que notre système réponde à une requête utilisateur formulée en langage naturel.  \\

\parindent =0.5cm
Dans tous les cas de mesures, nous allons évaluer la performance de notre méthode en utilisant \textit{Hermit} et sans l'utilisation de \textit{Hermit} (\textit{i.e.} raisonnement par défaut de OWL API); avec des requêtes simples et d'autres compliquées; avec \textit{SPARQL-DL} tout court et avec \textit{Cypher} (tout en passant par l'équivalence décrite dans le précédent chapitre). Enfin, nous allons analyser l'impact de la taille du graphe sur les performances de la BD graphique \textit{Neo4j}, en particulier sur le temps du parsing (de l'ontologie) et du chargement (du graphe RDF), ainsi que sur le temps de réponse à une requête utilisateur.

\section{Évaluation de l'approche}
 
\parindent =0.5cm
Afin d'évaluer la performance et la faisabilité de notre algorithme, nous allons utiliser différentes ontologies de différentes tailles. Dans un premier temps, nous décrivons les caractéristiques des ontologies utilisées. Après, nous notons les résultats obtenus et nous les discutons afin de mettre en relief les points forts et les points faibles de notre méthode.

\subsection{Avec de petites à moyennes ontologies}

\subsubsection{Pizza et la base « Anatomy »}

\parindent =0cm
$\bullet$ \emph{Pizza} est une ontologie bien connue dans la communauté du Web sémantique. Elle a été développée pour des fins éducatives, par l'université de Manchester, qui est une université leader dans le développement des technologies sémantiques. \\
\parindent =0.5cm
$\bullet$ La base \emph{« Anatomy »} est composée de deux ontologies de taille moyenne : « Mouse » qui décrit l'anatomie de la souris adulte, et « Human » qui décrit l'anatomie humaine, (Tableau \ref{Tableau 4.1}) \\

\begin{table}[!ht]
\begin{center}
\begin{tabular}{|l|c|c|c|c|}

\hline
\emph{Ontologies} & \textbf{Classes} & \textbf{Propriétés d'objet} & \textbf{Propriétés de type de données} & \textbf{Instances} \\
\hline \hline \hline
Pizza & 100 & 8 & 0 & 5 \\
\hline \hline
Mouse & 2744 & 3 & 0 & 0 \\
\hline \hline
Human & 3304 & 2 & 0 & 0 \\
\hline \hline

\end{tabular}
\caption{Présentation de Pizza et la base « Anatomy »  \label{Tableau 4.1}}
\end{center}
\end{table}

\subsection{Avec de grandes ontologies}
\subsubsection{Présentation de la base « Large Biomedical Ontologies »}

\parindent = 0.5cm
Dans le contexte du passage à l'échelle (scalabilité), nous avons choisi d'implémenter les ontologies appartenant à la base \textit{« Large Biomedical Ontologies »}, afin d'étudier l'influence de la taille des ontologies sur l'efficacité et la performance de notre système. \\
Cette base contient des ontologies qui sont sémantiquement riches et contiennent de dizaines de milliers de concepts (classes), tableau \ref{Tableau 4.2}. \\ La base \textit{LargeBio} se catégorise en trois classes de taille croissante : \\
- \textit{FMA-NCI \textbf{small} overlapping} ; \\
- \textit{FMA-NCI \textbf{extended} (medium) overlapping} ; \\
- \textit{FMA-NCI \textbf{whole} overlapping}  (la catégorie la plus complète). \\

\begin{table}[!ht]
\begin{center}
\resizebox{\textwidth}{!}{%
\begin{tabular}{|l|c|c|c|c|}

\hline
\emph{Ontologies} & \textbf{Classes} & \textbf{Prop objet} & \textbf{Prop data} & \textbf{Instances} \\
\hline \hline \hline
FMA (\textbf{small}) & 3696 & 0 & 24 & 0 \\
\hline 
NCI (\textbf{small}) & 6488 & 63 & 0 & 0 \\
\hline 
SNOMED (\textbf{small}) & 13 412 & 18 & 0 & 0 \\
\hline \hline \hline
FMA (\textbf{medium}) & 10 157 & 0 & 24 & 0 \\
\hline
NCI (\textbf{medium}) & 23 958 & 82 & 0 & 0 \\
\hline 
SNOMED (\textbf{medium}) & 51 128 & 51 & 0 & 0 \\
\hline \hline \hline
FMA (\textbf{whole}) & 78 988 & 0 & 54 & 0 \\
\hline 
NCI (\textbf{whole}) & 66 724 & 123 & 67 & 0 \\
\hline 
SNOMED (\textbf{whole}) & 122 464 & 55 & 0 & 0 \\
\hline \hline \hline

\end{tabular}}
\caption{Présentation de la base « LargeBio »  \label{Tableau 4.2}}
\end{center}
\end{table}

\subsection{Résultats et interprétations}

\subsubsection{Interrogation d'une seule ontologie - retour de toutes les informations}

\parindent = 0.5cm
Nous allons tout d'abord dégager les temps d'exécution suite à l'interrogation d'une seule ontologie avec le retour de toutes les informations des mots-clés saisis (tableau \ref{Tableau 4.5}, page \pageref{Tableau 4.5}; tableau \ref{Tableau 4.6}, page \pageref{Tableau 4.6}; tableau \ref{Tableau 4.7}, page \pageref{Tableau 4.7} et tableau \ref{Tableau 4.8}, page \pageref{Tableau 4.8}). Pour ce faire, nous avons effectué cinq exécutions simultanées, et nous avons calculé la moyenne de ces cinq mesures à chaque fois en milli-secondes et en minutes. \\

\begin{table}[!ht]
\begin{center}
\renewcommand {\arraystretch }{1.6}
\resizebox{\textwidth}{!}{%
\begin{tabular}{|l||l||l||l||l||l||}

\cline{2-6} \multicolumn{1}{l||}{} & & \multicolumn {2}{|c|}{\bf Temps parsing et stockage} & \multicolumn {2}{|c|}{\bf Temps de réponse} \\
\hline
Ontologie & \No exé & \textbf{avec} Hermit & \textbf{sans} Hermit & avec \textbf{SPARQL-DL} & avec \textbf{Cypher} \\
\hline \hline \hline
\multirow {2} {*} {\emph{Pizza}} & 1. & 2450 ms & 1688 ms & 63 ms & 2773 ms \\
                         & 2. & \textbf{2405 ms} & \textbf{1624 ms} & 64 ms & 2762 ms \\
                         & 3. & 2419 ms & 1682 ms & 79 ms & \textbf{2712 ms} \\
                         & 4. & 2559 ms & 1655 ms & \textbf{63 ms} & 2781 ms \\
                         & 5. & 2781 ms & 1701 ms & 93 ms & 2733 ms \\ \hline
                       \cline{2-6} \multicolumn{1}{l||}{} & \multirow {2}{*}{\bf Moy} & 2522 ms & 1670 ms & 72 ms & 2752 ms \\ \cline{3-6}
                                     \cline{3-6} \multicolumn{1}{l||}{}& & 0.042 min & 0.027 min & 0.0012 min & 0.045 min \\   
\hline \hline \hline
\multirow {2} {*} {\emph{Human}} & 1. & 3512 ms & 1983 ms & 47 ms & 3669 ms \\
                         & 2. & 3420 ms & \textbf{1764 ms} & 50 ms & 3506 ms \\
                         & 3. & \textbf{3325 ms} & 1888 ms & 32 ms & 3056 ms \\
                         & 4. & 3403 ms & 1889 ms & \textbf{16 ms} & \textbf{1359 ms} \\
                         & 5. & 3434 ms & 1904 ms & 63 ms & 2826 ms \\ \hline
                       \cline{2-6} \multicolumn{1}{l||}{} & \multirow {2}{*}{\bf Moy} & 3418 ms & 1885 ms & 41 ms & 2883 ms \\ \cline{3-6}
                                        \cline{3-6} \multicolumn{1}{l||}{} & & 0.056 min & 0.031 min & 0.0006 min & 0.048 min \\ 

\hline \hline \hline
\multirow {2} {*} {\emph{Mouse}} & 1. & 3200 ms & \textbf{1420 ms} & 47 ms & 3310 ms \\
                         & 2. & \textbf{3185 ms} & 1499 ms & 47 ms & \textbf{3154 ms} \\
                         & 3. & 3387 ms & 1624 ms & \textbf{31 ms} & 3208 ms \\
                         & 4. & 3262 ms & 1608 ms & 75 ms & 3230 ms \\
                         & 5. & 3231 ms & 1670 ms & 48 ms & 3294 ms \\ \hline
                       \cline{2-6} \multicolumn{1}{l||}{} & \multirow {2}{*}{\bf Moy} & 3253 ms & 1564 ms & 49 ms & 3239 ms \\ \cline{3-6}
                                   \cline{3-6} \multicolumn{1}{l||}{} & & 0.054 min & 0.026 min & 0.00082 min & 0.053 min \\ 

\cline{2-6}  

\end{tabular}}
\caption{Temps d'exécution de Pizza et de la base « Anatomy » avec retour de toutes les informations \label{Tableau 4.5}}
\end{center}
\end{table}

\begin{table}[!ht]
\begin{center}
\renewcommand {\arraystretch }{1.6}
\resizebox{\textwidth}{!}{%
\begin{tabular}{|l||l||l||l||l||l||}

\cline{2-6} \multicolumn{1}{l||}{} & & \multicolumn {2}{|c|}{\bf Temps parsing et stockage} & \multicolumn {2}{|c|}{\bf Temps de réponse} \\
\hline
Ontologie & \No exé & \textbf{avec} Hermit & \textbf{sans} Hermit & avec \textbf{SPARQL-DL} & avec \textbf{Cypher} \\
\hline \hline \hline
\multirow {2} {*} {\emph{1)}} & 1. &  5009 ms & 5836 ms & 47 ms & 5249 ms \\
                         & 2. & \textbf{4962 ms} & 5621 ms & 48 ms & 7539 ms \\
                         & 3. &  5367 ms &  5541 ms & \textbf{16 ms} & 7385 ms \\
                         & 4. &  5181 ms &  5561 ms & 48 ms & 7347 ms \\
                         & 5. & 5149 ms &  \textbf{5380 ms} &  50 ms & \textbf{5089 ms} \\ \hline
                       \cline{2-6} \multicolumn{1}{l||}{} & \multirow {2}{*}{\bf Moy} & 5133 ms & 27 939 ms & 41 ms & 6521 ms \\ \cline{3-6}
                                     \cline{3-6} \multicolumn{1}{l||}{}& & 0.085 min & 0.465 min & 0.00069 min & 0.108 min \\   
\hline \hline \hline
\multirow {2} {*} {\emph{2)}} & 1. & 4681 ms & \textbf{15 227 ms} & 50 ms & 5760 ms \\
                         & 2. & 2746 ms & 15 529 ms & 47 ms & 5245 ms \\
                         & 3. & 2621 ms & 16 085 ms & 32 ms & \textbf{5151 ms} \\
                         & 4. & \textbf{2527 ms} & 15 622 ms & 62 ms & 5446 ms \\
                         & 5. & 2574 ms & 16 087 ms & \textbf{31 ms} & 5229 ms \\ \hline
                       \cline{2-6} \multicolumn{1}{l||}{} & \multirow {2}{*}{\bf Moy} & 3029 ms & 15 710 ms & 44 ms & 5366 ms \\ \cline{3-6}
                                        \cline{3-6} \multicolumn{1}{l||}{} & & 0.050 min & 0.261 min & 0.00074 min & 0.089 min \\ 

\hline \hline \hline
\multirow {2} {*} {\emph{3)}} & 1. & \textbf{3868 ms} & 5119 ms & 32 ms & 8220 ms \\
                         & 2. & 5429 ms & 4665 ms & \textbf{31 ms }& \textbf{7560 ms} \\
                         & 3. & 7348 ms & 4979 ms & 48 ms & 7586 ms \\
                         & 4. & 4712 ms & 4854 ms & 48 ms & 7773 ms \\
                         & 5. & 3947 ms & \textbf{4433 ms} & 31 ms & 8632 ms \\ \hline
                       \cline{2-6} \multicolumn{1}{l||}{} & \multirow {2}{*}{\bf Moy} & 5060 ms & 4810 ms & 38 ms & 7954 ms \\ \cline{3-6}
                                   \cline{3-6} \multicolumn{1}{l||}{} & & 0.084 min & 0.080 min & 0.00063 min & 0.132 min \\ 

\cline{2-6}  

\end{tabular}}
\caption{Temps d'exécution de la base « LargeBio(\textbf{small}) » avec retour de toutes les informations \label{Tableau 4.6}}
\end{center}
\end{table}

\parindent = 0.5cm
Nous allons, dans ce qui suit, utiliser les ontologies selon ces numérotations. Nous les avons numéroté par ordre croissant : \\
\parindent = 0cm
1) \textit{Oei-FMA-small-overlapping-nci.owl \textbf{(3696 classes)} ;} \\
2)\textit{ Oei-NCI-small-overlapping-fma.owl \textbf{(6488 classes)} ;} \\
3) \textit{SNOMED-small-overlapping-fma.owl \textbf{(13 412 classes)} ;} \\
4) \textit{Oei-FMA-small-overlapping-snomed.owl \textbf{(10 157 classes)} ;} \\
5) \textit{Oei-NCI-small-overlapping-snomed.owl \textbf{(23 958 classes)} ;} \\
6) \textit{Oei-SNOMED-small-overlapping-nci.owl \textbf{(51 128 classes)} ;} \\
7) \textit{Oei-FMA-whole-ontology.owl \textbf{(78 988 classes)} ;} \\
8) \textit{Oei-NCI-whole-ontology.owl \textbf{(66 724 classes)}.} \\ \\

\newpage
\parindent = 0.5cm
Les valeurs en \textbf{gras} représentent les mesures minimales de tous les temps des cinq exécutions. \\ 

\parindent = 0.5cm
Pour les ontologies de petites à moyennes tailles (\emph{Pizza, Human et Mouse}), le temps du parsing et du chargement avec le raisonneur par défaut, est visiblement plus petit, dans tous les essais d'exécution, que celui en utilisant le raisonneur \textit{Hermit}. Le temps du parsing et du chargement sont différents (par rapport au raisonneur) dus aux résultats du parsing différents. Nous constatons que \textit{Hermit} est un raisonneur très performant qui donne des résultats inférés, pour laquelle il prend largement du temps (il déduit des connaissances, \textit{i.e.} la hiérarchie inférée de l'ontologie, les classes équivalentes, \textit{etc}). Par contre, le raisonneur hiérarchique d'\textit{OWL API} est beaucoup plus rapide car il donne des résultats non inférés, \textit{i.e.} des résultats en se basant uniquement sur les assertions écrites dans l'ontologie et il déduit juste la hiérarchie de l'ontologie telle qu'elle est définie.  \\ 

\parindent = 0.5cm
Pour les grandes ontologies, et d'après la comparaison entre le "parsing et le chargement \textbf{avec} \textit{Hermit}" et le "parsing et le chargement \textbf{sans} \textit{Hermit}", nous remarquons que le raisonneur n'affecte pas le temps d'analyse et du stockage, c'est une différence aussi minime et négligeable que dans la plupart du temps, le parsing et le stockage sans \textit{Hermit} dépasse ceux avec \textit{Hermit} (comme pour l'ontologie \no 2). Alors, c'est comme s'il n'y avait aucune différence si nous utilisons un raisonneur ou pas. \\

\begin{figure}[H]
\begin{center}

\includegraphics[scale=0.28]{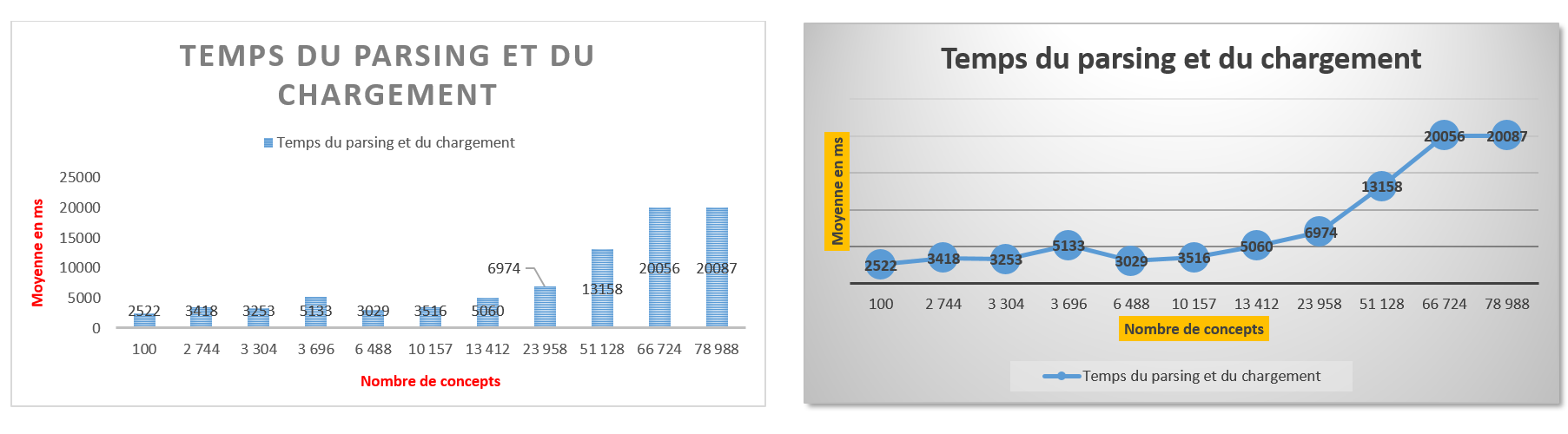}

\end{center}
\caption{Temps du parsing et du chargement \label{figure 4.60}}
\end{figure}

\parindent=0.5cm
Nous pouvons dire aussi que le temps du chargement dans \textit{Neo4j} dépend du nombre de triplets chargés (\textit{i.e.} la taille du graphe RDF à stocker), comme le prouve l'ontologie \no 7 et l'ontologie \no 8, qui contiennent \textit{78 988} et \textit{66 724} concepts (respectivement), et qui ont pris le plus du temps pour le parsing et le chargement, comme l'indique plus clairement la figure \ref{figure 4.60} (tableau \ref{Tableau 4.8}, page \pageref{Tableau 4.8}). Cela est probablement lié à la réorganisation des données sur disque, qui commence à un moment spécifique lors du chargement de l'ensemble de triplets RDF. \\

\begin{table}[H]
\begin{center}
\renewcommand {\arraystretch }{1.7}
\resizebox{\textwidth}{!}{%
\begin{tabular}{|l||l||l||l||l||l||}

\cline{2-6} \multicolumn{1}{l||}{} & & \multicolumn {2}{|c|}{\bf Temps parsing et stockage} & \multicolumn {2}{|c|}{\bf Temps de réponse} \\
\hline
Ontologie & \No exé & \textbf{avec} Hermit & \textbf{sans} Hermit & avec \textbf{SPARQL-DL} & avec \textbf{Cypher} \\
\hline \hline \hline
\multirow {2} {*} {\emph{4)}} & 1. & 3714 ms & \textbf{3091 ms} & \textbf{31 ms} & 3825 ms \\
                         & 2. & 3554 ms & 3214 ms & 31 ms & \textbf{3670 ms} \\
                         & 3. & 3448 ms & 3185 ms & 31 ms & 3934 ms \\
                         & 4. & \textbf{3417 ms} & 3347 ms & 31 ms & 3950 ms \\
                         & 5. & 3449 ms & 3214 ms & 31 ms & 3887 ms \\ \hline
                       \cline{2-6} \multicolumn{1}{l||}{} & \multirow {2}{*}{\bf Moy} & 3516 ms & 3210 ms & 31 ms & 3853 ms \\ \cline{3-6}
                                     \cline{3-6} \multicolumn{1}{l||}{}& & 0.058 min & 0.053 min & 0.00051 min & 0.064 min \\   
\hline \hline \hline
\multirow {2} {*} {\emph{5)}} & 1. & \textbf{6755 ms }& 7020 ms & 125 ms & 25 628 ms \\
                         & 2. & 6914 ms & 6991 ms & 49 ms & 14 227 ms \\
                         & 3. & 7005 ms & 7022 ms & \textbf{31 ms} & \textbf{13 496 ms} \\
                         & 4. & 7144 ms & 6989 ms & 31 ms & 13 928 ms \\
                         & 5. & 7052 ms & \textbf{6848 ms} & 31 ms & 16 739 ms \\ \hline
                       \cline{2-6} \multicolumn{1}{l||}{} & \multirow {2}{*}{\bf Moy} & 6974 ms & 6974 ms & 53 ms & 16 803 ms \\ \cline{3-6}
                                        \cline{3-6} \multicolumn{1}{l||}{} & & 0.116 min & 0.116 min & 0.00089 min & 0.280 min \\ 

\hline \hline \hline
\multirow {2} {*} {\emph{6)}} & 1. & 17 116 ms & 30 579 ms & 62 ms &  3164 ms \\
                         & 2. & \textbf{11 903 ms} & 32 826 ms & \textbf{47 ms} & \textbf{3059 ms} \\
                         & 3. & 12 717 ms & \textbf{29 670 ms} & 47 ms & 3075 ms \\
                         & 4. & 11 981 ms & 30 249 ms & 64 ms & 3090 ms \\
                         & 5. & 12 074 ms & 30 144 ms & 173 ms & 7787 ms \\ \hline
                       \cline{2-6} \multicolumn{1}{l||}{} & \multirow {2}{*}{\bf Moy} & 13 158 ms & 30 693 ms & 78 ms & 4035 ms \\ \cline{3-6}
                                   \cline{3-6} \multicolumn{1}{l||}{} & & 0.219 min & 0.511 min & 0.0013 min & 0.067 min \\ 

\cline{2-6}  

\end{tabular}}
\caption{Temps d'exécution de la base « LargeBio(\textbf{medium}) » avec retour de toutes les informations \label{Tableau 4.7}}
\end{center}
\end{table}

\parindent = 0.5cm
Il y a une autre ontologie qu'on n'a pas mentionné ici (nommée \textit{oaei-SNOMED-extended-overlapping-fma.owl}), elle contient \textit{122 464} concepts et elle est la plus grande de toutes les ontologies de la base "LargeBio", et qu'on n'a pas parvenu à l'exécuter. Nous avons estimé que Neo4j demande plus de \textit{heap space} afin de fonctionner sur de grands ensembles de données (\textit{i.e.} \textit{Neo4j} n'arrive pas à charger tous les triplets). Plusieurs réglages ont été faits sur les options de \textit{JVM} afin d'allouer plus de \textit{heap space}. Cette allocation a permis Neo4j de fonctionner au moins aussi longtemps avant de s'arrêter en raison d'une erreur de mémoire insuffisante.  \\

\parindent=0cm
En fait, Noe4j stocke les données en mémoire, mais au cas où la mémoire est insuffisante, Neo4j fera de son mieux pour charger les données (qui sont fréquemment consultées) en mémoire et le reste sur disque. En outre, puisque Neo4j est écrit en Java, et comme toutes les autres applications écrites en Java, nous devons configurer la JVM pour assurer des performances optimales. \\
\textbf{Remarque :} fournir trop de mémoire pour la JVM peut impliquer des dégradations de performances qui sont dus aux pauses causées par les longs cycles du GC. Cela se produit également avec un espace mémoire trop petit. Plus le heap space est grand, plus le GC prendra du temps pour s'exécuter. \\

\parindent = 0.5cm
En somme, nous pouvons conclure que l'inférence augmente, le plus souvent, le temps du parsing. La taille du graphe a un impact sur les performances de la BD graphique \textit{Neo4j} (elle fonctionne bien avec des données de petites quantités, mais elle reste restreinte avec des données de grandes tailles). L'inférence est un support efficace pour le traitement des requêtes, mais elle nécessite beaucoup d'espace de stockage pour le sauvegarde des données inférées.  \\

\parindent = 0.5cm
Concernant la comparaison entre le temps d'interrogation en utilisant \textit{SPARQL-DL} et celui en utilisant \textit{Cypher} (par équivalence), pour des requêtes simples ainsi que pour des requêtes compliquées, nous constatons que le temps de réponse à des requêtes \textit{Cypher} est toujours plus grand que celui à des requêtes \textit{SPARQL-DL}. 
Cela est à cause de la correspondance des fragments de la requête \textit{Cypher}, qui consiste à déterminer s'il existe une correspondance entre ces fragments (sous-graphes) et le graphe global dans la base de données RDF. De même, les parcours nécessitent l'analyse de grandes parties du graphe, ce qui explique les valeurs plus ou moins élevées des temps de réponse aux requêtes Cypher. \\ 
Nous pouvons dire aussi que ces mesures plus ou moins élevées sont dues à l'équivalence et à la transformation des requêtes \textit{SPARQL-DL} en des requêtes \textit{Cypher}. \\

Nous remarquons également que la performance ne dégrade pas toujours par rapport à la taille du graphe interrogé parce que, pour l'ontologie \no 5,  le temps de réponse à des requêtes \textit{Cypher}, est plus grand que celui des ontologies \no 6, \no 7 et \no 8, qui sont plus grandes qu'elle (figure \ref{Tableau 4.2}). Alors, nous pouvons dire que peut être la performance d'une requête \textit{Cypher} dépend de la façon dont elle est formulée. Cela peut être due, également, au fait que les requêtes \textit{Cypher} prennent du temps à choisir un plan d'exécution favorable.  \\

\begin{figure}[H]
\begin{center}

\includegraphics[scale=0.266]{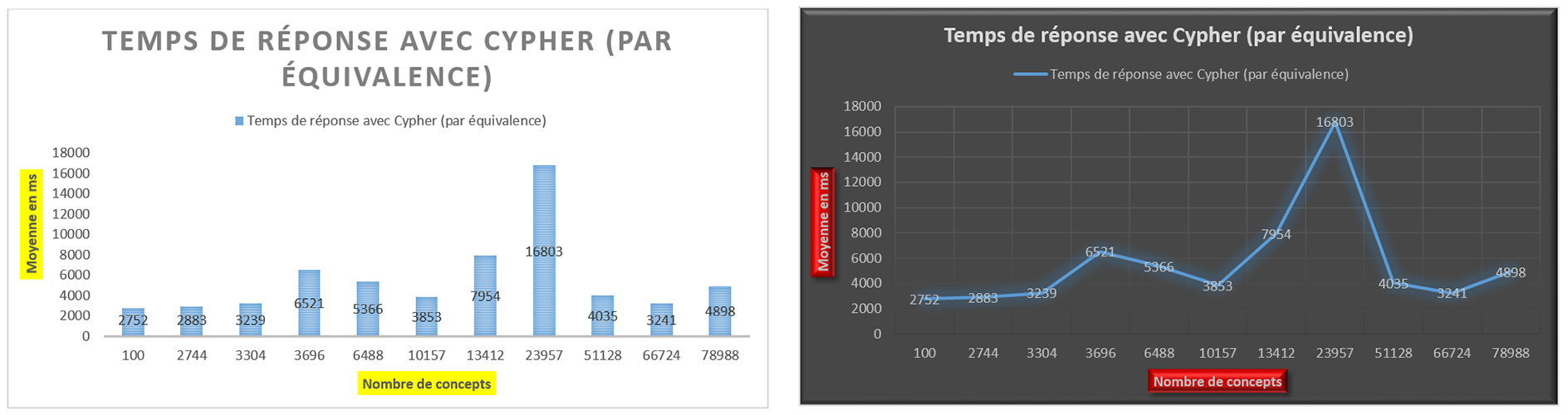}

\end{center}
\caption{Temps de réponse aux requêtes Cypher (par équivalence) \label{figure 4.2}}
\end{figure}

Pour toutes les ontologies, nous avons utilisé le raisonneur \textit{Hermit}, sauf pour les ontologies \no 6, \no 7 et \no 8, parce qu'elles nous posaient des problèmes de mémoire durant le stockage, vu que le raisonneur infère beaucoup de nouvelles données. Cela peut expliquer les valeurs retrouvées des temps de réponse en utilisant Cypher de ces ontologies, qui sont très courtes en les comparant à des valeurs d'autres ontologies plus petites. \\ 

En ce qui concerne la génération et la réponse à des requêtes \textit{SPARQL-DL}, les résultats observés sont prévisibles. N'oublions pas que la génération de requêtes \textit{SPARQL-DL} est logiquement très coûteuse car elle s'adresse à résoudre le problème de segmentation et de correspondance de mots-clés. Cela n'est pas le cas dans notre travail, tel que les résultats des temps de réponse aux requêtes \textit{SPARQL-DL} sont très encourageants et ils ne dépassaient pas les milli-secondes (figure \ref{Tableau 4.5}). \\ 

\parindent=0.5cm
Finalement, nous pouvons déduire que l'augmentation de l'ensemble de données à interroger, n'entraine que des diminutions de performance relativement faibles. \\ 

\begin{figure}[H]
\begin{center}

\includegraphics[scale=0.266]{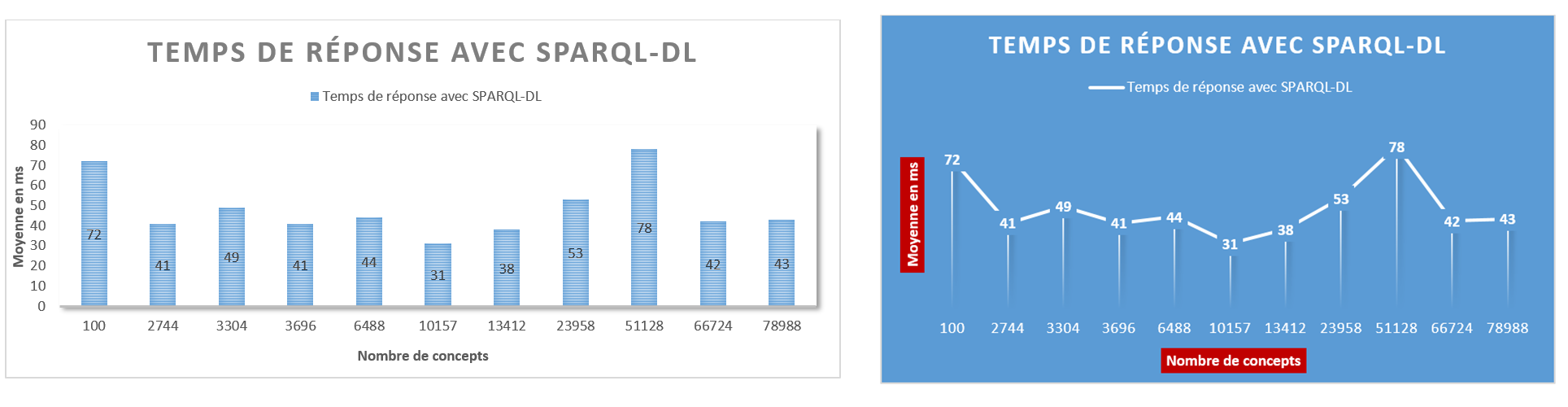}

\end{center}
\caption{Temps de réponse aux requêtes SPARQL-DL \label{figure 4.5}}
\end{figure}

\begin{table}[H]
\begin{center}
\renewcommand {\arraystretch }{1.5}
\resizebox{\textwidth}{!}{%
\begin{tabular}{|l||l||l||l||l||l||}

\cline{2-6} \multicolumn{1}{l||}{} & & \multicolumn {2}{|c|}{\bf Temps parsing et stockage} & \multicolumn {2}{|c|}{\bf Temps de réponse} \\
\hline
Ontologie & \No exé & \textbf{avec} Hermit & \textbf{sans} Hermit & avec \textbf{SPARQL-DL} & avec \textbf{Cypher} \\
\hline \hline \hline
\multirow {2} {*} {\emph{7)}} & 1. & \textbf{18 565 ms} & 26 581 ms & 47 ms &  6044 ms \\
                         & 2. & 20 263 ms &\textbf{18 659 ms} & 31 ms & \textbf{2311 ms} \\
                         & 3. & 20 426 ms & 19 035 ms & \textbf{15 ms} & 2311 ms \\
                         & 4. & 20 216 ms & 19 009 ms & 95 ms & 11 498 ms \\
                         & 5. & 20 965 ms & 18 833 ms &  31 ms & 2326 ms \\ \hline
                       \cline{2-6} \multicolumn{1}{l||}{} & \multirow {2}{*}{\bf Moy} & 20 087 ms & 20 423 ms & 43 ms & 4898 ms \\ \cline{3-6}
                                     \cline{3-6} \multicolumn{1}{l||}{}& & 0.334 min & 0.340 min & 0.00073 min & 0.081 min \\   
\hline \hline \hline
\multirow {2} {*} {\emph{8)}} & 1. & 21 252 ms & 19 768 ms & \textbf{32 ms} &  3246 ms \\
                         & 2. & 21 652 ms & 19 798 ms & 48 ms & 3343 ms \\
                         & 3. & 19 582 ms & \textbf{19 456 ms} & 43 ms & 3213 ms \\
                         & 4. & 19 446 ms & 19 534 ms & 34 ms & \textbf{3163 ms} \\
                         & 5. & \textbf{18 352 ms} & 19 907 ms & 56 ms & 3243 ms \\ \hline
                       \cline{2-6} \multicolumn{1}{l||}{} & \multirow {2}{*}{\bf Moy} & 20 056 ms & 19 692 ms & 42 ms & 3241 ms \\ \cline{3-6}
                                        \cline{3-6} \multicolumn{1}{l||}{} & & 0.334 min & 0.328 min & 0.00071 min & 0.054 min \\ 

\cline{2-6}  

\end{tabular}}
\caption{Temps d'exécution de la base « LargeBio(\textbf{whole}) » avec retour de toutes les informations \label{Tableau 4.8}}
\end{center}
\end{table}

\subsubsection{Une seule ontologie - raffinement du résultat}

\parindent = 0.5cm 
Le tableau récapitulatif (tableau \ref{Tableau 4.9}, page \pageref{Tableau 4.9}) englobe toutes les mesures minimales trouvées des temps de réponse pour chaque ontologie (\textit{i.e.} \textit{« pizza »}, la base « Anatomy » (\textit{« mouse » }et \textit{« human »}), et la base LargeBio (\textit{« FMA », « NCI »}, et \textit{« SNOMED »})). \\

Nous n'avons pas recalculé les temps du parsing et du chargement parce qu'ils restent les mêmes (que ce soit avec ou sans raisonneur). La seule chose qui diffère ici, c'est le temps de réponse. L'utilisateur peut interagir avec le système, sans connaître la façon / le schéma dont les données sont stockées dans la BD graphique Neo4j, il peut réduire l'espace de recherche. Au lieu d'exécuter toutes les requêtes, ce qui entraîne le parcours d'une grande partie du graphe RDF, l'utilisateur et en choisissant son besoin en informations, peut réduire la traversée du graphe et donc le temps de réponse. \\

Ceci explique les mesures trouvées et qui ne diffèrent pas trop de celles de la première version. Cependant, cela n'empêche qu'il y a une amélioration, comme par exemple le temps de réponse à des requêtes Cypher à partir du graphe RDF de l'ontologie \no 5 (NCI-medium), a réduit de 16 803 ms à 9107 ms, et c'est parce que le moteur d'interrogation ne traverse que la partie du graphe contenant le choix de l'utilisateur et non pas tout le graphe ni une grande partie de ce dernier. \\

\begin{table} [H]
\begin{center}
\renewcommand {\arraystretch }{1}
\resizebox{\textwidth}{!}{%
\begin{tabular}{|l||c||c|}
\hline \hline
\backslashbox {ontologie}{ms / min} & \bf avec SPARQL-DL & \bf avec Cypher \\
\hline \hline \hline
\multirow {2} {*} {\emph{Pizza}} & 63 ms & 2169 ms  \\
                         & 0.001 min & 0.036 min  \\
\hline \hline 
\multirow {2} {*} {\emph{Human}} & 31 ms & 2341 ms  \\
                         & 0.0005 min & 0.039 min  \\
\hline \hline 
                         
\multirow {2} {*} {\emph{Mouse}} & 47 ms & 2216 ms  \\
                         & 0.0007 min & 0.036 min  \\
\hline \hline 

\multirow {2} {*} {\emph{1. FMA-\textbf{small}}} &  48 ms & 2376 ms  \\
                         & 0.0008 min & 0.039 min  \\
\hline \hline 

\multirow {2} {*} {\emph{2. NCI-\textbf{small}}} & 25 ms & 3798 ms  \\
                         & 0.0004 min & 0.063 min  \\
\hline \hline 

\multirow {2} {*} {\emph{3. SNOMED-\textbf{small}}} & 47 ms & 2591 ms  \\
                         & 0.0007 min & 0.043 min  \\
\hline \hline 

\multirow {2} {*} {\emph{4. FMA-\textbf{medium}}} & 16 ms & 2514 ms  \\
                         &0.0002 min & 0.041 min  \\
\hline \hline 

\multirow {2} {*} {\emph{5. NCI-\textbf{medium}}} & 78 ms & 9107 ms  \\
                         & 0.0013 min & 0.151 min  \\
\hline \hline 
\multirow {2} {*} {\emph{6. SNOMED-\textbf{medium}}} & 94 ms & 2745 ms  \\
                         & 0.0015 min & 0.045 min  \\
\hline \hline 

\multirow {2} {*} {\emph{7. FMA-\textbf{whole}}} & 48 ms & 3090 ms  \\
                         & 0.0008 min & 0.051 min  \\
\hline \hline 

\multirow {2} {*} {\emph{8. NCI-\textbf{whole}}} & 79 ms & 7488 ms  \\
                         & 0.0013 min & 0.124 min  \\
\hline \hline

\end{tabular}}
\caption{Temps de réponse minimums en utilisant les interfaces à facettes \label{Tableau 4.9}}
\end{center}
\end{table}

\subsubsection{Traitement et interrogation de plusieurs ontologies}

\parindent = 0.5cm
Pour les deux ontologies \textit{FMA-small} et \textit{Mouse}, nous avons fait le même travail (\textit{i.e.} le parsing et le chargement dans deux BD-RDF Neo4j différentes).
Nous constatons d'après le tableau \ref{Tableau 4.10} (page \pageref{Tableau 4.10}), que le parsing et le chargement de l'ontologie \textit{FMA-small} de la base \textit{LargeBio}, prennent plus du temps. Cela est visiblement dû au fait que \textit{FMA-small} est une ontologie plus grande que \textit{Mouse}, et peut être qu'elle demande plus d'inférence. \\

Ces mesures ont été prises suite à l'exécution de la requête en langage naturel suivante : "What are Sacrum MA-0001480 and Lobe-of-prostate"; dont MA-0001480 appartient à l'ontologie Mouse, ainsi que Lobe-of-prostate et Sacrum sont deux entités qui appartiennent à l'ontologie FMA-small. \\

D'un autre côté, le temps de génération de graphe RDF diffère d'une ontologie à une autre. Du coup, nous pouvons trouver deux ontologies qui ont la même taille mais il y a une qui nécessite plus du temps pour la génération de son graphe RDF. C'est parce que la première contient, par exemple, des classes qui utilisent plusieurs propriétés d'objet et la deuxième contient des classes qui utilisent que quelques propriétés d'objet. La génération des triplets de la première demande forcément plus du temps que la deuxième. Cela a un impact sur le temps global du parsing et du chargement. \\

\parindent=0.5cm
Pour l'interrogation, et en comparant les temps de réponse, nous trouvons que la réponse à partir du premier graphe (contenant les données RDF de l'ontologie \textit{FMA-small}) prend plus du temps par rapport à la réponse à partir du deuxième graphe (contenant les triplets de \textit{Mouse}). Cela est apparemment dû au fait que le premier graphe nécessite plus de parcours, vu qu'il est plus grand. \\

\parindent=0.5cm
Nous avons expérimenté plusieurs exemples de requêtes afin d'examiner l'efficacité de notre méthode et l'exactitude des résultats, et nous avons estimé que l'efficacité (en terme du temps) augmente légèrement surtout quand les requêtes contiennent un nombre important d'entités qui appartiennent à différentes ontologies (ou graphes RDF). Cela n'empêche pas que le temps d'exécution global est satisfaisant. \\

\begin{table}[H]
\begin{center}
\renewcommand {\arraystretch }{2}
\resizebox{\textwidth}{!}{%
\begin{tabular}{|c||c||c||c||c||}

\cline{2-5} \multicolumn{1}{l||}{} & \multicolumn {2}{|c||}{\bf Temps Parsing et stockage} & \multicolumn {2}{|c||}{\bf Temps de réponse} \\
\hline
\backslashbox{\No éxecution}{Ontologie} & \textbf{FMA-small} & \textbf{Mouse} & \textbf{Graphe 1} & \textbf{Graphe 2} \\
\hline \hline \hline

1. & 8488 ms & 2856 ms & 16 449 ms & 515 ms \\
2. &  8331 ms & 3434 ms & 16 401 ms & 577 ms \\
3. & 8303 ms & 2888 ms & 12 408 ms & 858 ms \\
4. & 7036 ms & 2904 ms & 7709 ms & 687 ms \\
5. & 8441 ms & 2841 ms & 22 241 ms & 703 ms \\ \hline

\multirow {2}{*}{\bf Moy} & 8119 ms & 2984 ms & 15 041 ms & 668 ms \\ \cline{2-5}
                     
 & 0.135 min & 0.049 min & 0.250 min & 0.011 min \\ 
                  
\hline \hline \hline

\end{tabular}}
\caption{Temps minimums de l'interrogation des deux ontologies Mouse et «FMA-small» \label{Tableau 4.10}}
\end{center}
\end{table}

\section{L'apport de notre travail}

\parindent=0cm
- Utilisation de Neo4j ce qui assure une gestion persistante et efficace des données RDF. Avec Neo4j, le fonctionnement en cluster est possible afin de répartir la charge sur plusieurs machines. De plus, Neo4j respecte les propriétés ACID qui garantissent une transaction informatique fiable. Avec Neo4j, nous pouvons gérer des milliards de données grâce à sa structure en graphe qui s'accompagne d'une rapidité de traitement des requêtes beaucoup plus importante que les SGBD classiques. En outre, Neo4j fournit un moteur puissant accessible facilement, ainsi qu'un langage d'interrogation (Cypher) assez visuel. \\

- Utilisateur non-expert en langage d'interrogation, peut requêter des données structurées, en utilisant le langage naturel. \\

- Notre framework est automatique tel qu'il prend en entrée n'importe quelle ontologie, avec laquelle il produira son graphe RDF et le stocke dans Neo4j. Le graphe RDF résultant est complet et reflète fidèlement la sémantique décrite dans l'ontologie. \\

- Notre méthode est solide et exhaustive, dans le sens qu'elle est complète car elle retourne toutes les réponses correctes à une requête en langage naturel, et elle est également saine parce qu'elle ne retourne que des réponses correctes à une requête utilisateur. \\

- Nous donnons la possibilité à l'utilisateur de choisir son besoin en information grâce à un dialogue / une interaction Homme-Machine, et également de choisir avec quel langage il souhaite avoir ses résultats. \\

- Nous donnons la possibilité aux développeurs de pouvoir changer de base de données. Ceci se fait grâce à l'équivalence qui nous permet d'implémenter l'architecture de notre système au-dessus d'une grande variété de BD-RDF implémentant leurs langages d'interrogation, sans changer aucun des autres composants du système. \\ 

- Nous avons étudié sur les interprétations possibles et souhaitées de la requête utilisateur, afin de mieux le satisfaire. Cela se fait en effectuant toutes les différentes liaisons possibles entre les entités initialement fournies par l'utilisateur. \\

- Possibilité d'interroger l'ontologie directement (avec des requêtes SPARQL-DL), et / ou son graphe RDF dans Neo4j (avec des requêtes Cypher). \\

- Temps d'exécution très court et résultats très encourageants.

\section{Conclusion}

\parindent = 0.5cm
Dans ce chapitre, nous avons décrit une étude expérimentale menée avec notre méthode proposée pour l'interrogation des graphes \textit{RDF}. La solution proposée cherche à contourner le problème de la taille des ensembles de concepts pris comme une entrée pour l'outil d'interrogation. La solution offerte par \textit{Neo4j} dépasse parfois sa capacité de stockage. Les mesures des temps d'exécution prouvent l'efficacité de notre méthode et l'exactitude des résultats retournés.

\addcontentsline{toc}{chapter}{Conclusion Générale}
\chapter*{Conclusion Générale}

\parindent=0.5cm
La nécessité de gérer des données structurées et annotées sémantiquement, d'une manière efficace, se fait de plus en plus pressante. \\
La gestion des données RDF, leur maintenance et le raisonnement qui se fait dessus, sont devenus des travaux coûteux et même irréalisables. Alors, il est devenu indispensable de fournir des solutions pour le stockage des données \textit{RDF} et l'interrogation pertinente de ces dernières. \\

\parindent =0.5cm
Une revue sur quelques méthodes existantes, nous a permis de mettre en relief les exigences des utilisateurs non-experts en termes de stockage et de formulation de requêtes significatives avec des langages d'interrogation hautement spécifiques tels que \textit{SPARQL}.\\ 

\parindent =0.5cm
Dans ce contexte, nous avons développé une méthode pour un stockage et une interrogation efficaces des données RDF. Cette approche permet aux utilisateurs finaux d'interroger des bases de connaissances et des graphes RDF. L'approche guide l'interprétation de la question utilisateur formulée en langage naturel, ainsi que sa traduction en langage formel.\\

\parindent =0.5cm
Le but des travaux, qui ont été réalisés dans le cadre de ce mémoire, est de proposer une nouvelle approche d'interrogation d'ontologies (qui sont elle-mêmes stockées sous forme de graphes RDF, dans la base de données graphique \textit{Neo4j}). Cette méthode devrait surpasser certains handicaps tels que le volume de données, la langue naturelle, l'hétérogénéité qui explique que les informations et les connaissances peuvent provenir de sources différentes, \textit{etc}. \\

\parindent =0.5cm
L'utilisation de bases de données graphiques montre plusieurs avantages, tels qu'elles implémentent leurs langages d'interrogation, des interfaces de programmation (API), \textit{etc}. Elles fournissent également un moyen efficace permettant de représenter des informations inter-connectées ainsi de décrire des structures de données complexes. \\

\parindent=0.5cm
Nous avons mené une étude expérimentale afin de démontrer la pertinence et l'efficacité de notre approche en termes du temps d'exécution et d'exactitude de résultats. \\

\textcolor{red}{\textbf{Perspectives :} }\\

\parindent=0.5cm
- Bien que les résultats expérimentaux soient très encourageants, le problème du passage à l'échelle a un impact sur la performance des systèmes qui réalisent les travaux de stockage et d'interrogation, en particulier Neo4j. De même, puisque la quantité de données RDF continue à augmenter, il n'est plus possible de stocker ni d'accéder à toutes ces données sur une seule machine, tout en assurant des performances raisonnables. Alors, il est nécessaire d'améliorer les capacités de notre algorithme de stockage et d'interrogation, par l'adoption de paradigmes de distribution \textit{(Spark Streaming)}. \\

\parindent=0.5cm
- Améliorer la phase d'intégration de \textit{WordNet} par l'ajout des hyperonymes et des hyponymes.

\nocite{ref1}\nocite{ref2}\nocite{ref3}\nocite{ref4}\nocite{ref5}\nocite{refa}\nocite{ref6}\nocite{ref7}\nocite{ref8}\nocite{ref9}\nocite{ref10}\nocite{ref11}\nocite{refb}\nocite{ref12}\nocite{ref13}\nocite{ref14}\nocite{ref15}\nocite{ref16}\nocite{ref17}\nocite{refc}\nocite{ref18}\nocite{ref19}\nocite{ref20}

\bibliographystyle{apalike}
\bibliography{biblioo}
\addcontentsline{toc}{chapter}{Bibliographie}

\newpage
\thispagestyle{empty}

\parindent=0cm
\textbf{Résumé : } \\

\parindent=0.5cm
Ce travail est accompli dans le cadre d'un projet de mémoire de recherche. L'objectif est de générer des requêtes SPARQL en se basant sur des mots-clés fournis par l'utilisateur, afin d'interroger des graphes RDF. Pour ce faire, nous avons, tout d'abord, transformé l'ontologie d'entrée en un graphe RDF qui reflète la sémantique représentée dans l'ontologie. Par la suite, nous avons stocké ce graphe RDF dans la base de données graphique Neo4j, afin d'assurer une gestion efficace et persistante des données RDF. Au moment de l'interrogation, nous avons étudié les différentes interprétations possibles et souhaitées de la requête initialement formulée par l'utilisateur. Nous avons proposé également d'effectuer une sorte de transformation entre les deux langages d'interrogation SPARQL et Cypher, qui est propre à Neo4j. Ceci nous permet d'implémenter l'architecture de notre système au-dessus d'une grande variété de BD-RDF fournissant leurs langages d'interrogation, sans changer aucun des autres composants du système. Enfin, nous avons testé et évalué notre outil à l'aide de différentes bases de test, et il s'est avéré que notre outil est exhaustive, efficace, et assez performant. \\

\textbf{Mot clés :} Ontologie, Graphe RDF, Interrogation des ontologies, Interrogation des graphes RDF, Langages d'interrogation, Bases de données graphiques, Génération de requêtes, Triples Stores. \\ \\

\rule{\linewidth}{.5pt} \\ \\

\parindent=0cm
\textbf{Abstract : } \\

\parindent=0.5cm
This work is done as part of a research master's thesis project. The goal is to generate SPARQL queries based on user-supplied keywords to query RDF graphs. To do this, we first transformed the input ontology into an RDF graph that reflects the semantics represented in the ontology. Subsequently, we stored this RDF graph in the Neo4j graphical database to ensure efficient and persistent management of RDF data. At the time of the interrogation, we studied the different possible and desired interpretations of the request originally made by the user. We have also proposed to carry out a sort of transformation between the two query languages SPARQL and Cypher, which is specific to Neo4j. This allows us to implement the architecture of our system over a wide variety of BD-RDFs providing their query languages, without changing any of the other components of the system. Finally, we tested and evaluated our tool using different test bases, and it turned out that our tool is comprehensive, effective, and powerful enough. \\

\parindent=0.5cm
\textbf{Keywords :} Ontology, RDF Graph, Query ontologies, Query RDF Graphs, Query Languages, Graphical Databases, Query Generation, Triples Stores. \\ \\

\rule{\linewidth}{.5pt}

\end{document}